\RequirePackage{ifpdf}
\documentclass[12pt,letterpaper,bibliography=totoc,
    listof=totoc,
    final]{JHEP3}

\usepackage{amscd,amsmath,amssymb,amsfonts,xspace,mathrsfs,caption, tabularx,url}
\usepackage[utf8]{inputenc}
\usepackage{graphicx}

 \usepackage{lineno}
 \usepackage{enumerate}
 \usepackage{url}

\usepackage[notref]{showkeys}

\newcolumntype{L}[1]{>{\raggedright\arraybackslash}p{#1}}
\newcolumntype{C}[1]{>{\centering\arraybackslash}p{#1}}
\newcolumntype{R}[1]{>{\raggedleft\arraybackslash}p{#1}}

\newcommand{\be}{\begin{equation}}
\newcommand{\ee}{\end{equation}}
\newcommand{\bes}{\begin{equation*}}
\newcommand{\ees}{\end{equation*}}
\newcommand{\sgn}{\mathrm{sgn}}

\newcommand{\CB}{\mathcal{B}}

\newcommand{\CF}{\mathcal{F}}

\newcommand{\CH}{\mathcal{H}}
\newcommand{\CI}{\mathcal{I}}

\newcommand{\CK}{\mathcal{K}}
\newcommand{\CL}{\mathcal{L}}
\newcommand{\CM}{\mathcal{M}}
\newcommand{\CN}{\mathcal{N}}
\newcommand{\CO}{\mathcal{O}}

\newcommand{\CQ}{\mathcal{Q}}

\newcommand{\CU}{\mathcal{U}}

\newcommand{\IC}{\mathbb{C}}

\newcommand{\BR}{\mathbb{R}}

\newcommand{\BH}{\mathbb{H}}

\newcommand{\BZ}{\mathbb{Z}}
\newcommand{\BQ}{\mathbb{Q}}

\newcommand{\bra}{\langle}
\newcommand{\ket}{\rangle}

\newcommand{\bfb}{{\boldsymbol b}}
\newcommand{\bfc}{{\boldsymbol c}}
\newcommand{\bfe}{{\boldsymbol e}}

\newcommand{\bfmu}{{\boldsymbol \mu}}

\newcommand{\bfx}{{\boldsymbol x}}
\newcommand{\bfy}{{\boldsymbol y}}
\newcommand{\bfrho}{{\boldsymbol \rho}}
\newcommand{\bfk}{{\boldsymbol k}}

\newcommand{\bfp}{{\boldsymbol p}}

\newcommand{\bfz}{{\boldsymbol z}}

\newcommand{\TB}{\bar{\tau}}
\newcommand{\TT}{\tau}

\newcommand{\non}{\nonumber}

\title{Mocking the $u$-plane integral\\
}

\author{Georgios Korpas$^{1,2}$, Jan Manschot$^{1,2}$, Gregory W. Moore$^3$ and Iurii Nidaiev$^3$\\
{\it $^1$ School of Mathematics, Trinity College, Dublin 2, Ireland}\\
{\it $^2$ Hamilton Mathematical Institute, Trinity College, Dublin 2, Ireland}\\
{\it $^3$ NHETC and Department of Physics and Astronomy, Rutgers University, 126 Frelinghuysen Rd., Piscataway NJ 08855, USA}}

\abstract{The $u$-plane integral is the contribution of the Coulomb
  branch to correlation functions of $\CN=2$ gauge theory on a compact four-manifold. We consider the
  $u$-plane integral for correlators of point and surface observables of
  topologically twisted theories with gauge group
  ${\rm SU}(2)$, for an arbitrary
  four-manifold with $(b_1,b_2^+)=(0,1)$.  The $u$-plane contribution equals
  the full correlator in the absence of Seiberg-Witten contributions
  at strong coupling,
  and coincides with the mathematically defined Donaldson invariants in
  such cases. We demonstrate that the $u$-plane
  correlators are efficiently determined using mock modular
  forms for point observables, and Appell-Lerch sums for surface
  observables. We use these results to discuss the asymptotic behavior of
  correlators as function of the number of observables. Our findings
  suggest that the vev of exponentiated point and surface
  observables is an entire function of the fugacities.
}

\preprint{}

\begin{document}

\section{Introduction}
A powerful approach to understand the dynamics of supersymmetric field
theories is to consider such theories on a compact four-manifold
without boundary \cite{Witten:1994ev, Witten:1994cg, Moore:1997pc, LoNeSha,
  Marino:1998bm, Pestun:2007rz, Shapere:2008zf}. We consider in this paper the topologically twisted
counterpart of $\CN=2$ supersymmetric Yang-Mills theory with gauge
group ${\rm SU}(2)$ and in the presence of arbitrary 't Hooft flux \cite{Witten:1988ze}. The gauge group is broken to ${\rm
  U}(1)$ on
the Coulomb branch $\CB$, which is parametrized by the vacuum expectation value
$u=\frac{1}{16\pi^2}\left<\mathrm{Tr}[\phi^2]\right>_{\mathbb{R}^4}$,
where the subscript indicates that this is a vev in a vacuum state of the theory
on flat $\mathbb{R}^4$. The Coulomb branch, also known as the ``$u$-plane'' can be considered as a
three punctured sphere, where the punctures correspond to the weak
coupling limit, $u\to \infty$, and the two strong coupling
singularities for $u=\pm \Lambda^2$.

The contribution of the $u$-plane to a correlation function
$\left< \CO_1\CO_2\dots \right>$ is
non-vanishing if the four-manifold $M$ satisfies the topological condition
$b_2^+(M)\leq 1$, where $b_2^+$ is the number of positive definite
eigenvalues of the intersection form of two-cycles of $M$. For an observable $\CO=\CO_1\CO_2\dots$, the
vev $\left< \CO \right>$ can be expressed as a sum of two
contributions: the Seiberg-Witten
contribution $\left< \CO\right>_{\rm SW}$ from the strong
coupling singularities $u=\pm \Lambda^2$, and the contribution from
the $u$-plane $\Phi[\CO]$,
\be
\left< \CO\right>=\left< \CO\right>_{\rm SW}+\Phi[\CO],
\ee

This paper considers the $u$-plane contribution $\Phi[\CO]$ for compact four-manifolds with
$b_2^+=1$ known as the $u$-plane
integral \cite{Moore:1997pc}.\footnote{The $u$-plane integral also
  contributes for manifolds with $b_2^+=0$. The integrand is one-loop exact
  in this case \cite{Moore:1997pc}, but the one-loop determinants have never been worked out with great care.}
 The integrand of $\Phi[\CO]$ for $b_2^+=1$ does not receive perturbative
 corrections, such that the path integral reduces to a finite dimensional integral over the
 zero modes of the fields. After including the non-perturbative corrections to the integrand using the Seiberg-Witten solution \cite{Seiberg:1994rs},
the $u$-plane integral has been evaluated for some four-manifolds with $b_2=1$ or 2, namely for four-manifolds which are
rational or ruled complex surfaces \cite{Moore:1997pc,
  Gottsche:1996aoa, Marino:1998rg, Malmendier:2008db,
  Malmendier:2010ss, Griffin:2012kw}.
The final expressions appeared to be in terms of mock modular forms \cite{ZwegersThesis, MR2605321},
which could be traced to simplifying features, such as a vanishing chamber,
wall-crossing, or birational transformations. For generic four-manifolds with $b_2^+=1$, these simplifying
features are not available. Nevertheless, we will demonstrate that $u$-plane
integrals of arbitrary four-manifolds with $b_2^+=1$ can be readily evaluated
by integration by parts leading to expressions in terms of mock
modular forms and Appell-Lerch sums. For simplicity, we will restrict to
four-manifolds with $(b_1,b_2^+)=(0,1)$, but not necessarily simply-connected.

\begin{figure}[h]
\includegraphics[scale=0.9]{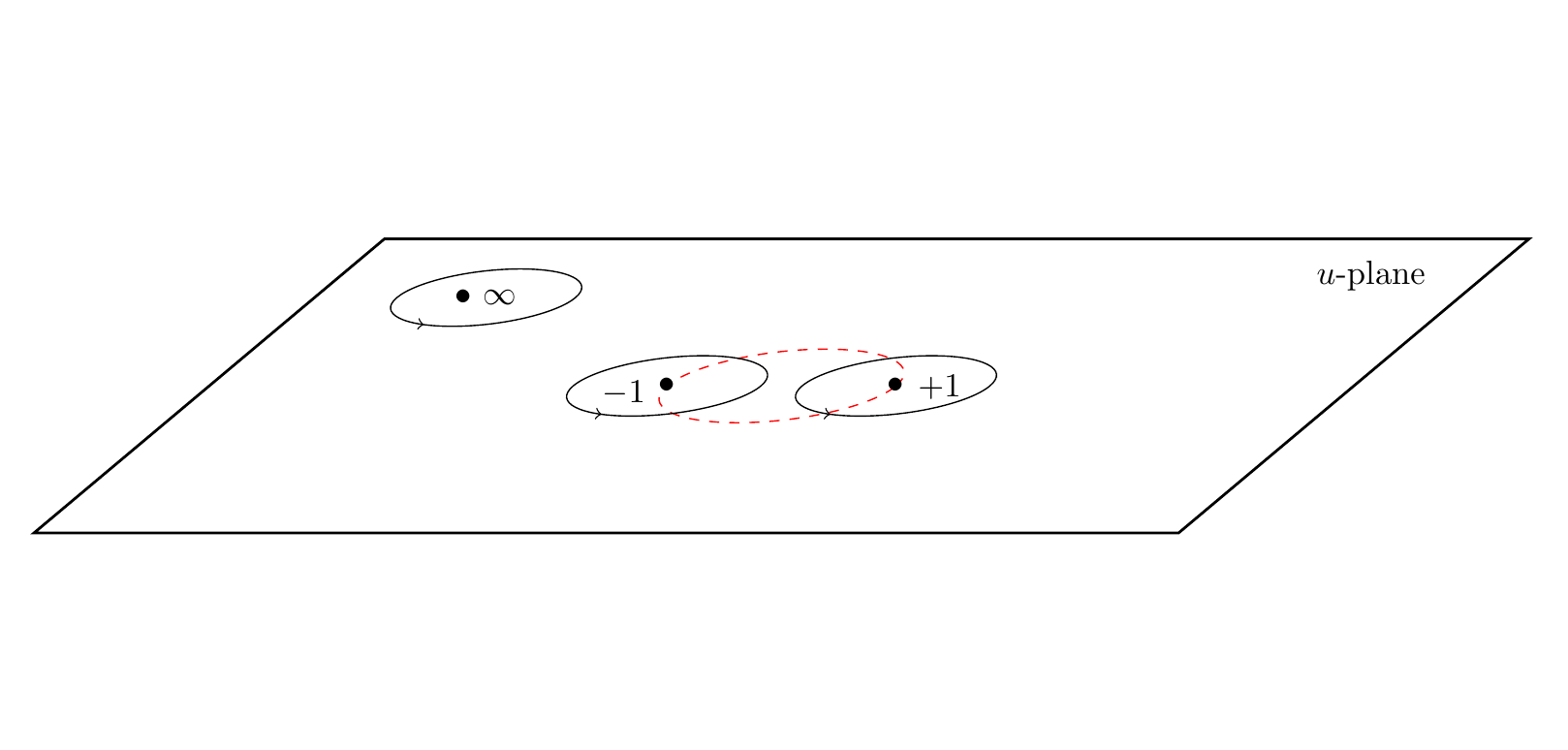}
\centering
  \caption{Schematic representation of the $u$-plane, with the
    singularities $\{\infty, -1, +1 \}$. The black circles indicate
    the boundaries $\partial_j\CB$ of the $u$-plane after removing neighborhoods of the
    singularities, while the dashed circle denotes the wall of
    marginal stability between the strong and weak coupling regions of the $u$-plane.   \label{uPlanePic}}
\end{figure}

To achieve the evaluation of these $u$-plane integrals, we change
variables from $u$ to the effective coupling constant $\tau$, such
that $\Phi$ becomes an integral over the modular
fundamental domain $\mathbb{H}/\Gamma^0(4)$, where $\Gamma^0(4)$ is
the duality group of the theory. We are able to
express the integrand as a total derivative $d\tau\wedge d\bar
\tau\,\partial_{\bar \tau} (\frac{du}{d\tau} H_\CO)$, for some $H_\CO$ which
depends on the observable $\CO$. Reversing the change of variables,
this demonstrates that the integrand takes the form $du\wedge d\bar
u\,\partial_{\bar u} H_\CO$, and the integral is thus reduced to
integrals over the boundaries $\partial_j \CB$, $j=1,2,3$ in
the vicinity of each singularity $\{-1, +1, \infty \}$ by Stokes'
theorem. See Figure \ref{uPlanePic}. More explicitly, we have
\be \label{contour}
\Phi[\CO]= \int_{\CB} du\wedge d\bar u\, \partial_{\bar u}
H_\CO(u,\bar u)= \sum_{j=1}^3 \oint_{\partial_j \CB}  du~H_\CO(u,\bar
u).
\ee
In order for this expression to be useful, it is necessary that $H_\CO(u,\bar u)$,
when expressed in terms of $\tau, \bar\tau$, has good modular properties allowing
one to make the required duality transformation near strong coupling singularities.
We will find, for a special choice of metric, that $H_\CO(\tau, \bar\tau)$ can be expressed
in terms of mock modular forms. Then, given the expression for the wall-crossing formula
using indefinite theta functions \cite{Korpas:2017qdo, Moore:2017cmm} the same result
follows for general metric.

The expression for the $u$-plane integral as a modular integral over
$\mathbb{H}/\Gamma^0(4)$ paves the way for its evaluation. Earlier work has demonstrated that
such modular integrals evaluate to the constant term of a $q$-series, or more
specifically, the $q^0$ term of a mock modular form \cite{Korpas:2019ava, 1603.03056}. We
thus establish a close connection between $u$-plane correlation
functions and mock modular forms. Said more mathematically, we have
established a connection between Donaldson invariants for general manifolds
with $b_2^+=1$ and mock modular forms. The explicit expressions are \eqref{PhiepuFinal}
for manifolds with odd intersection form and just point observables inserted, \eqref{eq:OddLattSurfaceFinal}
for manifolds with odd intersection form and just surface observables inserted, and
\eqref{eq:EvenLattSurfaceFinal} for manifolds with even intersection form and
just surface observables inserted. These expressions hold for a particularly nice choice of
metric. The metric dependence only enters through the choice of \emph{period point}, i.e.  the unique
self-dual degree two cohomology class in the forward light-cone in $H^2(M;\mathbb{R})$. Using the
expression for the wall-crossing formula in terms of indefinite theta functions \cite{Korpas:2017qdo, Moore:2017cmm}
one can produce analogous mock modular forms
relevant to other chambers. Expressions \eqref{PhiepuFinal},
\eqref{eq:OddLattSurfaceFinal} and \eqref{eq:EvenLattSurfaceFinal}  (or close cousins thereof) have appeared before in
\cite{Gottsche:1996aoa}. The derivations in \cite{Gottsche:1996aoa}
relied on the existence of a vanishing chamber and applied wall-crossing
formulae. By contrast, in this paper we evaluate the $u$-plane integral directly,
and do not rely on the existence of a vanishing chamber. Consequently, our
formulae are justified for a larger class of manifolds.

Using the expression for $\Phi[\CO]$ in terms of mock modular forms
(see for example Equation (\ref{PhiepuFinal})), we can address
analytic properties of the correlators for $b_2^+=1$, analogously to
the structural results for manifolds with $b_2^+>1$
\cite{kronheimer1995}. We study the asymptotic
behavior of $\Phi[u^\ell]$ for large $\ell$, and find experimental evidence that
$\Phi[u^\ell]\sim 1/ (\ell\,\log(\ell))$ for any four-manifold with
$(b_1,b_2^+)=(0,1)$. Remarkably, the
asymptotic behavior of $\Phi[u^\ell]$ suggests that
$\Phi[e^{2p\,u}]=\sum_{\ell\geq 0} (2p)^\ell \,\Phi[u^\ell]/\ell!$ is an
entire function of $p$ rather than a formal expansion. We find similar
experimental evidence that the $u$-plane contribution to the
exponentiated surface observable $\Phi[e^{I_-(\bfx)}]$ is an entire
function of $\bfx\in H_2(M,\mathbb{C})$. We leave a more rigorous analysis of these
aspects for future work. The questions we address here would seem to be related to the
analysis of correlation functions of large charge that have recently been
studied in \cite{Grassi:2019txd} and again we leave the investigation of this
potential connection for future work.

One can change variables from $q$ to the complex electric mass $a$ in $\Phi[\CO]$, and express
the $u$-plane integral as a residue of $a$ around $\infty$ and $0$. One may in
this way connect to other techniques for the evaluation of Donaldson
invariants, for examples those using toric localization
\cite{Nekrasov:2003vi, Bershtein:2015xfa}. Our results may also be useful for the evaluation
of Coulomb branch integrals of different theories, such as those including
matter and superconformal theories \cite{Moore:2017cmm}, and for
four-manifolds with $b_1\neq 0$.

The outline of the paper is as follows. Section \ref{SWreview} reviews
Seiberg-Witten theory and its topological twist. Section
\ref{4manifolds} gives a lightning overview of compact four-manifolds with
$b_2^+=1$. Section \ref{PathCor} continues with introducing the path integral and
correlation functions of the theory on these manifolds, which are
evaluated in Section \ref{sec:evaluation}. We close in Section
\ref{asympbehavior} with an analysis on the asymptotic behavior of
correlation functions with a large number of fields inserted.

\section{Seiberg-Witten theory and Donaldson-Witten theory}
\label{SWreview}
We give a brief review of pure Seiberg-Witten theory
\cite{Seiberg:1994rs, Seiberg:1994aj}, and its topologically twisted counterpart aka
Donaldson-Witten theory \cite{Witten:1988ze}. See \cite{Laba05, MooreNotes2017} for a detailed introduction to both of
these theories.

\subsection{Seiberg-Witten theory}
\begin{figure}[h]
\begin{center}
\includegraphics[width=16cm]{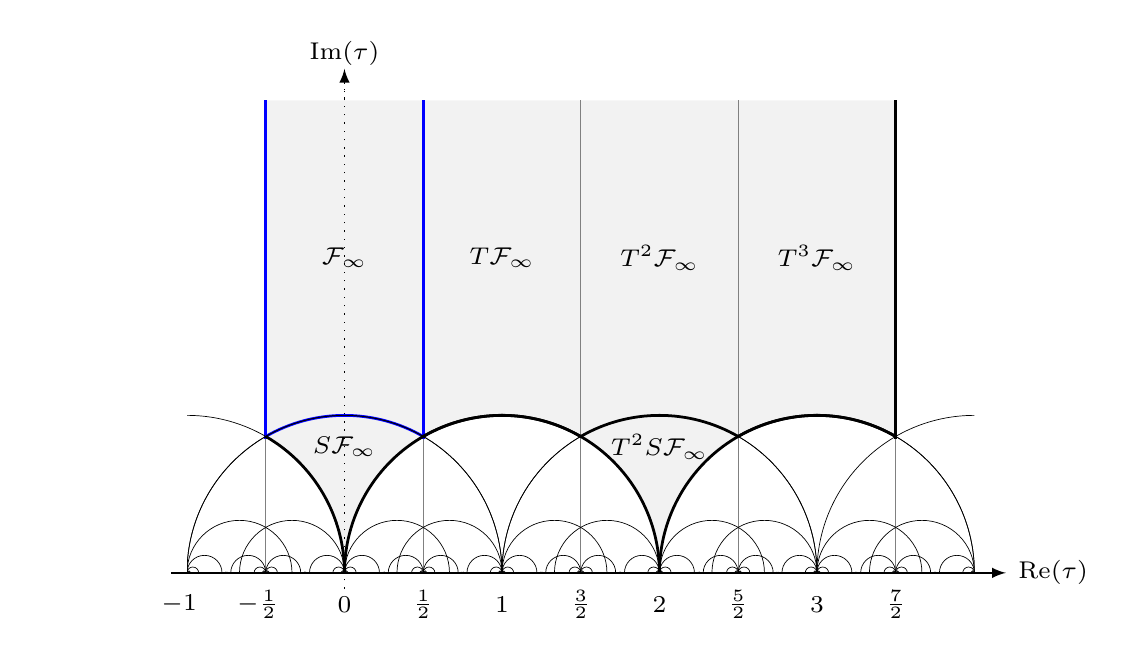}
\caption{Upper-half plane $\mathbb{H}$ with the area bounded by blue
   ($\CF_\infty$) a fundamental domain of $\mathbb{H}/{\rm
     SL}(2,\BZ)$, and the shaded area a fundamental domain of
   $\mathbb{H}/\Gamma^0(4)$. \label{fund_domain}}
\end{center}
\end{figure}

Seiberg-Witten theory is the low energy effective theory of $\CN = 2$
supersymmetric Yang-Mills theory with gauge group $G = {\rm SU}(2)$ or ${\rm SO}(3)$ and Lie algebra $\mathfrak{su}(2)$. The building blocks of the theory contain a $\CN=2$ vector multiplet which consists of a
gauge field $A $, a pair of (chiral, anti-chiral) spinors $\psi$ and  $\bar \psi$, a complex scalar Higgs field $\phi$ (valued in $\mathfrak{su}(2)\otimes \IC$), and an auxiliary scalar field $D_{ij}$ (symmetric in
${\rm SU}(2)_R$ indices $i$ and $j$, which run from $1$ to
$2$). $\CN=2$ hypermultiplets can be included in general. Here we
will consider pure Seiberg-Witten theory with gauge group as above, so
we assume no  hypermultiplets. The gauge group is spontaneously broken
to $\rm{U}(1)$ on the Coulomb branch $\CB$. The pair $(a,a_D)\in
\mathbb{C}^2$ are the central charges for a
unit electric and magnetic charge.
The parameters $a$ and $a_D$ are expressed in terms of the holomorphic
prepotential $\CF$ of the theory
\begin{equation}
a_D=\frac{\partial \CF(a)}{\partial a}.
\end{equation}
Its second derivative equals the effective coupling constant
\be
\tau = \frac{\partial^2 \CF(a)}{\partial a^2}=\frac{\theta}{\pi} + \frac{8\pi i }{g^2} \in \BH,
\ee
where $\theta$ is the instanton angle with periodicity $4\pi$, $g$ is
the Yang-Mills coupling and $\BH$ is the complex upper half-plane. The Coulomb branch $\CB$ is parametrized by the order parameter,
\be
\label{uorder}
u = \frac{1}{16\pi^2}\left\bra \text{Tr}\,[\phi^2] \right\ket_{\mathbb{R}^4},
\ee
where the trace is in the $2$-dimensional representation of $\rm{SU}(2)$. The
renormalization group flow relates the Coulomb branch parameter $u$ and the effective coupling
constant $\tau$. Using the Seiberg-Witten geometry \cite{Seiberg:1994rs}, the order
parameter $u$ can be exactly expressed as a function
of $\tau$ in terms of modular forms,
\be
\label{utau}
\frac{u(\tau)}{\Lambda^2} = \frac{\vartheta_2^4 + \vartheta_3^4}{2 \vartheta_2^2 \vartheta_3^2} = \frac{1}{8}\,q^{-\frac{1}{4}} + \frac{5}{2}\,q^{\frac{1}{4}} - \frac{31}{4}\,q^{\frac{3}{4}} + O(q^{\frac{5}{4}}),
\ee
where $\Lambda$ is a dynamically generated scale, $q=e^{2\pi i
  \tau}$, and $\vartheta_i(\tau)$ are the Jacobi theta functions, which are explicitly given in Appendix
\ref{app_mod_forms}. The function $u(\tau)$ is invariant under
transformations $\tau \mapsto \frac{a\tau+b}{c\tau+d}$ given by
elements of the congruence subgroup $\Gamma^{0}(4) \subset
\mathrm{SL}(2,\BZ)$.\footnote{One way to understand this duality group
  is that the Seiberg-Witten family of curves is the universal family
  of family of elliptic curves with a distinguished order 4 point \cite{Seiberg:1994rs}.}
 See Equation (\ref{Gamma04}) in Appendix \ref{app_mod_forms} for the
 definition of this group. A change of variables from $u$ to $\tau$ maps the $u$-plane to a fundamental domain of
$\Gamma^0(4)$ in the upper-half plane $\mathbb{H}$. We choose the
fundamental domain as the union of the images of the familiar key-hole
fundamental domain of $\mathrm{SL}(2,\BZ)$ under $\tau \mapsto \tau+1$,
$\tau + 2$, $\tau +3$, $-1/\tau$ and $2-1/\tau$, which
is displayed in Figure \ref{fund_domain}. Let $u_D$ be the
vector-multiplet scalar for the dual photon vector multiplet with
coupling constant $\tau_D=-1/\tau$. Then
\be
\label{uD}
\frac{u_D(\tau_D)}{\Lambda^2}=\frac{u(-1/\tau_D)}{\Lambda^2}=\frac{\vartheta_4^4+\vartheta_3^4}{2\vartheta_4^2\vartheta_3^4}=1+32\,q_D+256\,q_D^2+1408\,q_D^3+O(q_D^4).
\ee

At the cusp $\tau\to 0$ (respectively $\tau\to 2$) a monopole (respectively a dyon) becomes massless, and the effective theory
breaks down since new additional degrees of freedom need to be taken
into account. Another quantity which we will frequently encounter is
the derivative $\frac{da}{du}$. It is expressed as function of $\tau$ as
\be
\label{dadu}
\Lambda\, \frac{da}{du}(\tau) = \frac{1}{2}\,\vartheta_2(\tau)\vartheta_3(\tau),
\ee
and transforms under a standard pair of generators of $\Gamma^0(4)$ as
\be
\label{dautrafos}
\begin{split}
&\frac{da}{du}(\tau+4)=-\frac{da}{du}(\tau),\\[0.4em]
&\frac{da}{du}\!\left( \frac{\tau}{\tau+1} \right)=(\tau+1) \frac{da}{du}(\tau).
\end{split}
\ee
Let us also give the expression of the dual of this quantity $(\frac{da}{du})_D$
\be
\label{daduD}
\Lambda \left(\frac{da}{du}\right)_{\!\! D}\!\!(\tau_D)=\tau_D^{-1}\frac{da}{du}(-1/\tau_D)=-\frac{i}{2}\,\vartheta_3(\tau_D)\vartheta_4(\tau_D),
\ee

\subsection{Donaldson-Witten theory}
\label{DonWitten}
Donaldson-Witten theory is the topologically twisted version of Seiberg-Witten theory with gauge group $\mathrm{SU}(2)$ or
$\rm{SO}(3)$, and contains a
class of observables in its $\mathcal{Q}$-cohomology, which famously
provide a physical realization of the mathematically defined Donaldson invariants
\cite{DONALDSON1990257, Donaldson90}.

Topological twisting
preserves a scalar fermionic symmetry $\CQ$ of $\CN=2$ Yang-Mills on an
arbitrary four-manifold\footnote{Note that in \cite{Moore:1997pc,
    Korpas:2017qdo} this operator is denoted as $\overline{\CQ}$.}
\cite{Witten:1988ze}. The twisting involves a
choice of an isomorphism of an associated bundle to the SU(2) $R$-symmetry
bundle with an associated bundle to the frame bundle. Namely, we choose
an isomorphism of the adjoint bundle of the ${\rm SU}(2)_{R}$
${R}$-symmetry bundle with the
bundle of anti-self-dual 2-forms, and we choose a connection on the
$R$-symmetry
bundle, which under this isomorphism becomes the Levi-Civita connection on
the bundle of anti-self-dual 2-forms. In practice, this allows us to replace the
quantum numbers of fields under the $\rm{SU(2)}_{-}\times \rm{SU}(2)_{+}$ factor of the
$\CN=2$ supergroup by the quantum numbers of a diagonally embedded $SU(2)$
group.

The original supersymmetry generators transform as
the $(\bf{1,2,2})\oplus (2,1,2)$ representation of $\mathrm{SU}(2)_{+}
\times \mathrm{SU}(2)_{-} \times \mathrm{SU}(2)_R$ group. Their
representation under the twisted rotation group $\mathrm{SU}(2)'_{+}
\times \mathrm{SU}(2)_{-} \times \rm{U}(1)_{\rm R}$ is $({\bf{1,1}})^{+1} \oplus
({\bf{2,2}})^{-1} \oplus ({\bf{1,3}})^{+1} $. The first
term $({\bf{1,1}})^{+1}$ corresponds to the BRST-type operator ${\CQ}$, whose cohomology
provides operators in the topological field theory. The second
term $({\bf{2,2}})^{-1}$ corresponds to the one-form operator
$K$, which provides a canonical solution to the descent equations
\be
\label{CQO}
\{ \CQ, \CO^{(i+1)} \} = d \CO^{(i)}, \qquad i=0,\dots,3,
\ee
by setting $\CO^{(i)} = K^i \CO^{(0)}$  \cite{Moore:1997pc, LoNeSha, Labastida:1991qq}. Integration of the operators $\CO^{(i)}$ over
$i-$cycles gives topological observables since $\{ \CQ, K \} = d$.

The field content of the topologically twisted theory is a one-form gauge
potential $A$, a complex scalar $a$, together with anti-commuting (Grassmann valued) self-dual
two-form $\chi$, one-form $\psi$ and zero-form $\eta$. The auxiliary
fields of the non-twisted theory combine to a self-dual two-form $D$. The action of the BRST operator $\CQ$ on these fields is given by
\be
\label{barQcomm}
\begin{split}
&[\mathcal{ Q},A]=\psi, \hspace{55pt} [\mathcal{Q},a]=0, \hspace{50pt} [\mathcal{ Q},\bar a]=\sqrt{2}i \eta,  \\
&   [\mathcal{ Q},D]=(d_A\psi)_+, \qquad  \{\mathcal{ Q},\psi\}=4\sqrt{2}\,da, \\
& \{\mathcal{Q},\eta\}=0, \hspace{54pt}    \{\mathcal{ Q},\chi\}=i(F_+-D).
\end{split}
\ee

The low energy Lagrangian of the Donaldson-Witten theory is given by \cite{Moore:1997pc}
\be
\label{Lagr}
\begin{split}
\mathcal{L}_{\rm DW}=&\frac{i}{16 \pi} (\bar \tau F_+ \wedge F_++\tau F_-\wedge F_-)+\frac{y}{8\pi} da\wedge * d\bar a-\frac{y}{8\pi} D\wedge * D\\
&-\frac{1}{16\pi} \tau \psi \wedge * d\eta+\frac{1}{16\pi}\bar \tau \eta \wedge d*\psi+\frac{1}{8\pi}\tau \psi \wedge d\chi-\frac{1}{8\pi}\bar \tau \chi \wedge d\psi\\
&+\frac{i\sqrt{2}}{16\pi} \frac{d\bar \tau}{d\bar a} \eta \chi \wedge (F_++D)-\frac{i\sqrt{2}}{2^7\pi}\frac{d\tau}{da}\psi\wedge \psi\wedge (F_-+D)\\
&+ \frac{i}{3\cdot 2^{11}} \frac{d^2\tau}{da^2} \psi\wedge
\psi\wedge\psi\wedge\psi-\frac{i\sqrt{2}}{3\cdot 2^5 \pi}\left\{\CQ,
  \chi_{\mu\nu} \chi^{\nu\lambda} \chi_{\lambda}^{\,\,\mu}
\right\}\sqrt{g}\,d^4x,
\end{split}
\ee
where $y=\mathrm{Im}(\tau)>0$.


\section{A survey of four-manifolds with $b_2^+=1$}
\label{4manifolds}
We aim to evaluate and analyze the $u$-plane integral for compact
four-manifolds with $(b_1,b_2^+)=(0,1)$ (and without
boundary\footnote{We will only consider four-manifolds without
  boundary in this paper.}). This is a large class of
manifolds which includes among others complex rational surfaces and
examples of symplectic manifolds. The $u$-plane integral is well-defined and can be
evaluated for all these four-manifolds. This section gives a
brief review of the standard geometric aspects of these four-manifolds.

\subsection{Four-manifolds and lattices}

Let $M$ be a compact four-manifold, and let $b_j=\dim(H^j(M,\mathbb{R}))$ be the Betti numbers of
$M$. For simplicity we restrict to manifolds with $b_1=0$, we do not require
them to be simply connected. The torsion subgroups of $H_1(M,\mathbb{Z})$
and   $H^2(M,\mathbb{Z})$ are naturally dual by Poincar\'e duality. They will not play an
important role here, since they simply lead to an overall factor (the order) from the
addition of flat connections.

We denote by $L$ the image of the Abelian group
$H^2(M,\mathbb{Z})\in H^2(M,\mathbb{R})$, which effectively mods out the
torsion in $H^2(M,\mathbb{Z})$. As a result, $L$ is a lattice in a real vector
space, and we can divide elements of $L$ without ambiguity. If the context allows, we will occasionally use $H^2(M,\mathbb{Z})$ and $L$ interchangeably.
The intersection form on $H^2(M,\mathbb{Z})$ provides a
natural non-degenerate bilinear form $B: (L\otimes \mathbb{R})\times (L\otimes \mathbb{R})
\to \mathbb{R}$ that pairs degree two co-cycles,
\be
B(\bfk_1, \bfk_2) := \int_M \bfk_1\wedge \bfk_2,
\ee
and whose restriction to $L \times L$
is an integral bilinear form. The bilinear form provides the quadratic
form $Q(\bfk) :=  B(\bfk, \bfk)\equiv \bfk^2$, which is uni-modular
and possibly indefinite. For later use, recall that a characteristic element of $L$ is an element $\bfc\in
L$, such that
\be
Q(\bfk)+B(\bfc,\bfk)\in 2\mathbb{Z}.
\ee
We let furthermore $H^2(M,\BR)^{\pm}$ be the positive definite and negative definite subspaces
 of $H^2(M,\BR)$, and set $b_2^{\pm}= {\rm
   dim}(H^2(M,\BR)^\pm)$. Van der Blij's Lemma states that a
 characteristic element $\bfc$ of a lattice $L$ satisfies $Q(\bfc)=\sigma_L \mod 8$, where
$\sigma_L=b_2^+-b_2^-$ is the signature of $L$.

The second Stiefel-Whitney class $w_2(T_M)$ is a class in
$H^2(M,\mathbb{Z}_2)$, which distinguishes spinnable from
non-spinnable manifolds. A smooth, spinnable manifold has $w_2(T_M)=0$, while $w_2(T_M)\neq 0$ for non-spinnable manifolds. The class $w_2(T_M)$ has implications for the intersection form of the lattice $L$. In four (but not in higher) dimensions the Stiefel-Whitney
class always has an integral lift. Any integral lift of the Stiefel-Whitney class defines a characteristic
vector in $L$.
Therefore,  $w_2(T_M)=0$ implies that $L$ is an even lattice.
The converse is however only true if $M$ is simply connected due to
the possibility that $w_2(T_M)$ is represented by a torsion class in $H^2(M,\mathbb{Z})$. An even
stronger statement for the intersection form of smooth, simply connected, spinnable
four-manifolds is Rokhlin's theorem, which states that the signature of such manifolds
satisfies $\sigma_L=0\mod 16$. Note that the Enriques surface is smooth while it has intersection
form $\mathbb{I}^{1,1}\oplus L_{E_8}$, where $\mathbb{I}^{1,1}$ is the two-dimensional lattice with quadratic form
$\textstyle{\left( \begin{smallmatrix} 0 & 1 \\ 1 & 0  \end{smallmatrix}\right)}$, and $L_{E_8}$ is minus the $E_8$ root
lattice.  This does not contradict
Rokhlin's theorem since the Enriques surface is not simply connected.
It is also worth noting that for complex manifolds the canonical class $K$ is an integral
lift of the Stiefel-Whitney class and therefore any other integral lift differs by twice
a lattice vector in $L$.

Any closed, orientable four-manifold admits a Spin$^\mathbb{C}$
structure. To a Spin$^\mathbb{C}$ structure one attaches a first Chern class of a
certain line bundle, which we refer to as the first Chern class of the
 Spin$^\mathbb{C}$ structure. The first Chern class of a Spin$^\mathbb{C}$ structure
 is an integral lift of $w_2(T_M)$ and is therefore  a characteristic vector $\bfc \in
L$. Interestingly, the existence of an almost complex structure for a smooth
four-manifold $M$ is related to the existence of a characteristic
vector $\bfc$ with fixed norm. Note that an almost complex structure ensures that
the tangent bundle $T_M$ is complex, such that its Chern class
$c_1(T_M)\in H^2(M,\mathbb{Z})$ and canonical class $K=-c_1(T_M)$ are well-defined. The Riemann-Roch
theorem for four-manifolds with an almost complex structure
demonstrates that its canonical class $K$
is a characteristic element of $L$. Moreover:
\begin{itemize}
\item The modulo 2 reduction of $K$ satisfies
\be
\label{w2mod2}
w_2(T_M)=K \mod 2.
\ee
\item By the Hirzebruch signature theorem
\be
\label{QK}
Q(K)=2\chi+3\sigma,
\ee
where $\chi=2-2\,b_1+b_2$ is the Euler number of $M$, and
$\sigma=b_2^+-b_2^-$ is the signature of $M$.
\end{itemize}
In fact the converse
holds as well: any
characteristic vector $\bfc\in L$, which satisfies (\ref{w2mod2}) and (\ref{QK}),
gives rise to an almost complex structure \cite{Wu:1952, Scorpan}.
Combination of this statement with Van der Blij's Lemma demonstrates that if $M$ admits an
almost complex structure, then $b_2^++b_1$ must be odd.

\subsection{Four-manifolds with $b_2^+=1$}
We will specialize in the following to $b_2^+=1$. In this case, the quadratic form $Q$ can be brought to
a simple standard form \cite[Section 1.1.3]{Donaldson90}, which
will be instrumental to evaluate the $u$-plane integral in Section
\ref{sec:evaluation}. The standard form depends on whether the lattice
is even or odd:
\begin{itemize}
\item If $Q$ is odd, an integral change of basis can bring
the quadratic form to the diagonal form
\be
\label{lattice_odd}
\left< 1 \right> \oplus m \left<-1\right>,
\ee
with $m=b_2-1$. This has an important consequence for characteristic
elements of such lattices. If $K$ is a characteristic element, $\bfk^2+B(K,\bfk)\in 2\mathbb{Z}$ for any $\bfk\in
L$. In the diagonal basis (\ref{lattice_odd}) this equivalent to
$\sum_{j=1}^{b_2} k_j^2 +K_j k_j\in 2\mathbb{Z}$ with
$K=(K_1,K_2,\dots, K_{b_2})$. This can only be true for all $\bfk\in L$
if $K_j$ is odd for all $j=1,\dots, b_2$.
\item If $Q$ is even,  the quadratic form
$Q$ can be brought to the form
\be
\label{lattice_even}
\mathbb{I}^{1,1} \oplus n\, L_{E_8},
\ee
where $\mathbb{I}^{1,1}$ and $L_{E_8}$ as defined above and $n=(b_2-2)/8$. The components $K_{j}$, $j=1,2$ must therefore be even in this basis.
\end{itemize}

Another important aspect of $M$ is its period point $J\in
H^2(M,\mathbb{R})$, which is the generator of  $H^2(M,\mathbb{R})^+$,
normalized such that $Q(J)=1$. The period point depends on the metric
due to the self-duality condition. In fact, the metric dependence in
the expressions below only enters through a choice of $J$.
Using $J$, we can project $\bfk\in L$ to the positive and negative definite subspaces $H^2(M,\mathbb{R})^\pm$: $\bfk_+=B(\bfk,{J})\,{J}$ is the projection of $\bfk$ to
$H^2(M,\mathbb{R})^+$, and $\bfk_-=\bfk-\bfk_+$ is the projection to $H^2(M,\mathbb{R})^-$. Note that these projections are also the
self-dual and anti-self-dual parts of $\bfk$ with
respect to the Hodge $*$-operation.

\subsubsection*{Complex four-manifolds with $b_2^+=1$}
Complex four-manifolds with $b_2^+=1$ are well-studied and classified by
the Enriques-Kodaira classification. This classification starts with
the notion of a minimal complex surface. This is a non-singular surface which can not be obtained from another
non-singular surface by blowing up a point. This is equivalent to the
statement that the surface does not contain rational curves with
self-intersection $-1$ (or $(-1)$-curves).  The Enriques-Kodaira
classification classifies minimal surfaces using the so-called Kodaira dimension.

The relevant surfaces for us are those with $(b_1,b_2^+)=(0,1)$, whose Kodaira dimension is either $-\infty$, 0, 1 or 2:
\begin{itemize}
\item Surfaces with Kodaira dimension $-\infty$ are surfaces whose
  canonical bundle does not admit holomorphic sections. These surfaces are birational
  to more than one minimal surface. The simply connected surfaces with
  $b_2^+=1$ in this family are the rational surfaces, i.e. the complex projective plane
$\mathbb{P}^2$, Hirzebruch surfaces and blow-ups of these
surfaces. A special property of these surfaces are vanishing
chambers where the moduli spaces of instantons are empty. This has been useful for the explicit determination
of partition functions on these geometries, including the $u$-plane integral
\cite{Moore:1997pc, Gottsche:1996aoa, Yoshioka1994, Manschot:2014cca}.
\item Surfaces with Kodaira dimension $0$ are surfaces for which the
  canonical class $K$ satisfies $Q(K)=0$ and $B(K,C)=0$ for any curve
  $C$. If they satisfy in addition $(b_1,b_2^+)=(0,1)$, they are known as
  Enriques surfaces. Their intersection form is $\mathbb{I}^{1,1}\oplus L_{E_8}$. Note that
  this four-manifold is not simply-connected, and that $w_2(T_M)$ is
  represented by a torsion class in $H^2(M,\mathbb{Z})$.
\item Surfaces with Kodaira dimension 1 are surfaces for which the
  canonical class $K$ satisfies $Q(K)=0$, and $B(K,C)>0$ for any curve $C$. Such surfaces are elliptic (but the
  converse is not always true). The Dolgachev surfaces are a
  family of simply-connected surfaces with Kodaira dimension 1.
\item Surfaces with Kodaira dimension $2$ are surfaces of general
  type. If a surface in this class is simply connected with $b_2^+=1$,
  its holomorphic Euler character $\chi_{h}$ equals 1. Their Euler numbers lie between 3 and 11, and there are
  examples for each integer in this set such as the Godeaux and Barlow
  surfaces which both have Euler number 11. See for example \cite{Barth} for a more
  comprehensive list and details.
\end{itemize}

\subsubsection*{Beyond complex four-manifolds}
Although many four-manifolds with $b_2^+=1$ admit an almost complex
structure, most four-manifolds are not complex and their classification is
an important open problem. A distinguished class of
four-manifolds with $b_2^+=1$ are symplectic ones, which partially
overlap with the complex four-manifolds. For a four-manifold to be
symplectic, its period point $J$ must provide a symplectic structure.\footnote{A
  symplectic structure of a four-manifold is given by a two-form
  $\omega$, which satisfies $d\omega=0$ and $\omega\wedge \omega>0$
  for every point on $M$. In other words, $\omega$ is closed and non-degenerate.}
Reference \cite{mcduff:1996} provides a survey of such manifolds. Examples of
symplectic four-manifolds which are not complex are four-manifolds
denoted by $E(1)_N$, that is four-manifolds which are homotopy
equivalent to a rational elliptic surface and whose construction
relies on a fibered knot $N$ in $S^3$ \cite{park_2004, Park2005}. The manifolds in this class
have $b_1=0$. For recent progress on symplectic, non-complex manifolds with Kodaira dimension $1$ ($Q(K)=0$
and $B(K,C)>0$) with $b_1\neq
0$, see \cite{Baldridge:2005}.

\subsection{Donaldson invariants}
Donaldson invariants have been of crucial importance for the
classification of four-manifolds, since they can distinguish among smooth
structures on four-manifolds \cite{Donaldson90, kronheimer1995}. These invariants are based on ASD equations and via the
Donaldson-Uhlenbeck-Yau theorem to semi-stable vector bundles. We
briefly recall the definition of the Donaldson invariants in the formalism of
topological field theory. Let $\CM_{\gamma}$ be the moduli space
of solutions to the ASD equations for gauge group 
${\rm SU}(2)$ or ${\rm SO}(3)$, where $\gamma=(c_1,k)$ represents the topological numbers of the solution, that
is to say $c_1=\frac{i}{2\pi}\mathrm{Tr}(F)\in H^2(M,\mathbb{Z})$ and
$k=\frac{1}{8\pi^2}\int_M \mathrm{Tr}[F^2]$. The map $\mu^\gamma_D:
H_i(M,\mathbb{Q})\to H^{4-i}(\CM_\gamma,\mathbb{Q})$ maps an $i$-cycle on $M$ to a
$(4-i)$-form on $\CM_{\gamma}$.

This map is constructed using the 
universal curvature $\CF$ of the universal
bundle $\CU$ over $M\times \CM_\gamma$ if it exists, which can be
expressed as a formal sum of the fields
of the topological theory $\CF=F+\psi+\phi$ \cite{baulieu1989}. The
class $\mu_D$ is defined in terms of the first Pontryagin class of the universal bundle,
\be
\mu_D=-\frac{1}{4}\,p_1(\CU)=\frac{1}{8\pi^2}\,\mathrm{Tr}[\CF^2],
\ee
where the trace is in the two-dimensional representation of the gauge group, i.e. the fundamental representation
for SU$(2)$ and the spinor representation for SO$(3)$.

We will only consider the image of $\mu_D$ for 0- and 2-cycles of $M$.
Let $\{r_j\}$ be a finite set of points of $M$ and $\bfp = [r_1] + [r_2] + \dots\in
H_0(M,\mathbb{Z})$ the corresponding 0-cycle. Then $\mu_D(\bfp)$
evaluates to
\be
\label{muDs}
\mu_D(\bfp)=\frac{1}{8\pi^2}\,\sum_{j} \mathrm{Tr}[\phi(r_j)^2],
\ee
which we interpret here as a four-form on $\CM_\gamma$. Since the cohomology
class of $\mathrm{Tr}[\phi^2]$ is independent of
position, we can express $\mu_D(\bfp)$ equivalently as
\be
\label{muDs2}
\mu_D(\bfp)= 2\,p(\bfp)\,u,
\ee
with $u$ as in (\ref{uorder}) and $p:H_0(M,\mathbb{Z})\to \mathbb{R}$ the unique linear map
satisfying $p(\bfe)=1$, where $\bfe$ is a generator of
$H_0(M,\mathbb{Z})$. For $\bfx\in H_2(M,\mathbb{Z})$, $\mu_D(\bfx)$
provides similarly a two-form on $\CM_\gamma$. See Equation (\ref{I-})
for the precise expression in terms of the physical fields. Using the linearity of
the map $\mu_D$, we extend the definition of $\mu_D$ from
$H_*(M,\mathbb{Z})$ to $H_*(M,\IC)$.

Using the map $\mu_D$, we can define the Donaldson invariant
$D^\gamma_{\ell,s}(\bfp,\bfx)\in \mathbb{Q} $ as the intersection number
\be
\label{Dinvariant}
D^\gamma_{\ell,s}(\bfp,\bfx)=\int_{\CM_{\gamma}} \mu_D(\bfp)^\ell\wedge
\mu_D(\bfx)^s\in \mathbb{Q}.
\ee
The number $D^{\gamma}_{\ell,s}$ is only non-vanishing if
$4\ell+2s=\mathrm{dim}_\mathbb{R}(\CM_\gamma)$. For smooth
four-manifolds the virtual dimension of the moduli space is 
\be
\dim_\mathbb{R}(\CM_\gamma)=-2p_1- \frac{3}{2}(\chi+\sigma)=8k - \frac{3}{2}(\chi+\sigma).
\ee
and in general this is in fact the dimension. 
For complex surfaces, we can write $2\ell+s
=4k-c_1^2-3\chi_h$ with $k\in \mathbb{Z}$ and $\chi_h$ the
holomorphic Euler characteristic, $\chi_h=(\chi+\sigma)/4$.

\section{Path integral and correlation functions} \label{PathCor}
This section reviews general properties the $u$-plane integral. We will
treat the partition function in Subsection \ref{path_int} and
correlation functions in Subsection \ref{subsec_correlation}.

\subsection{Path integral}
\label{path_int}

We consider Donaldson-Witten theory on a four-manifold $M$ with
$b_2^+=1$ as discussed in the previous section. For the case of pure
SYM with no hypermultiplets we are always free to consider the case where the principal $SO(3)$
gauge bundle has a nontrivial 't Hooft flux $w_2(P) \in H^2(M;\mathbb{Z}_2)$. We choose
an integral lift $\overline{w_2(P)}$ (and we assume such a lift exists) and embed it in $H^2(M;\mathbb{R})$,
and we denote   $\bfmu := \frac{1}{2} \overline{w_2(P)} \in L\otimes \mathbb{R}$. The dependence on the choice
of lift will only enter through an overall sign. The path
integral over the Coulomb branch of Donaldson-Witten
theory, denoted by $\Phi_{\bfmu}^J$, is an integral over the infinite
dimensional field space, which reduces to a finite dimensional integral over
the zero modes \cite{Moore:1997pc}. We restrict for simplicity to
four-manifolds with $b_1=0$, such that there are no zero modes for the
one-form fields $\psi$. The path integral of the effective theory on the Coulomb branch then becomes
\begin{equation} \label{PartitionFunction}
\Phi_{\bfmu}^J=\Lambda^{-3}\sum_{{\rm U}(1)\,\,{\rm fluxes } } \int_{\CB} da\wedge
d{\bar{a}}\wedge dD\wedge d\eta\wedge d\chi \, \,A(u)^{\chi}\,B(u)^{\sigma}\,e^{-\int_{M} \CL_0},
\end{equation}
where $\mathcal{L}_0$ is the Lagrangian (\ref{Lagr}) specialized to
the zero modes including the ones of the gauge field.
The functions $A(u)$ and $B(u)$ are curvature
couplings; they are holomorphic functions of $u$, given by
\cite{Moore:1997pc, Witten:1995gf}
\begin{equation}
\begin{split}
A(u) &= \alpha \left( \frac{du}{da} \right)^{\frac{1}{2}}, \\[0.5em]
B(u) &= \beta\, (u^2 - \Lambda^4)^{\frac{1}{8}}. \
\end{split}
\end{equation}
The coefficients $\alpha$ and $\beta$ are numerical
factors, which we choose to match with results on Donaldson invariants
from the mathematical literature. Note that
$A(u)^{\chi}\,B(u)^{\sigma}$ has dimension $\Lambda^2$ since
$\chi+\sigma=4$. Moreover, $da\wedge
d{\bar{a}}\wedge dD\wedge d\eta\wedge d\chi$ has dimension
$\Lambda$, such that $\Phi_{\bfmu}^J$ (\ref{PartitionFunction}) is
dimensionless.\footnote{The dimensons of $a$, $A_\mu$, $D_{\mu\nu}$,
  $\eta$, $\psi_\mu$ and $\chi_{\mu\nu}$ are respectively $1$, 1, 2,
$\frac{3}{2}$, $\frac{3}{2}$ and $\frac{3}{2}$ in powers of $\Lambda$. The dimension of differential form fields is reduced by their form degree. For
  example, the dimension of $F=dA$ is 0. The dimensions of the differentials $da$, $dD$, $d\eta$ and $d\chi$ are respectively $1$, $0$, $-\frac{3}{2}$ and $\frac{1}{2}$. }
We denote the contribution of the Coulomb
branch to a correlation function $\left< \CO_1 \CO_2\dots \right>_\bfmu^J$ by
$\Phi_\bfmu^J[\CO_1 \CO_2\dots]$. This corresponds to an insertion of
$\CO_1\CO_2\dots$ in the rhs of (\ref{PartitionFunction}) plus possible contact
terms depending on the $\CO_j$.

We proceed by reviewing the evaluation of $\Phi_\bfmu^J$. Integration over $D$, and the fermions $\eta$ and $\chi$ gives
\be
\label{intfermionzeros}
\int dD\wedge d\eta \wedge d\chi\,e^{\frac{y}{8\pi} \int_M D\wedge *D-\frac{\sqrt{2}i}{16\pi} \frac{d\bar \tau}{d\bar a} \eta
\chi\wedge (F_++D)} =- \frac{\pi}{\sqrt{y}}
\frac{d\bar \tau}{d\bar a} B(\bfk,J).
\ee
where the vector $\bfk$ equals $[F]/4\pi$ and represents a class in
$L+\bfmu$ with $\bfmu\in  L/2$. The factor $\frac{d\bar \tau}{d\bar
  a}$ suggests that it is natural to change variables from $a$ to
the effective coupling constant $\tau\in \BH/\Gamma^0(4)$ in
(\ref{PartitionFunction}). To this end, we define the holomorphic ``measure factor"
\begin{equation}  \label{measureterm}
\tilde{\nu}(\tau) := \Lambda^{-3}\,2\sqrt{2}\pi i \, A^\chi\, B^\sigma\, \frac{da}{d\tau},
\end{equation}
so that Equation (\ref{PartFun}) below will hold. Using Matone's relation \cite{Matone:1995rx}
\begin{equation}
\frac{du}{d\tau} = \frac{4\pi }{i}(u^2-\Lambda^4)\left( \frac{da}{du} \right)^2,
\end{equation}
and (\ref{utau}) and (\ref{dadu}), we can express $\tilde \nu$ in terms
of modular functions
\be
\label{tnumod}
\tilde \nu(\tau)=-\frac{i}{8} \frac{\vartheta_4^{13-b_2}(\tau)}{\eta^{9}(\tau)},
\ee
where we fixed the constants $\alpha$ and $\beta$.\footnote{The values
of $\alpha$ and $\beta$ are slightly different from those quoted in
\cite{Korpas:2017qdo}, since we have used a different normalization for the
integral over $D$.} The modular transformations of $\tilde \nu$ for the two generators $ST^{-1}S: \tau
\mapsto \frac{\tau}{\tau+1}$ and $T^4: \tau \mapsto \tau+4$ of
$\Gamma^0(4)$ are:
\begin{equation}
\label{nutrafos}
\begin{split}
&\tilde {\nu}\!\left(\frac{\tau}{\tau+1}\right)=(\tau+1)^{2-b_2/2} e^{-\frac{\pi i \sigma}{4}} \tilde {\nu}(\tau),\\[0.5em]
&\tilde {\nu}(\tau+4)=-\tilde {\nu}(\tau).
\end{split}
\end{equation}
The measure $\tilde{\nu}(\tau)$ behaves near the weak coupling cusp
$\tau\to i\infty$ as $\sim q^{-\frac{3}{8}}$. Near the monopole
cusp, we have $\tilde \nu(-1/\tau_D)=(-i\tau_D)^{2-b_2/2}\,\tilde
\nu_D(\tau_D)$ with
\be
\tilde \nu_D(\tau_D)=-\Lambda^{-3}\,8i(u_D^2-\Lambda^4) \left(\frac{da}{du}\right)_D\,\vartheta_2(\tau_D)^\sigma,
\ee
whose $q_D$-series starts at $q_D^{1+\frac{\sigma}{8}}$.

The photon path integral takes the form of a Siegel-Narain theta function with kernel $\mathcal{K}$
\begin{equation} \label{PsiK}
\Psi^J_\bfmu\left[ \CK \right](\tau,\bar \tau)=\sum_{\bfk\in L +
  \bfmu}\,\CK(\bfk)\, (-1)^{B(\bfk,K)}\,q^{-\frac{\bfk_-^2}{2}} \bar q^{\frac{\bfk_+^2}{2}},
\end{equation}
where $K$ is a characteristic vector for $L$ corresponding to an almost
complex structure or Spin$^\mathbb{C}$ structure.\footnote{Note that, compared to
 equation $(3.13)$ of
   \cite{Moore:1997pc} there is an overall phase difference. This phase difference
   can be written as $\exp[ \frac{i \pi}{2} \left( \bfk_0^2 -
    B(\bfk_0, K)\right)]$, where $\bfk_0$ is a  lift of
 $w_2(P)$ to $H^2(M,\mathbb{Z})$.   Because $K$ is a characteristic
 vector this factor is a $\bfk_0$-dependent sign. The choice of sign is related to a choice of
   orientation of instanton moduli space.}
If one considers correlation functions rather than the partition function, the sum over
$\rm{U}(1)$ fluxes can be expressed as $\Psi^J_\bfmu\left[ \CK
\right]$, with the kernel $\CK$ dependent on the fields in the
correlation function \cite{Korpas:2019ava}. For the partition function,
the factor (\ref{intfermionzeros}) leads to $\Psi^J_\bfmu\left[ \CK_0
\right]$ with
\begin{equation}
\label{CK0}
\CK_0(\bfk)=\tfrac{i}{2\sqrt{2y}}\,B(\bfk, {J}),
\end{equation}
where we have left out the factor $\frac{d\bar \tau}{d\bar a}$, which
provides the change of variables from the Coulomb branch parameters to
a fundamental domain of $\Gamma^0(4)$ in $\mathbb{H}$. Combining all ingredients, we arrive at the following expression for $\Phi_\bfmu^J$,
\begin{equation} \label{PartFun}
\Phi_\bfmu^J=\int_{\BH/\Gamma^0(4) } d\tau \wedge d\bar{\tau} \, \tilde{\nu}(\tau) \Psi_{\bfmu}^J[\CK_0](\tau, \bar{\tau}).
\end{equation}

An important requirement for (\ref{PartFun}) is the modular
invariance of the integrand under $\Gamma^0(4)$ transformations. We can easily
determine the modular transformations of $\Psi_{\bfmu}^J[\CK_0]=:\Psi_{\bfmu}^J$ from
those of $\Psi_{\bfmu}^J[1]$ (\ref{Psitrafos}). The effect of
replacing 1 by $\CK_0$ in $\Psi_{\bfmu}^J[1]$  is to increase
the weight by $(\frac{1}{2},\frac{3}{2})$. (The factor $1/\sqrt{y}$
contributes $(\frac{1}{2},\frac{1}{2})$ and $B(\bfk, J)$
contributes $(0,1)$ to the total weight.) We then arrive at
\begin{equation}
\label{PsiJtrafos}
\begin{split}
&\Psi_{\bfmu}^J\!\left(\frac{\tau}{\tau+1},\frac{\bar \tau}{\bar
    \tau+1} \right) = (\tau+1)^{\frac{b_2}{2}} (\bar{\tau}+1)^2
e^{\frac{\pi i}{4}\sigma}\, \Psi_{\bfmu}^J(\tau,\bar \tau),\\[0.5em]
&\Psi_{\bfmu}^J(\tau+4,\bar \tau+4) =e^{2\pi i B(\bfmu,K)}\, \Psi_{\bfmu}^J(\tau,\bar \tau),
\end{split}
\end{equation}
where we used that $Q(K)=\sigma \mod 8$. Combining (\ref{nutrafos})
and (\ref{PsiJtrafos}), we deduce that the integrand of
(\ref{PartFun}) is invariant under the $\tau\mapsto
\frac{\tau}{\tau+1}$ transformation. Moreover, the integrand is
invariant under $\tau \mapsto \tau+4$ if $B(\bfmu, K)=\frac{1}{2} \mod
\mathbb{Z}$. However, if $B(\bfmu, K)=0 \mod
\mathbb{Z}$, the integrand is multiplied by $-1$ for $\tau\mapsto
\tau+4$. Since $\Psi_{\bfmu}^J$ vanishes identically in the latter case
case, there is no violation of the duality.

We conclude therefore that the Coulomb branch
integral (\ref{PartFun}) is well defined since the measure $d\tau
\wedge d\bar{\tau}$ transforms as a mixed modular form of weight $(-2,-2)$ while
the product $\tilde{\nu}\, \Psi_{\bfmu}^J$ is a mixed modular
form of weight (2,2) for the group $\Gamma^0(4)$ making the integrand
modular invariant. We close this subsection with Table \ref{Table} that collects the weights
of the various modular forms that appear in the context of $u$-plane integrals.
Evaluation of the integral is postponed to Section
\ref{sec:evaluation}.

\begin{center}
\begin{table}[]
\centering
\begin{tabular}{| l | r |}   \hline
Ingredient & Mixed weight   \\ \hline
$d \tau \wedge d\bar{\tau}$ & $(-2,-2)$  \\
 $y$ & $(-1, -1)$ \\
  $\partial_{\bar{\tau}}$ & raises $(\ell,0)$ to $(\ell,2)$ \\
 $\tilde{\nu}(\tau)$ & $ (2-b_2/2,0) $ \\
 $ \Psi_{\bfmu}^{J}[\CK_0] $ &  $(b_2/2,2)$   \\
\hline
\end{tabular}
\caption{Modular weights of various ingredients for the $u$-plane
  integral. Transformations are in
  $\mathrm{SL}(2,\mathbb{Z})$ for the first three rows, while in
  $\Gamma^0(4)$ for the last two rows.}
\label{Table}
\end{table}
\end{center}

\subsection{Correlators of point and surface observables}
\label{subsec_correlation}

Much more information about the theory is obtained if we include observables in the
path integral \cite{Moore:1997pc, Witten:1988ze}, which contain
integrals over positive degree homology cycles of the four-manifold $M$.
Since we restrict to four-manifolds with $b_1=b_3=0$, we will focus in this article on surface
observables involving integrals over elements of $H_0(M,\BQ)$ and $H_2(M,\BQ)$.

The Donaldson invariants are correlation functions of observables in Donaldson-Witten
theory. The canonical UV surface observable of Donaldson-Witten theory is
defined using the descent operator $K$ mentioned below Eq. (\ref{CQO}),
\begin{equation}
\label{I-}
I_-(\bfx)= \int_\bfx K^2 u =\frac{1}{4\pi^2} \int_{\bfx} \mathrm{Tr}\!\left[ \frac{1}{8} \psi\wedge \psi -\frac{1}{\sqrt{2}}\phi F \right],
\end{equation}
with $\bfx \in H_2(M,\BQ)$. The Donaldson invariant
$D^{\gamma}_{\ell,s}$  (\ref{Dinvariant}) can be expressed as a correlation function of
the twisted Yang-Mills theory,
\be
D^{\gamma}_{\ell,s}(\bfp,\bfx)=\Lambda^{-2\ell-s} \left< (2\,p(\bfp)\,u)^\ell\,
  (I_-(\bfx))^s  \right>^J_\bfmu
\ee
where on the rhs, $\gamma=(2\bfmu,k)$ with $k\in
\mathbb{Z}-2\bfmu^2$. The map $p:H_0(M,\mathbb{Z})\to \mathbb{R}$ was
introduced below (\ref{muDs}).

Note that $D^{\gamma}_{\ell,s}(\bfp,\bfx)\in \mathbb{Z}$ if $\bfp/4\in H_0(M,\mathbb{Z})$
and $\bfx /2\in H_2(M,\mathbb{Z})$, since the coefficients of $u(\tau)$
are in $\mathbb{Z}/8$ and the flux $[F]/2\pi\in H^2(M,\mathbb{Z})$. It is natural to form a generating function of correlation functions by
including exponentiated observables in the path integral
\be
\left< e^{2p(\bfp)\,u/\Lambda^2+\,I_-(\bfx)/\Lambda}  \right>^J_\bfmu=\sum_{k,\ell,s}\frac{D^{\gamma}_{\ell,s}(\bfp,\bfx)}{\ell!\,s!}.
\ee
We will often suppress the argument of $p$, and consider it simply as
a fugacity in which we can make a (formal) series expansion.

In the effective theory in the infrared, the operator $I_-(\bfx)$ becomes
\begin{equation} \label{Imin}
\tilde I_-(\bfx)=\frac{i}{\sqrt{2}\,\pi}\int_\bfx \left( \frac{1}{32} \frac{d^2 u}{d a^2} \psi\wedge \psi -\frac{\sqrt{2}}{4} \frac{du}{da} (F_-+D)\right).
\end{equation}
Inclusion of this operator in the path integral gives rise to a
contact term in the IR, $e^{\bfx^2\,T(u)}$ \cite{Moore:1997pc, LoNeSha}, with
\begin{equation}
\begin{split}
T(u)&=-\frac{1}{2\pi i\,\Lambda^2}\left(\frac{du}{da}
\right)^2\,\partial_\tau\, \mathrm{\log}\,\vartheta_4\\
& = q^{\frac{1}{4}}-2\,q^{\frac{3}{4}}+O(q^{\frac{5}{4}}),
\end{split}
\end{equation}
where $\vartheta_4$ is the fourth classical Jacobi theta
function. The dual contact term reads
\be
\begin{split}
T_D(u_D)&=-\frac{1}{2\pi i\,\Lambda^2}\left(\frac{du}{da}
\right)_{\!\!D}^2\,\partial_\tau\, \mathrm{\log}\,\vartheta_2\\
&=\frac{1}{2}+8\,q_D+48\,q_D^2+O(q_D^3).
\end{split}
\ee

We include moreover the $\CQ$-exact operator $I_+(\bfx)$ \cite{Korpas:2017qdo},
\begin{equation} \label{I+}
I_{+}(\bfx) = -\frac{1}{4\pi} \int_{\bfx} \{ \CQ, {\rm Tr}[\bar \phi \chi]\},
\end{equation}
which can aid the analysis in the context of mock modular forms. As
explained in \cite{Korpas:2019ava}, addition of this observable to
$I_-(\bfx)$ does not change the answer, once the integrals over the
$u$-plane are suitably defined. And more generally, if we add
$\alpha\,I_{+}(\bfx)$, the integral is independent of $\alpha$. Nevertheless, the
{\it integrand} depends in an interesting way on $\alpha$. We will discuss this in 
more detail in Subsection \ref{asymp_behavior}. Here we will continue with $\alpha=1$.
In the effective infrared theory, $I_{+}(\bfx)$ becomes
\begin{equation} \label{Iplus}
\tilde I_+(\bfx)=-\frac{i}{\sqrt{2}\,\pi} \int_\bfx \left(
  \frac{1}{2} \frac{d^2\bar u}{d\bar a^2}\,\eta\,
  \chi+\frac{\sqrt{2}}{4}\,\frac{d\bar u}{d\bar a} (F_+-D) \right).
\end{equation}

With (\ref{Imin}) and (\ref{Iplus}), we find that the contribution of
the Coulomb branch to $\left< e^{I_-(\bfx)+I_+(\bfx)}\right>^J_\bfmu$ reads
\begin{equation}
\label{path_int_surf}
\begin{split}
\Phi_{\bfmu}^J\!\left[e^{I_-(\bfx)/\Lambda+I_+(\bfx) /\Lambda}\right]&=\Lambda^{-3}\sum_{{\rm U}(1)\,\,{\rm fluxes } } \int_{\CB} da\wedge
d{\bar{a}}\wedge dD\wedge d\eta\wedge d\chi ~
A(u)^{\chi}\,B(u)^{\sigma}\\
&\quad \times \,e^{-\int_{M} \CL_0  + \tilde{I}_-(\bfx)/\Lambda +
  \tilde{I}_+(\bfx)/\Lambda + \bfx^2\,T(u)} ,
\end{split}
\end{equation}

As a first step towards evaluating this integral, we carry out the integral over $D$. If we just consider the terms in (\ref{path_int_surf}) that
depend on $D$, this gives
\begin{equation}
\label{intoutD}
2\pi i \sqrt{\frac{2}{y}} \exp\!\left(-2\pi y\,
  \bfb_+^2+\frac{i\sqrt{2}}{4} \frac{d\bar \tau}{d\bar a} \int_M
  \bfb_+\wedge \eta\chi  \right).
\end{equation}
where we have defined $\bfb\in L\otimes \mathbb{R}$ through
\begin{equation}
\bfrho=\frac{\bfx}{2\pi\,\Lambda}\,\frac{du}{da},\qquad \bfb=\frac{\mathrm{Im}(\bfrho)}{y}.
\end{equation}
The variable $\bfrho$ transforms with weight $-1$. With this
normalization, it will appear as a natural elliptic variable in the sum over
fluxes. The dual variable  is
\be
\bfrho_D(\tau_D)= \tau_D\, \bfrho(-1/\tau_D)=
\frac{\bfx}{2\pi\,\Lambda}\,\left(\frac{du}{da}\right)_{\!\!D},
\ee
where $\left(\frac{du}{da}\right)_{\!\!D}$ is given in (\ref{daduD}).

Substitution of (\ref{intoutD}) into the path integral and integration
over the $\eta$ and $\chi$ zero modes modifies the sum over the ${\rm U}(1)$
fluxes  to $\Psi^J_\bfmu[\CK_{\rm s}]$ (\ref{PsiK}) where the kernel
$\CK_{\rm s}$ given by \cite{Moore:1997pc, Korpas:2017qdo}
\be
\CK_{\rm s}=\exp\!\left(-2\pi y\, \bfb_+^2 -2\pi i B(\bfk_-,\bfrho)-2\pi i B(\bfk_+,\bar \bfrho)\right) \partial_{\bar \tau}
\left(\sqrt{2y}\,B(\bfk+\bfb,J)\right).
\ee
This gives the standard generalization of $\Psi_\bfmu^J(\tau,\bar
\tau)$ to a theta series with an elliptic variable $\bfrho$. The
holomorphic part couples to $\bfk_-$ and the anti-holomorphic part to $\bfk_+$
We will therefore also denote $\Psi^J_\bfmu[\CK_{\rm s}]$ as
\begin{equation}
\label{FluxesIpm}
\begin{split}
\Psi_\bfmu^J(\tau,\bar \tau,\bfrho, \bar \bfrho)&=\exp\!\left(-2\pi
  y\, \bfb_+^2\right)\sum_{\bfk\in L+\bfmu} \partial_{\bar \tau} \left(\sqrt{2y}\,B(\bfk+\bfb,J)\right)\, (-1)^{B(\bfk,K)}\,q^{-\frac{\bfk_-^2}{2}}\,\bar q^{\frac{\bfk_+^2}{2}}\\
&\quad \times \exp\Big(-2\pi i B(\bfk_-,\bfrho)-2\pi i B(\bfk_+,\bar \bfrho)\Big),
\end{split}
\end{equation}
Note that $\Psi_\bfmu^J(\tau,\bar \tau,0,0)=\Psi_\bfmu^J [\CK_0](\tau,\bar \tau)$ (\ref{PsiK}).
We postpone the remaining steps of the evaluation to Section \ref{ev_surface_obs}.

After describing the $u$-plane integrand, we can also give the
Seiberg-Witten contribution of the strong coupling singularities $u=\pm\Lambda^2$ to
$\left< e^{2p\,u+I_-(\bfx)} \right>^J_\bfmu$. Setting $\Lambda=1$, the contribution for
$u=1$ from a Spin$^{\mathbb{C}}$ structure $\bfk$ is \cite{Moore:1997pc}
\be
\label{SWcontribution}
\begin{split}
\left< e^{2p\,u+I_-(\bfx)} \right>^J_{\rm SW,\bfk,+}&= 2\,{\rm
  SW}(\bfk)_+\,\mathrm{Res}_{a_D=0}\left[ \frac{da_D}{a_D^{1+n}}\, C(u)^{\bfk^2/2}\,P(u)^{\sigma/8}\,L(u)^{\chi/4}
\right.\\
& \left. \quad  \times\,
\,\exp\!\left(
    2p\,u+i \,\frac{du}{da}\, B(\bfx,\bfk) +\bfx^2\,T(u) \right) \right],
\end{split}
\ee
with $n=-(2\chi+3\sigma)/8+\bfk^2/2$ and
the functions $C(u)$, $P(u)$, $L(u)$ given by
\be
\begin{split}
C(u)&=\frac{a_D}{q_D},\\
P(u)&=-\frac{64\,\vartheta_2(\tau_D)^8}{\vartheta_3(\tau_D)^4\vartheta_4(\tau_D)^4}\,
a_D^{-1},\\
L(u)&=\frac{8\,i}{\vartheta_3(\tau_D)^2\vartheta_4(\tau_D)^2}.
\end{split}
\ee
For four-manifolds of SW-simple type, the only $\bfk$ for which the
(\ref{SWcontribution}) is non-vanishing have $n=0$. The expression
then simplifies considerably. For the contribution from $u=1$,
\be
\label{SWsimpletype}
\left< e^{2p\,u+I_-(\bfx)}\right>^J_{\rm
  SW,\bfk,+}={\rm
  SW}(\bfk)\, 2^{1+K^2-\chi_h} \,
e^{2p+\bfx^2/2+2iB(\bfx,\bfk)}.
\ee
and for the contribution from $u=-1$,
\be
\label{SWsimpletypeminus}
\left< e^{2p\,u+I_-(\bfx)}\right>^J_{\rm
  SW,\bfk,-}={\rm
  SW}(\bfk)\, 2^{1+K^2-\chi_h} \,
e^{-2p-\bfx^2/2+2\,B(\bfx,\bfk)}.
\ee
The full correlation function for manifolds of simple type therefore reads
\be
\left< e^{2p\,u+I_-(\bfx)}\right>^J_{\bfmu}=\Phi_\bfmu^J[e^{2p\,u+I_-(\bfx)}]+\sum_{\pm} \sum_{\bfk\in
  L+\frac{1}{2}\overline{w_2(T_M)} \atop \bfk^2=(2\chi+3\sigma)/4} \left< e^{2p\,u+I_-(\bfx)}\right>^J_{\rm
  SW,\bfk,\pm}.
\ee
Manifolds with $b_2^+=1$ are however rarely of SW-simple type
\cite{mcduff:1996}. These manifolds may give rise to
SW moduli spaces of arbitrarily high dimension. The SW
contributions will then be more involved, but are entire functions of
$p$ and $\bfx$ as is the case for (\ref{SWsimpletype}) and (\ref{SWsimpletypeminus}).

\subsection{Summary}
For compact four-manifolds with $(b_1,b_2^+)=(0,1)$, the contribution of the
$u$-plane to the vev of an observable $\CO$, is given by
\be
\label{Phi_summ}
\Phi_\bfmu^J[\CO]=\int_{\BH/\Gamma^0(4)} \tilde
\nu(\tau)\,\Psi_\bfmu^J[\CK_\CO](\tau,\bar \tau).
\ee
Besides the choice of $\CO$, it depends on the following data of the four-manifolds
\begin{itemize}
\item the lattice $L$ with signature $(1,b_2-1)$,
\item a period point $J \in L \otimes \mathbb{R}$, normalized to $Q(J)=1$,
\item An integral lift $K\in L$ of $w_2(T_M)$,
\item An integral lift $\overline{w_2(P)}$ of the 't Hooft flux so
  that $\bfmu = \frac{1}{2}\overline{w_2(P)}\in H^2(M,\mathbb{R})$.
\end{itemize}

\section{Evaluation of $u$-plane integrals}
\label{sec:evaluation}
This section discusses the evaluation of $u$-plane integrals
using mock modular forms. Subsection \ref{regularization} reviews the
evalulation and renormalization of integrals over a modular
fundamental domain \cite{Moore:1997pc, Korpas:2019ava, 1603.03056}. Section \ref{strategy}
explains the strategy for arbitrary correlation functions. Subsection
\ref{subsec_part} factors the sum over fluxes into holomorphic and
non-holomorphic terms for a specific choice of $J$. We apply this
result to the evaluation of the
partition function and topological correlators in Subsections
 \ref{uplanemock} and \ref{ev_surface_obs}.

\subsection{Integrating over $\mathbb{H}/  {\rm SL}(2,\mathbb{Z})$}
\label{regularization}

In the previous section we arrived at the general form
(\ref{Phi_summ}) for the contribution of the $u$-plane to the correlators.
Order by order in $\bfx$ we encounter modular integrals of the form
\be
\label{If}
\CI_f=\int_{\mathcal{F_\infty}} d\tau\wedge d\bar \tau \,y^{-s}\,f(\tau,\bar
\tau),
\ee
where $f$ is a non-holomorphic modular form of weight $(2-s,2-s)$, and
$\CF_\infty$ is the standard keyhole fundamental domain for the modular group, $\CF_\infty =
\BH/\mathrm{SL}(2,\BZ)$. The integral is naturally independent of the
choice of fundamental domain due to the modular properties of $f$.
We assume that $f$ has a convergent Fourier series expansion
\be
\label{expansion_f}
f(\tau,\bar\tau)=\sum_{m,n\gg -\infty }\, c(m,n)\, q^m \bar
q^n,
\ee
where the exponents $m, n$ are bounded below. They may be real and negative,
but $m-n \in \mathbb{Z}$ by the requirement that $f$ is a modular
form. Since $m$ and $n$ can be both negative, the integral $\CI_f$ is
in general divergent and needs to be properly defined
\cite{Korpas:2019ava, 1603.03056, Dixon:1990pc,
  Harvey:1995fq}. While the definition of
the regularized and renormalized integral $\CI^{\rm r}_f$ is quite
involved, the final result is quite elegant and compact, at least if
$f$ can be expressed as a total anti-holomorphic derivative,
\be
\label{dh=f}
\partial_{\bar \tau} \widehat h(\tau,\bar \tau)=y^{-s} f(\tau,\bar \tau),
\ee
where $\widehat h$ transforms as a modular form of weight $(2,0)$. In
this case, the integrand of (\ref{If}) is exact and equal to $-d(d\tau\,\widehat h)$.
If only terms with $n>0$ contribute to the sum in
(\ref{expansion_f}), $\widehat h$ is a mock modular form and can be expressed as
\be
\label{widehath}
\widehat h(\tau,\bar\tau)=h(\tau) + 2^s  \int_{-\bar \tau}^{i\infty}
\frac{f(\tau,-v)}{(-i(v+\tau))^s}dv,
\ee
where $h$ is a (weakly) holomorphic function with Fourier expansion
\be
\label{holoh}
h(\tau)=\sum_{m\gg -\infty \atop m\in \mathbb{Z}} d(m)\,q^m.
\ee
Note that the two terms on the rhs of (\ref{widehath}) are separately
invariant under $\tau\to \tau+1$, while the transformation of
the integral under $\tau\to -1/\tau$ implies for $h(\tau)$,
\be
\label{h-1/tau}
h(-1/\tau)=\tau^2 \left( h(\tau)+2^s \int_{0}^{i\infty} \frac{f(\tau,-v)}{(-i(v+\tau))^s}\,dv\right).
\ee
Reference \cite{Korpas:2019ava} gives a  definition of the integral $\CI^{\rm r}_f$ 
such that the value turns out to be
\be
\begin{split} \label{Constant}
\CI^{\rm r}_f&= d(0).
\end{split}
\ee
As a result the only contribution to the integral arises from the
constant term of $h(\tau)$. The definition in \cite{Korpas:2019ava} 
reduces to the older definition for $\CI_f$ if
either $m$ or $n$ is non-negative \cite{Moore:1997pc,Harvey:1995fq} 
but is new if both $n,m$ are negative. It is shown in \cite{Korpas:2019ava} 
that, at least for Donaldson-Witten theory, the new definition is physically sensible
in the sense that $\cal{Q}$-exact operators decouple. 

Note that the absence of holomorphic modular forms of weight two for
$\mathrm{SL}(2,\BZ)$ implies that $h(\tau)$ is uniquely determined by the polar
coefficients, that is to say those $d(m)$ with $m<0$. The ambiguity in polar coefficients
gives thus rise to an ambiguity in the anti-derivative
$h(\tau)$. Different choices for $h(\tau)$ differ by a weakly holomorphic
modular form of weight 2. However, this ambiguity does not lead to an
ambiguity in the final result, $d(0)$, since the constant term of such weakly
holomorphic modular forms vanishes. This can be understood from the cohomology of $\CF_\infty$.
Since the first cohomology of $\CF_\infty$ is trivial, any closed one-form
$\xi$ is necessarily exact. Such a one-form $\xi$ can be expressed as
$C(\tau)\,d\tau$, with $C(\tau)$ a (weakly holomorphic) modular form of weight two. Since
$\xi$ is exact, the period $\int^{Y+1}_{Y}C(\tau)\,d\tau$ vanishes,
which implies that the constant term of $C(\tau)$ vanishes. Indeed, a basis of weakly
holomorphic modular forms of weight 2 is given by derivatives of
powers of the modular invariant $J$-function, $\partial_{\tau}\!
\left(J(\tau)^\ell\right)$, $\ell \in \mathbb{N}$, which all have
vanishing constant terms.

\subsection{General strategy}
\label{strategy}
Recall that in Section \ref{PathCor} we
analyzed the partition function of Donaldson-Witten theory, which
led to an integrand of the form $\tilde \nu(\tau)\,\Psi_\bfmu^J[\CK_0](\TT,\TB)$,
with a specific kernel $\CK_0$ (\ref{CK0}). For more general
correlation functions, the integrand takes a similar form,
\be \label{uPlane}
\Phi^J_{\bfmu}[\CO]=\int_{\mathbb{H}/\Gamma^0(4)} d\tau\wedge d\bar \tau\, \tilde \nu(\tau) \,\Psi_\bfmu^J[\CK_\CO].
\ee
where the kernel $\CK_\CO$ depends on the insertion
$\CO=\CO_1\CO_2\dots$. This can be expressed as an integral of
the form (\ref{If}), whose integrand could consist of several terms
$\sum_{j} y^{-s_j} f_j $. Moreover, one can express the integral over
$\Gamma^0(4)$ as the sum of six integrals over
$\CF_\infty$ using modular transformations. As explained in the previous subsection,
an efficient technique to evaluate these integrals is to express the integrand as a total
derivative with respect to $\bar \tau$, which has indeed been used in
a few special cases to evaluate the $u$-plane integral
\cite{Moore:1997pc, Malmendier:2008db,
  Malmendier:2010ss, Griffin:2012kw}. We express the integrand of the
generic integral (\ref{uPlane}) as
\be \label{Hproperty}
\frac{d}{d \bar \tau}\widehat \CH^J_{\bfmu}[\CO](\TT, \TB)=\tilde \nu(\TT)\,\Psi_\bfmu^J[\CK_\CO](\TT,\TB),
\ee
which by a change of variables is equivalent to an anti-holomorphic
derivative in $u$ as discussed in the Introduction. The inverse map $u^{-1}: \CB\to \BH/\Gamma^0(4)$ maps each of the boundaries $\partial_j \CB$ to arcs in $\BH/\Gamma^0(4)$
in the vicinity of the cusps $\{i\infty, 0 ,2\}$ displayed in Figure \ref{fund_domain}.

The function $\widehat \CH_\bfmu^J[\CO](\TT, \TB)$ is required to transform as a
modular form of weight $(2,0)$ with trivial multiplier system, which one may hope to determine explicitly
using methods from analytic number theory, especially the theory of
mock modular forms \cite{ZwegersThesis, MR2605321}. To derive a
suitable $\widehat \CH^J_\bfmu[\CO]$, we will choose a convenient period point $J$.
Once $\widehat \CH^J_\bfmu[\CO]$ is known it is straightforward to apply the discussion of Section
\ref{regularization}. To relate the integral over
$\mathbb{H}/\Gamma^0(4)$ to an integral over $\CF_\infty$, we use coset representatives of 
$\mathrm{SL}(2,\mathbb{Z})/\Gamma^0(4)$  to map the six different images of $\CF_\infty$
within $\mathbb{H}/\Gamma^0(4)$, displayed in Figure
\ref{fund_domain}, back to $\CF_\infty$.  After
this inverse mapping, we use the modular properties of the integrand to express
each of the six integrands as a series in $q$ and $\bar q$, after
which the techniques of Section \ref{regularization} can be
applied. To this end, one can use the relations (\ref{SL2Z}) for
$\Psi_\bfmu^J$, while the $q$-series for $\tilde \nu(\tau)$ follows
from the standard relations for Jacobi theta functions.

Since the maps $\tau\mapsto \tau - n$, $n=1,2,3$ do not change the
constant part of the integrand, we find that $\Phi_{\bfmu}^J[\CO]$ evaluates to
\be
\label{sumofcusps}
\Phi_{\bfmu}^J[\CO]=4\left[\widehat \CH^J_\bfmu[\CO](\TT, \TB)\,\right]_{q^0}+\Big[ \tau\mapsto
-\tfrac{1}{\tau}\Big]_{q^0}+\Big[ \tau \mapsto \tfrac{2\tau-1}{\tau} \Big]_{q^0},
\ee
where for the second and third brackets on the rhs, one makes
the indicated modular transformation for $\tau$, $S$ and $T^2S$, and then determines the $q^0$ coefficient
of the resulting Fourier expansion.

An important point is the possibility to add to $\widehat \CH^J_\bfmu[\CO]$
a holomorphic integration ``constant'' $s_\CO$, which is required to be a weight 2
modular form for $\Gamma^0(4)$. Of course, $\Phi^J_\bfmu[\CO]$
should be independent of $s_\CO$, since definite integrals do not
depend on the integration constant. To see the independence of $\Phi^J_\bfmu[\CO]$ on $s_\CO$, note that $s_\CO$ will be mapped to a
weight 2 form for $\mathrm{SL}(2,\mathbb{Z})$ by the inverse
mapping. As discussed in Section \ref{regularization}, there are no
holomorphic $\mathrm{SL}(2,\mathbb{Z})$ modular forms with weight 2,
and the weakly holomorphic ones have a vanishing constant term. There
is therefore no ambiguity arising from the holomorphic integration
constant.

On the other hand, the integration constant $s_\CO$ can modify the contribution from each
cusp, since a non-vanishing holomorphic modular form of weight 2 for $\Gamma^0(4)$
exists. It is explicitly given by
$\vartheta_2(\tau)^4+\vartheta_3(\tau)^4$, and while it contributes
4 at the cusp at infinity, the contributions of the three cusps together
add up to 0. We can make a natural choice of the integration constant
by requiring that the exponential behavior of $\CH^J_{\bfmu}$ for $\tau\to
i\infty$ matches the behavior of $\tilde \nu\,\Psi^J_\bfmu$ in this limit.

Once we have determined $\Phi_{\bfmu}^J[\CO]$ for a specific period
point $J$, one can change to an arbitrary $J$ quite easily using
indefinite theta functions as discussed in \cite{Gottsche:1996aoa,Korpas:2017qdo,Moore:2017cmm}. 
The integrand can thus be expressed as a total derivative (\ref{uPlane}) for any $J$.

\subsection{Factoring $\Psi^J_\bfmu$}
\label{subsec_part}

To evaluate the partition function $\Phi_\bfmu^J$, we will choose a 
convenient period point $J$ so that $\Psi^J_\bfmu$, as a function of $\tau$,
has a simple factorization as a holomorphic times an anti-holomorphic function. 
In this way, we can easily determine an anti-derivative using
the theory of mock modular forms. Using the classification of the uni-modular
lattices, Equations (\ref{lattice_odd}) and (\ref{lattice_even}), a
convenient factorisation is possible for any intersection form.

\subsubsection*{Odd intersection form}
\label{oddLattice}
Let us first assume that the intersection lattice $L$ is odd, such that its quadratic form can be
brought to the standard form in Equation (\ref{lattice_odd}). Since
the wall-crossing formula for Donaldson invariants is known \cite{Moore:1997pc},
it suffices to determine $\Phi^J_\bfmu$ for a convenient choice of
$J$. To this end, we choose the polarization
\be
\label{Jodd}
J=(1,\boldsymbol 0),
\ee
where ${\bf 0}$ is the $(b_2-1)$-dimensional 0-vector.
For this choice of $J$, the orthogonal decomposition of the lattice,
$L=L_+\oplus L_-$ into a 1-dimensional positive definite lattice $L_+$ and
$(b_2-1)$-dimensional negative definite sublattice $L_-$, implies that
the sum over the ${\rm U}(1)$ fluxes $\Psi_{\bfmu}^{J}(\tau,
\bar{\tau})$ factors. To see this explicitly, we let $\bfk =
(k_1, \bfk_{-})\in L$, and $k_1
\in \BZ + \mu_1$, $\bfk_{-} \in L_{-} +\bfmu_{-}$ and
$\bfmu=(\mu_1,\bfmu_-)$. The
Siegel-Narain theta function
$\Psi_{\bfmu}^{J}=\Psi_{\bfmu}^{J}[\CK_0]$ (\ref{PsiK}) now factors as
\be
\label{Psiodd}
\Psi_{\bfmu}^{J}(\tau, \bar{\tau}) =-i\,(-1)^{\mu_1 (K_1-1)}\,f_{\mu_1}(\tau,\bar \tau)\,\Theta_{L_-,\bfmu_-}(\tau),
\ee
with
\be
\label{ThetaLambda-}
\begin{split}
f_{\mu}(\tau,\bar \tau)& :=-\frac{e^{\pi i \mu}}{2\sqrt{2y}}\sum_{k\in \mathbb{Z}+\mu}\, (-1)^{k-\mu}\, k\, \bar q^{k^2/2}, \\
\Theta_{L_-,\bfmu_-}(\tau)& = \sum_{\bfk_{-}\in L_{-}+\bfmu_-} (-1)^{B(\bfk_{-}, K_{-} )}  q^{-\bfk_{-}^2/2},
\end{split}
\ee
We used also that $K_{1}$ is odd since $K$ is a characteristic vector, as discussed below Eq. (\ref{lattice_odd}). Using
$$\sum_{k\in \mathbb{Z}+\frac{1}{2}}
(-1)^{k-\frac{1}{2}}\,k\, q^{k^2/2}=\eta(\tau)^3,$$
we can express $f_\mu$ in terms of the Dedekind eta function $\eta$,
\be
\label{fmu}
f_\mu(\tau,\bar \tau)=\left\{ \begin{array}{rl}
 0,
    & \qquad\mu=0 \mod \mathbb{Z},\\
    -\frac{ i}{2\sqrt{2y}}\,\overline{\eta(\tau)^3}, & \qquad \mu=\frac{1}{2} \mod \mathbb{Z}.\\  \end{array}\right.
\ee
We can similarly evaluate $\Theta_{L_-,\bfmu_-}$. Since all
$K_j$ are odd, $\Theta_{L_-,\bfmu_-}(\tau)$ vanishes, except if
$\bfmu_-={\bf 0} \mod \mathbb{Z}^{b_2-1}$. In that case,
$\Theta_{L_-,\bfmu_-}$ is a power of the Jacobi theta function $\vartheta_4$,
\be
\Theta_{L_-,\bfmu_-}(\tau)=\left\{ \begin{array}{rl}
 \vartheta_4(\tau)^{b_2-1} ,
    & \qquad \bfmu_-={\bf 0} \mod \mathbb{Z}^{b_2-1},\\
  0, & \qquad \mathrm{else}.\\  \end{array}\right.
\ee

After substitution of $\tilde \nu$ (\ref{tnumod}), we find for the integrand
\be
\label{tnuPsi}
\tilde \nu\,\Psi_\bfmu^J=\left\{ \begin{array}{rl}
   \frac{(-1)^{(K_1+1)/2}}{8}\,f_{\frac{1}{2}}(\tau,\bar \tau) \,
\frac{\vartheta_4(\tau)^{12}}{\eta(\tau)^{9}},  & \qquad \bfmu=({\tfrac{1}{2},\bf 0}) \mod \mathbb{Z}^{b_2},\\
  0, & \qquad \mathrm{else}.\\  \end{array}\right.
\ee
Note that the dependence of the integrand on $b_2$ has disappeared, and that the
integrand diverges for $\tau \to \infty$.

\subsubsection*{Even intersection form}
\label{evenLattice}
We continue with the even lattices, whose quadratic form can be
brought to the form given in Equation (\ref{lattice_even}),
$L=\mathbb{I}^{1,1}\oplus n\,L_{E_8}$.  We choose for the period point
\be
\label{Jeven}
J=\frac{1}{\sqrt{2}}(1,1,{\bf 0}),
\ee
where the first two components correspond to $\mathbb{I}^{1,1}\subset
L$, and ${\bf 0}$ is now the $(b_2-2)$-dimensional 0-vector. We have then for the positive and negative definite components of $\bfk\in L$,
\be
\bfk_+^2=\frac{1}{2}(k_1+k_2)^2,\qquad
\bfk_-^2=-\frac{1}{2}(k_1-k_2)^2+\bfk_{n}^2,
\ee
where $\bfk_n\in nL_{E_8}$. Note $\bfk_n^2\leq 0$, since $L_{E_8}$ is the negative $E_8$ lattice.

The sum over fluxes $\Psi_\bfmu^J$ factors for this choice of $J$,
\be
\label{Psievenfact}
\Psi_\bfmu^J(\tau,\bar \tau)=\Psi_{\mathbb{I},(\mu_+,\mu_-)}(\tau,\bar \tau)\,\Theta_{n L_{E_8},\bfmu_n}(\tau),
\ee
where the subscript is $\bfmu=(\mu_+,\mu_-,\bfmu_n)$, and $\Psi_{\mathbb{I},(\mu_+,\mu_-)}(\tau,\bar
\tau)$ is given by
\be\label{eq:Psi-but-zero}
\Psi_{\mathbb{I},(\mu_+,\mu_-)}(\tau,\bar\tau)=\frac{i}{4\sqrt{y}}\sum_{\bfk\in \mathbb{I}^{1,1}+(\mu_+,\mu_-)}
(k_1+k_2)\,(-1)^{k_1 K_2+k_2K_1} \,q^{(k_1-k_2)^2/4}\, \bar q^{(k_1+k_2)^2/4}.
\ee
Moreover, the theta series $\Theta_{nE_8,\bfmu}$ for the negative
definite lattice equals
\be
\Theta_{nE_8,\bfmu_n}(\tau)=\sum_{\bfk_n\in nL_{E_8}+\bfmu_n} q^{-\bfk_n^2/2}.
\ee

As before,  the $K_{j}$ are components of the characteristic element
$K\in L$, this time in the basis (\ref{lattice_even}). Recall $K_1$ and $K_2\in 2\mathbb{Z}$
since they are components of a characteristic vector of
$\mathbb{I}^{1,1}$. Changing the sign of $k_1$ and $k_2$ in the
summand gives
$\Psi_{\mathbb{I},(\mu_+,\mu_-)}=-\Psi_{\mathbb{I},(\mu_+,\mu_-)}$, hence
$\Psi_{\mathbb{I},(\mu_+,\mu_-)}$ vanishes identically. Nevertheless, it is
instructive to evaluate the integral using the approach of Section
\ref{strategy}, to set up notation for working with
the closely analogous function in Equation \eqref{eq:Psi-mu-nonzero}, which is
definitely nonzero.

To express $\Psi_{\mathbb{I},(\mu_+,\mu_-)}$ as an anti-holomorphic
derivative, we split the lattice into a positive and negative definite
one, by changing summation variables to
\be
n_+=k_1+k_2,\qquad n_-=k_1-k_2,
\ee
and similarly for the 't Hooft flux and the canonical class,
\be
\begin{split}
\mu_+& =\mu_1+\mu_2,\qquad \mu_-=\mu_1-\mu_2,\\
K_{+} & = \frac{1}{2}(K_{1}+K_{2}), \qquad K_{-}=\frac{1}{2}(K_{1}-K_{2}),\\
\end{split}
\ee
where $\mu_j\in \mathbb{Z}/2$ as before. Given $\mu_\pm$, the
summation over $n_\pm$ runs over two sets, namely $n_\pm\in
2\mathbb{Z} +\mu_\pm+j$ with $j=0,1$.
We can now express the sum over fluxes as
\be
\label{PsiI11}
\Psi_{\mathbb{I},\bfmu}(\tau,\bar \tau)=-i\,(-1)^{\mu_+K_+-\mu_-K_-}\sum_{j=0,1} h_{\mu_++j}(\tau,\bar \tau)\, t_{\mu_-+j}(\tau),
\ee
where $\bfmu=(\mu_+,\mu_-)$ and
\be
\label{gtheta}
\begin{split}
&h_{\nu}(\tau, \bar \tau)=-\frac{1}{4\,\sqrt{y}} \sum_{n\in 2\mathbb{Z}+\nu}
n\,  \bar{q}^{n^2/4},\\
&t_{\nu}(\tau)=\sum_{n\in 2\mathbb{Z}+\nu} q^{n^2/4},
\end{split}
\ee
with $\nu\in \mathbb{Z}/2 \mod 2\mathbb{Z}$. For the four conjugacy classes of
$\nu$,  we find that $h_{\nu}$ equals
\be
\label{gmu}
h_{\nu}(\tau, \bar \tau)=\left\{ \begin{array}{rl}
 0,
    & \qquad\nu = 0 \mod \mathbb{Z},\\
    \frac{i}{8\sqrt{y}}\,e^{\pi i \nu}\,\overline{\eta(\tau/2)^3}, & \qquad
    \nu=\frac{1}{2} \mod \mathbb{Z}.\\  \end{array}\right.
\ee
We have similarly for $t_{\nu}$
\be
\label{thetanu}
t_\nu(\tau)=\left\{ \begin{array}{rl}
 \vartheta_3(2\tau),
    & \qquad\nu=0 \mod 2\mathbb{Z},\\
  \vartheta_2(2\tau),  & \qquad\nu=1 \mod 2\mathbb{Z},  \\
   \frac{1}{2}\,\vartheta_2(\tau/2), & \qquad \nu=\frac{1}{2} \mod \mathbb{Z}.\\  \end{array}\right.
\ee
Substitution of the expressions (\ref{gmu}), (\ref{thetanu}) in (\ref{PsiI11}) confirms the vanishing of
$\Psi_{\mathbb{I},\bfmu}$.

\subsection{$u$-plane integrands and mock modular forms}
\label{uplanemock}
Our next aim is to express the integrand as an anti-holomorphic
derivative of a non-holomorphic modular form. We will determine
functions $\widehat F_\mu$ (respectively $\widehat H_\mu$), which
transform as weight $\frac{1}{2}$ modular forms, and such that
\be
\label{partF}
\partial_{\bar \tau} \widehat F_\mu=f_\mu,\qquad \partial_{\bar \tau} \widehat H_\nu=h_\nu,
\ee
for odd and even lattices respectively. The holomorphic parts of
$\widehat F_\mu$ and $\widehat H_\nu$ are known as mock modular forms and contain
interesting arithmetic information \cite{ZwegersThesis, MR2605321}.

We consider first the case that the lattice $L$ is odd. We
deduce from Equation (\ref{tnuPsi}) that for $\mu_1\in \mathbb{Z}$, we
can take $\widehat F_{\mu_1}=0$.  We thus only need to be concerned with finding an anti-derivative $\widehat F_\frac{1}{2}$ of
$-\frac{i}{2\sqrt{2y}}\,\overline{\eta}^3$. Let us reduce notation by setting
$\widehat F=\widehat F_\frac{1}{2}$, then $\widehat F$ takes the general form
\be
\label{hatF}
\widehat F(\tau,\bar \tau)=F(\tau)-\frac{i}{2}\,\int_{-\bar
\tau}^{i\infty} \frac{\eta(w)^3}{\sqrt{-i(w+\tau)}}\,dw,
\ee
and is required to transform as a $\Gamma^0(4)$ modular form with
(holomorphic) weight $\frac{1}{2}$. The first term on the rhs is
holomorphic and is a mock modular form \cite{ZwegersThesis, MR2605321}, while the second term on the right hand side is known as a period integral
and transforms with a shift under transformations of
${\rm SL}(2,\mathbb{Z})$. The function $\eta^3$ is
known as the shadow of the mock modular form $F$. Similarly to the discussion above
Eq. (\ref{h-1/tau}), we deduce that the holomorphic part $F(\tau)$ must be non-vanishing to cancel the
shift.

The derivation of such a function is in general non-trivial.
The theory of indefinite theta functions provides a
constructive approach to derive a suitable $F(\tau)$.
Appendix \ref{Zwegers_theta} provides a brief introduction to these
functions, and derives an explicit expression for $F$:
\be
\label{feta3}
\begin{split}
F(\tau)&=\frac{-1}{\vartheta_4(\tau)}\sum_{n\in \mathbb{Z}}
\frac{(-1)^nq^{\frac{1}{2}n^2-\frac{1}{8}}}{1-q^{n-\frac{1}{2}}} \\
&=2\,q^{\frac{3}{8}}\left(1 +3\,q^{\frac{1}{2}}
  +7\,q+14\,q^{\frac{3}{2}}+\dots\right).
\end{split}
\ee

To evaluate $\Phi_\bfmu^J$ following (\ref{sumofcusps}), we need to
determine the $q-$expansion of $F$ at the other cusps. We introduce to
this end $F_D$ and $\widehat F_D$,
\be
\begin{split}
\widehat F_D(\tau,\bar \tau)&:=-(-i\tau)^{-\frac{1}{2}}\,\widehat F(-1/\tau,-1/\bar
\tau)\\
&=:F_D(\tau) -\frac{i}{2}\,\int_{-\bar
\tau}^{i\infty} \frac{\eta(w)^3}{\sqrt{-i(w+\tau)}}\,dw,
\end{split}
\ee
where $\tau$ is now the local coordinate which goes to $i\infty$ near the strong coupling cusp $u \to \Lambda^2$. 
Appendix \ref{Zwegers_theta} discusses how to derive the $q$-expansion of $F_D$ using the transformations
of the indefinite theta function (\ref{indef_theta_mod}). One finds
\be
\label{FDtau}
\begin{split}
F_D(\tau)&= \frac{-1}{\vartheta_2(\tau)} \sum_{n\in \mathbb{Z}}  \frac{q^{\frac{1}{2}(n+\frac{1}{2})^2-\frac{1}{8}}}{1+q^n}\\
&=\frac{1}{4}\,q^{-\frac{1}{8}}\left(-1 -3\,q+7\,q^2-14\,q^3+21\,q^4+\dots\right),
\end{split}
\ee
For the cusp $\tau\to 2$, the $q$-expansion is $-i\,F_D(\tau)$.

We leave the precise evaluation for later in this Subsection, and
continue with the even lattices, which can be treated
more briefly. We see from (\ref{gmu}) that $h_{\mu}(\tau, \bar
\tau)=\frac{1}{4}\, f_{\mu}(\tau/2,\bar \tau/2)$, with $f_\mu$ as in
(\ref{fmu}). We can thus easily determine a suitable anti-derivative
for $h_\mu$, namely
\be
\widehat H_\mu(\tau)=\left\{ \begin{array}{rl} 0, & \qquad \mu\in \mathbb{Z}, \\
   \frac{1}{2} \widehat F(\tau/2), & \qquad \mu\in \mathbb{Z} + \frac{1}{2},  \end{array}\right.
\ee
with $\widehat F$ as in (\ref{hatF}). We similarly define
$H(\tau/2)=\frac{1}{2}F(\tau/2)$ with $F$ as in (\ref{feta3}).\\

\noindent {\bf Remark}\\
Malmendier and Ono have emphasized the connection between $q$-series appearing in
the context of Mathieu moonshine and the $u$-plane integral for the complex projective plane $\mathbb{P}^2$
\cite{Malmendier:2012zz}. See \cite{HarveyINI2012, HarveyGGI2015, Kachru:2016nty,HarveyICTP2016, Anagiannis:2018jqf}
for overviews of the moonshine phenomenon. Our discussion above demonstrates that the
appearance of these $q$-series is quite generic for four-manifolds with
$b_2^+=1$. In particular, the function $F$ (\ref{feta3}) equals $1/8$ times the function
$H_{1A,2}^{(4)}$, which appears in the context of umbral moonshine on page 107 of \cite{Cheng:2013wca}.
Similarly, $F_D$ (\ref{FDtau}) equals $1/8$ times the function $H_{2A,1}^{(2)}$ on page 103 of
\cite{Cheng:2013wca}.

Moreover, $F$ and $F_D$ can be expressed in terms of the famous $q$-series
$H^{(2)}(\tau)$ of Mathieu moonshine \cite{Eguchi:2010ej}, whose coefficients are sums of dimensions of irreducible
representations of the finite sporadic group $M_{24}$, and which appeared in
the elliptic genus of the $K3$ sigma model with $(4,4)$
supersymmetry. We have for $F$,
\be
\begin{split}
F(\tau)=\frac{1}{24}\left( H^{(2)}(\tau)+2\frac{\vartheta_2(\tau)^4+\vartheta_3(\tau)^4}{\eta(\tau)^3}\right),
\end{split}
\ee
where \cite{Dabholkar:2012nd},
\be
\begin{split}
H^{(2)}(\tau)&=2\frac{\vartheta_2^4(\tau)-\vartheta_4^4(\tau)}{\eta(\tau)^3}-\frac{48}{\vartheta_3(\tau)}\sum_{n=1}^\infty
\frac{q^{\frac{1}{2}n^2-\frac{1}{8}}}{1+q^{n-\frac{1}{2}}}\\
 &=2\,q^{-\frac{1}{8}}(-1+45\,q+231\,q^2+
770 q^3 +\dots).
\end{split}
\ee
Whereas $F$ is a mock modular form for the subgroup
$\Gamma^0(4)\subset {\rm SL}(2,\BZ)$,
$H^{(2)}$ is a mock modular form for the full ${\rm
  SL}(2,\mathbb{Z})$. The completion
\be
\widehat H^{(2)}(\tau,\bar
\tau)=H^{(2)}(\tau)-12i\int_{-\bar
  \tau}^{i\infty}\frac{\eta(w)^3}{\sqrt{-i(w+\tau)}}dw,
\ee
transforms under the two generators of ${\rm SL}(2,\BZ)$ as
\be
\begin{split}
\widehat H^{(2)}(-1/\tau,-1/\bar \tau) &=-\sqrt{-i\tau}\,\widehat H^{(2)}(\tau), \\ 
\widehat H^{(2)}(\tau+1,\bar \tau+1)&=e^{-\frac{2\pi i}{8}} \widehat H^{(2)}(\tau). \
\end{split}
\ee
The holomorphic part $H^{(2)}$ therefore transforms as
\be
\label{H2trafo}
H^{(2)}(-1/\tau)=-\sqrt{-i\tau}\,\left( H^{(2)}(\tau)-12i\int_0^{i\infty} \frac{\eta(w)^3}{\sqrt{-i(w+\tau)}}dw\right).
\ee
We can express $F_D$ in terms of $H^{(2)}$ as
\be
F_D(\tau)=\frac{1}{24}\left(H^{(2)}(\tau)-2\frac{\vartheta_4(\tau)^4+\vartheta_3(\tau)^4}{\eta(\tau)^3} \right).
\ee

As a last example of a mock modular form with shadow $\eta^3$, we mention the function $Q^+$, which was introduced by Malmendier and Ono in the context of the $u$-plane integral
\cite{Malmendier:2008db, Malmendier:2012zz}
\be
\begin{split}
Q^+(\tau) &=\frac{1}{12}\, H^{(2)}(\tau)+\frac{7}{6}\,\frac{\vartheta_2(\tau)^4+\vartheta_3(\tau)^4}{\eta(\tau)^3}\\
&=q^{-\frac{1}{8}} (1+28\,q^{\frac{1}{2}}+39\,q+196\,q^{\frac{3}{2}}+161\,q^2+\dots).\\
\end{split}
\ee
Since $F$, $H^{(2)}$ and $Q^+$ are all weight $\frac{1}{2}$ mock
modular forms for $\Gamma^0(4)$ and have shadows proportional to
$\eta^3$, each of them can be used for the evaluation of the $u$-plane
integral. Note that among these functions, only $F$ vanishes in the limit $\tau\to i\infty$. Thus it
behaves similarly to its derivative $f_{\frac{1}{2}}$ in this limit.

\subsection*{Evaluation}
We continue by evaluating the $u$-plane integral for an arbitrary
four-manifold with $(b_1,b_2^+)=(0,1)$. As mentioned in
Subsection \ref{subsec_part}, the partition function $\Phi_\bfmu^J$ is only
non-vanishing for odd lattices, and with 't Hooft flux $\bfmu$ with
$\mu_1=\frac{1}{2}$ in the standard basis. We have then using (\ref{tnuPsi}) and (\ref{partF}),
\be
\label{PhiFinal}
\begin{split}
\Phi_\bfmu^J&=\tfrac{1}{8}(-1)^{(K_1+1)/2}\int_{\mathbb{H}/\Gamma^0(4)}
d\tau\wedge d\bar \tau\, \frac{\vartheta_4(\tau)^{12}}{\eta(\tau)^9}\, \partial_{\bar
  \tau}  \widehat F(\tau,\bar \tau)\\
&=\tfrac{1}{8}(-1)^{(K_1+1)/2}\left( 4\left[ F(\tau)\,\frac{\vartheta_4(\tau)^{12}}{\eta(\tau)^9}\right]_{q^0} +2\left[F_D(\tau)\,\frac{\vartheta_2(\tau)^{12}}{\eta(\tau)^9}\right]_{q^0}        \right),
\end{split}
\ee
where the first term in the straight brackets is due to the
contribution at $i\infty$ and the second term due to the two strong
coupling singularities, which contribute equally. The strong coupling singularities do not
contribute to the $q^0$ term. We finally arrive at
\be
\label{Phipart}
\Phi^J_{\bfmu}=\left\{ \begin{array}{rl}
   (-1)^{(K_1+1)/2},  & \qquad \bfmu=({\tfrac{1}{2},\bf 0}) \mod \mathbb{Z}^{b_2},\\
  0, & \qquad \mathrm{else}.\\  \end{array}\right.
\ee
This is in agreement with the results for $\mathbb{P}^2$ for which
$K_1=3$ \cite{ellingsrud1995wall}.

It is straightforward to include the exponentiated point observable
$e^{2p\,u}$ in the path integral. One then arrives at
\be
\label{Phiepu}
\begin{split}
\Phi^J_{\bfmu}[e^{2p\,u}]&=\tfrac{1}{8}(-1)^{(K_1+1)/2}\left( 4\left[
    F(\tau)\,\frac{\vartheta_4(\tau)^{12}}{\eta(\tau)^9}\,e^{2p\,u(\tau)}\right]_{q^0}\right.\\
&\quad    \left.+\left[F_D(\tau)\,\frac{\vartheta_2(\tau)^{12}}{\eta(\tau)^9}\,\left(e^{2p\,u_D}+e^{-2p\,u_D}\right)\right]_{q^0}
\right),
\end{split}
\ee
with $u_D$ given in (\ref{uD}). We deduce from the expansion of $F_D$
(\ref{FDtau}) and $u_D(\tau)=1+O(q)$ that the monopole cusps do not contribute
to the $q^0$-term for any power of $p$. The result is therefore
completely due to the weak coupling cusp,
\be
\label{PhiepuFinal}
\Phi^J_{\bfmu}[e^{2p\,u}]=\tfrac{1}{2}(-1)^{(K_1+1)/2}\left[
  F(\tau)\,\frac{\vartheta_4(\tau)^{12}}{\eta(\tau)^9}\,e^{2p\,u(\tau)}\right]_{q^0}.
\ee
Only even powers of $p$ contribute to the constant term, which is in agreement with the interpretation of the point observable as a four-form on
the moduli space of instantons. Except for the mild dependence of (\ref{Phiepu}) on
$K_1$, this equation demonstrates that the contribution from the $u$-plane
to $\left< e^{2p\,u} \right>_\bfmu^J$ is universal for any four-manifold with
odd intersection form and period point $J$ (\ref{Jodd}).

We list $\Phi_\bfmu^J[u^\ell]$ for small
$\ell$ in Table \ref{TableDon}. See  \cite{ellingsrud1995wall} for a
more extensive list. Section
\ref{asympbehavior} will discuss these numbers and the convergence of
$\Phi_\bfmu^J[e^{2p\,u}]=\sum_{\ell\geq 0}
\Phi_\bfmu^J[u^\ell]\,(2\,p)^\ell/\ell!$ in more detail.

\begin{center}
\begin{table}[]
\centering
\begin{tabular}{| l | r |}   \hline
$\ell$ & $8^\ell\,\Phi^J_\bfmu[u^\ell]$  \\ \hline
0 & 1 \\
2  & 19 \\
4 & 680 \\
6 & 29\,557\\
8 & 1\,414\,696   \\
\hline
\end{tabular}
\caption{List of non-vanishing $\Phi_\bfmu^J[u^\ell]$ with $0 \leq \ell\leq 8$ for smooth four-manifolds with $(b_1,b_2^+)=(0,1)$.
These numbers are universal for all four-manifolds with odd intersection form, $K_1=3
\mod 4$, period point $J$ (\ref{Jodd}), and $\mu_1=\frac{1}{2} \mod
\mathbb{Z}$. For $K_1=1\mod 4$, they differ by a sign, while they
vanish for any four-manifold with an even intersection form.  }
\label{TableDon}
\end{table}
\end{center}

\subsection{Evaluation of surface observables}
\label{ev_surface_obs}
This subsection continues with the evaluation of the contribution of the $u$-plane to vevs of surface
observables.

\subsubsection*{Odd intersection form}
We proceed as in Section \ref{oddLattice} choosing $J=(1, {\bf
  0})$ and set $\rho_+=B(\bfrho,J)$ and $\bfrho_-=\bfrho-\rho_+\,J$. Specializing (\ref{FluxesIpm}) gives
\be
\Psi^J_\bfmu(\tau,\bar \tau,\bfrho,\bar \bfrho)=
-i\,(-1)^{\mu_1(K_1-1)}\,f_{\mu_1}(\tau, \bar \tau,\rho_+,\bar \rho_+)\,\Theta_{L_-,\bfmu_-}(\tau,\bfrho_-).
\ee
with
\be
\label{fmusurface}
f_{\mu}(\tau,\bar \tau,\rho,\bar \rho)=i\,e^{\pi i \mu}\exp(-2\pi y\,
b^2)\sum_{k\in \mathbb{Z}+\mu} \partial_{\bar \tau}(\sqrt{2y}\,(k+b))\,(-1)^{k-\mu}\,\bar
q^{k^2/2}\,e^{-2\pi i\, \bar \rho\, k},
\ee
where $b=\mathrm{Im}(\rho)/y$, and we removed the subscript of $\mu_1$
as before in Section \ref{sec:evaluation}. Moreover,
$\Theta_{L_-,\bfmu_-}(\tau,\bfrho_-)$ is given by
\be
\Theta_{L_-,\bfmu_-}(\tau,\bfrho_-)=\sum_{\bfk_-\in
  L_-+\bfmu_-} (-1)^{B(\bfk_-,K_-)}\,q^{-\bfk_-^2/2}\,e^{-2\pi i
  B(\bfrho_-,\bfk_-)}.
\ee
The functions $f_{\mu}(\tau,\bar \tau,\rho_+,\bar \rho_+)$ and
$\Theta_{L_-,\bfmu_-}(\tau,\bfrho_-)$ specialize to $f_{\mu}(\tau,\bar
\tau)$ and $\Theta_{L_-,\bfmu_-}(\tau)$ (\ref{ThetaLambda-})
for $\bfrho=0$. The theta series
$\Theta_{L_-,\bfmu_-}(\tau,\bfrho_-)$ can be expressed as a product of
$\vartheta_1$ and $\vartheta_4$ (\ref{Jacobitheta}) depending on the precise value of
$\bfmu_-$. We define the dual theta series $\Theta_{D,L_-,\bfmu_-}$ as
\be
\begin{split}
\Theta_{D,L_-,\bfmu_-}(\tau,\bfrho_{D,-})&=(-i\tau)^{-(b_2-1)/2}\,e^{-\frac{\pi
  i}{\tau}\bfrho_{D,-}^2}\,\Theta_{L_-,\bfmu_-}(-1/\tau,\bfrho_D/\tau)\\
&=  \sum_{\bfk_-\in
  L_-+K_-/2} (-1)^{B(\bfk_-,\bfmu_-)}\,q^{-\bfk_-^2/2}\,e^{-2\pi i
  B(\bfrho_{D,-},\bfk_-)}.
\end{split}
\ee

We aim to write this as a total anti-holomorphic derivative of a
real-analytic function $\widehat F_\mu(\tau,\bar \tau,\rho,\bar \rho)$ of $\tau$ and
$\rho$. We can achieve this using the Appell-Lerch sum $M(\tau,u,v)$,
with $u,v\in \mathbb{C}\backslash \{\mathbb{Z}\tau+\mathbb{Z}\}$, which has appeared at many places in mathematics and
mathematical physics. See for examples \cite{ Dabholkar:2012nd, Semikhatov:2003uc,Manschot:2011dj}. The function $M(\tau,u,v)$
is meromorphic in $u$ and $v$, and weakly holomorphic in
$\tau$. More properties are reviewed in Appendix
\ref{subsecZwmu}. Equation (\ref{mucomplete}) is the main
property for us. It states that $\widehat M(\tau,\bar \tau,u,\bar
u,v,\bar v)=M(\tau,u,v)+\frac{i}{2}R(\tau,\bar \tau,u-v,\bar u-\bar v)$
transforms as a multi-variable Jacobi form of weight $\frac{1}{2}$,
where $R(\tau,\bar \tau,u,\bar u)$ is real-analytic in both $\tau$ and $u$. To
determine $\widehat F_\mu(\tau,\bar \tau,\rho,\bar \rho)$, we express $f_\mu(\tau,\bar
\tau,\rho,\bar \rho)$ in terms of $\partial_{\bar \tau} R(\tau,\bar
\tau,u,\bar u)$ (\ref{dbarR}). With $w=e^{2\pi i
  \rho}$, we find
\be
f_{\mu}(\tau,\bar \tau,\rho,\bar \rho)=\tfrac{1}{2}\,e^{\pi i \nu} \,q^{-\nu^2/2}\,w^{-\nu}\,\partial_{\bar
  \tau} R(\tau,\bar \tau,\rho+\nu \tau, \bar \rho+\nu \bar \tau),\qquad \nu=\mu-\tfrac{1}{2}.
\ee

We can thus determine the anti-derivative of $f_\mu$ in terms of the completion
$\widehat M$ (\ref{hatmuv}) in Appendix \ref{subsecZwmu} by choosing
$u$ and $v$ such that $u-v=\rho+\nu\tau$ while avoiding the poles in
$u$ and $v$. We will find below that the choice $u=\rho + \mu
\tau$ and $v=\frac{1}{2}\tau$ is particularly convenient. From Appendix \ref{subsecZwmu}, we find that a candidate for the
completed function is
\be
\label{hatFmu}
\begin{split}
\widehat F_\mu(\tau,\bar \tau,\rho,\bar \rho)=-i\,e^{\pi i \nu}\,
q^{-\nu^2/2}\,w^{-\nu}\left(M(\tau,\rho+\mu\tau,\tfrac{1}{2}\tau)+\tfrac{i}{2}R(\tau,\bar \tau,\rho+\nu\tau, \bar \rho+\nu\bar \tau)\,\right).
\end{split}
\ee
We will find that for this choice of $u$ and $v$, the holomorphic part $F_{\frac{1}{2}}(\tau,\rho)$ reduces to $F_\frac{1}{2}(\tau)=F(\tau)$
(\ref{feta3}) for $\rho=0$. Indeed, substitution of this choice in $M(\tau,u,v)$
gives for $F_{\frac{1}{2}}(\tau,\rho)$
\be
\label{F1candidate}
F_{\frac{1}{2}}(\tau,\rho)= -\frac{w^{\frac{1}{2}}}{\vartheta_4(\tau)} \sum_{n\in \mathbb{Z}}  \frac{(-1)^n\,q^{n^2/2-\frac{1}{8}}}{1-w\,q^{n-\frac{1}{2}}},
\ee
which satisfies $F_{\frac{1}{2}}(\tau,0)=F_{\frac{1}{2}}(\tau)$.

Let us move on to $\mu\in \mathbb{Z}$, or $\mu=0$ to be specific. The function
$\widehat F_0(\tau,\bar \tau,\rho,\bar \rho)$ (\ref{hatFmu}) evaluates then to
\be
\label{F0candidate}
\begin{split}
\widehat F_0(\tau,\bar \tau,\rho,\bar
\rho)&=-\frac{i\,w}{\vartheta_4(\tau)} \sum_{n\in \mathbb{Z}}
\frac{(-1)^n\,q^{n^2/2+n}}{1-w\,q^n}\\
&\quad -\tfrac{i}{2}\,q^{-\frac{1}{8}}\,w^{\frac{1}{2}}
R(\tau,\bar \tau,\rho-\tfrac{1}{2} \tau,\bar \rho-\tfrac{1}{2} \bar
\tau).
\end{split}
\ee
where we used
$\vartheta_1(\tau,\frac{1}{2}\tau)=-i\,q^{-\frac{1}{8}}\vartheta_4(\tau)$. Note
that (\ref{F0candidate}) contains a pole for $\rho=0$, since $n=0$ is included in the
sum. We will discuss this in more detail later. Let us mention first
that we have to be careful with singling out the holomorphic part of
(\ref{F0candidate}), i.e. the part which does not vanish in the
limit $y\to \infty$, $b\to 0$,  keeping $\rho$ and $\tau$ fixed.
Since the elliptic argument of $R$ is shifted by $-\frac{1}{2}\tau$,
we have $\lim_{y\to \infty} q^{-\frac{1}{8}}\,w^{\frac{1}{2}}
R(\tau,\bar \tau,-\frac{1}{2}\tau,-\frac{1}{2}\bar \tau)=1$. The holomorphic
part of (\ref{F0candidate}) is thus
\be
\label{FmuZ}
F_0(\tau,\rho)=\frac{i}{2}-\frac{i}{\vartheta_4(\tau)} \sum_{n\in \mathbb{Z}} \frac{(-1)^n\,q^{n^2/2}}{1-w\,q^n}.
\ee

To write the non-holomorphic part, we define for $\mu\in
\{0,\frac{1}{2}\} \mod \mathbb{Z}$,
\be
\begin{split}
R_\mu(\tau, \bar
\tau,\rho,\bar \rho)&=-e^{\pi
  i\nu}\delta_{\nu,\mathbb{Z}+\frac{1}{2}}+e^{\pi i\nu}\,q^{-\nu^2/2}w^{-\nu} R(\tau,\bar
\tau,\rho+\nu\tau,\bar \rho+\nu \bar \tau)\\
& =-i\,e^{\pi i \mu}\sum_{n\in \mathbb{Z}+\mu}\left(\sgn(n)-\mathrm{Erf}((n+b)\sqrt{2\pi
  y})\right)\,(-1)^{n-\mu}\,e^{-2\pi i \rho n} q^{-n^2/2}.
\end{split}
\ee
Note that $R_\mu$ vanishes in the limit $y\to\infty$ with $b=0$ for
any $\mu$. Using these expressions, we can write the completed
functions $\widehat F_\mu$ as
\be
\widehat F_\mu(\tau, \bar
\tau,\rho,\bar \rho)=F_\mu(\tau, \rho)+\tfrac{1}{2} R_\mu(\tau, \bar
\tau,\rho,\bar \rho).
\ee

We mentioned in the previous subsection the connection to dimensions of representations of sporadic groups. Other arithmetic
information that has appeared in the context of the $u$-plane integral are
 the Hurwitz class numbers  \cite{Moore:1997pc}, which count binary integral quadratic forms with
 fixed determinant. Using relations for the Appell-Lerch sum (\ref{mushifts}), we can
make this connection more manifest for the $F_\mu$. To this end, let us consider the functions
\be
\begin{split}
g_0(\tau,z)&=\frac{1}{2}+\frac{q^{-\frac{3}{4}}e^{10\pi iz}}{\vartheta_2(2\tau,2z)}\sum_{n\in
\mathbb{Z}} \frac{q^{n^2+n}e^{-4\pi i n z}}{1-e^{8\pi i z}q^{2n-1}}, \\
g_1(\tau,z)&=\frac{q^{-\frac{1}{4}}e^{6\pi i z}}{\vartheta_3(2\tau,2z)}\sum_{n\in
\mathbb{Z}} \frac{q^{n^2}e^{-4\pi i n z}}{1-e^{8\pi i z}q^{2n-1}}.
\end{split}
\ee
These functions appear in the refined partition function of $SU(2)$
and $SO(3)$ Vafa-Witten theory on $\mathbb{P}^2$ \cite{Vafa:1994tf,
  Yoshioka:1994, Bringmann:2010sd}. They vanish for $z\to
0$, while their first derivative give
generating functions of Hurwitz class numbers $H(n)$ \cite{Bringmann:2010sd}:
\be
\lim_{z\to 0}
\frac{1}{4\pi i}\frac{\partial g_j(\tau,z)}{\partial z} = 3\sum_{n\geq 0} H(4n-j)\,q^{n-\frac{j}{4}}.\ee Using
(\ref{mushifts}), we can express the functions $g_j$ in terms of $F_\mu$ as
\be
\begin{split}
g_0(\tau/2,z)&=-i\,F_0(\tau,-3z+\tfrac{1}{2})-\frac{i\,\eta(\tau)^3\,\vartheta_1(\tau,2z)\,\vartheta_3(\tau,z)}{\vartheta_2(\tau,3z)\,\vartheta_4(\tau)\,\vartheta_4(\tau,2z)\,\vartheta_2(\tau,z)},\\
g_1(\tau/2,z)&=-i\,F_\frac{1}{2}(\tau,-3z+\tfrac{1}{2})-\frac{i\,\eta(\tau)^3\,\vartheta_1(\tau,2z)\,\vartheta_2(\tau,z)}{\vartheta_3(\tau,3z)\,\vartheta_4(\tau)\,\vartheta_4(\tau,2z)\,\vartheta_3(\tau,z)},\\
\end{split}
\ee
which demonstrates the connection of the integrand to the class numbers.

It might come as a surprise that the expressions we have defined
give  well-defined power series in $\bfx$ after integration, since the integrand
involves expressions with poles in $\bfx$. This could be avoided
by the addition of a meromorphic Jacobi form of weight 2 and index 0,
with a pole at $\rho=0$ with opposite residue. The reason is that the
 addition of such a meromorphic Jacobi form does not alter the value
of the integral. To see this, note that a meromorphic Jacobi form
$\phi$ of weight 2 and index 0 has a Laurent expansion in $\bfx$ of the form
\be
\phi(\tau,\bfrho)=\sum_{{\boldsymbol \ell}} \phi_{\boldsymbol
  \ell}(\tau)\, \bfx^{\boldsymbol \ell},
\ee
where ${\boldsymbol \ell}=(\ell_1,\dots, \ell_{b_2})\in \mathbb{N}^{b_2}$ and $\bfx^{\boldsymbol
  \ell}=x_1^{\ell_1}\cdots x_{b_2}^{\ell_{b_2}}$. The
$\phi_{\boldsymbol \ell}$ are weakly holomorphic modular forms for $\Gamma^0(4)$ of weight $2$,
since $\bfx$ is invariant under $\Gamma^0(4)$. Mapping the six
images of $\CF_\infty$ in $\mathbb{H}/\Gamma^0(4)$ to $\CF_\infty$
gives us a meromorphic Jacobi form $\widetilde \phi$ for
$\operatorname{SL}(2,\mathbb{Z})$ with expansion
\be
\widetilde \phi(\tau,\bfrho)=\sum_{{\boldsymbol \ell}} \widetilde \phi_{\boldsymbol \ell} (\tau)\, \bfx^{\boldsymbol \ell},
\ee
where the $\widetilde \phi$ are modular forms for $\operatorname{SL}(2,\mathbb{Z})$ of
weight 2. These have a vanishing constant term as discussed before,
and thus do not contribute to $\Phi^J_\bfmu$.

To illustrate this, we present an alternative for $F_0(\tau,\rho)$
(\ref{FmuZ}),
\be
\frac{i}{2}-\frac{i}{\vartheta_4(\tau)} \sum_{n\in \mathbb{Z}}
\frac{(-1)^n\,q^{n^2/2}}{1-w\,q^n}-\frac{i}{\vartheta_4(\tau,\rho)}\,\partial_\rho\ln\!\left(\frac{\vartheta_1(\tau,\rho)}{\vartheta_4(\tau,\rho)}\right).
\ee
This series is analytic for $\rho\to 0$, and can be expressed as
\be
\begin{split}
&\frac{1}{\eta(\tau)^3}\sum_{k_1\in
  \mathbb{Z} \atop k_2\in \mathbb{Z}+\frac{1}{2}} \left(
  \sgn(k_1+k_2)-\sgn(k_1)\right)\, k_2\, (-1)^{k_1+k_2}\,e^{2\pi i \rho k_1}\,q^{\frac{1}{2}(k_2^2-k_1^2)}.
\end{split}
\ee
This is the series in terms of which G\"ottsche expressed the
Donaldson invariants of $\mathbb{P}^2$ \cite[Theorem
3.5]{Gottsche:1996}.

Let us return to the evaluation of $\Phi_\bfmu^J[e^{I_-(\bfx)}]$. To
this end, we also need to determine the magnetic dual
versions $F_{D,\mu}$. We let $w_D=e^{2\pi i \rho_D}$, and
define $\widehat F_{D,\mu}$ by
\be
\widehat F_{D,\mu}(\tau,\bar \tau,\rho_D,\bar
\rho_D)=-(-i\tau)^{-\frac{1}{2}}\,e^{\frac{\pi i
    \rho_D^2}{\tau}}\,\widehat F_{\mu}(-1/\tau,-1/\bar
\tau,\rho_D/\tau,\bar \rho_D/\bar\tau).
\ee
We evaluate the rhs using the transformation of $\widehat M$
(\ref{mucomplete}). Subtracting the subleading non-holomorphic part
gives for $F_{D,\mu}(\tau,\rho_D)$
\be
F_{D,\mu}(\tau,\rho_D)= - \frac{w_D^\frac{1}{2}}{\vartheta_2(\tau)}\sum_{n\in \mathbb{Z}} \frac{q^{n(n+1)/2}}{1-(-1)^{2\mu}\,w_D\,q^n},
\ee
which indeed reduces for $\mu=\frac{1}{2}$ and $\rho_D\to 0$ to $F_D$ (\ref{FDtau}).

Having determined $F_{D,\mu}(\tau,\rho_D)$, we can write down our
final expression for $\Phi^J_\bfmu[ e^{I_-(\bfx)}]$ for four-manifolds
with an odd intersection form. Similarly to Section \ref{uplanemock},
we express $\Phi^J_\bfmu[ e^{I_-(\bfx)}]$, as a sum of three terms, one from each
cusp,
\be\label{eq:OddLattSurfaceFinal}
\begin{split}
\Phi^J_\bfmu[ e^{I_-(\bfx)}]&=-i (-1)^{\mu_1(K_1-1)}\sum_{s=1}^3 \Phi^J_{s,\bfmu}[e^{I_-(\bfx)}],
\end{split}
\ee
with
\be
\non
\begin{split}
\Phi^J_{1,\bfmu}[e^{I_-(\bfx)}] & = 4\left[\tilde
\nu(\tau)\,e^{\bfx^2\,T(u)}\,F_{\mu_1}(\tau,\rho_1)\,\Theta_{L_-,\bfmu_-}(\tau,\bfrho_-)\right]_{q^0},  \\
\Phi^J_{2,\bfmu}[e^{I_-(\bfx)}]&= \left[\tilde
\nu_D(\tau)\,e^{\bfx^2\,T_D(u_D)}\,F_{D,\mu_1}(\tau,\rho_{D,1})\,\Theta_{D,L_-,\bfmu_-}(\tau,\bfrho_{D,-})\right]_{q^0},\\
\Phi^J_{3,\bfmu}[e^{I_-(\bfx)}]&=i\,e^{-2\pi i\bfmu^2}\left[\tilde
\nu_D(\tau)\,e^{-\bfx^2\,T_D(u_D)}\,F_{D,\mu_1}(\tau,-i\rho_{D,1})\,\Theta_{D,L_-,\bfmu_-}(\tau,-i\bfrho_{D,-})\right]_{q^0}.
\end{split}
\ee

Note that for this choice of $J$, $\Phi^J_\bfmu[e^{I_-(\bfx)}]$ only depends on $\bfmu$, $K$ and $b_2$ (assuming $b_1=0$). We list $\Phi^J_\bfmu[I^s_-(\bfx)]$ for the
first few non-vanishing $s$ in Table \ref{TableDon3}. If we specialize
to the four-manifold $\mathbb{P}^2$,
these results are in agreement with the results in Reference
\cite[Theorem 4.2 and Theorem 4.4]{ellingsrud1995wall}.

\begin{center}
\begin{table}[]
\centering
\begin{tabular}{| l | r |}   \hline
$s$ & $\Phi_\bfmu^J[I_-^s(\bfx)]$ \\ \hline
1 & $-\frac{3}{2}$ \\
5  & 1 \\
9 & $3$ \\
13 & $54$ \\
17 & $2\,540$   \\
\hline
\end{tabular}
\hspace{2cm}
\begin{tabular}{| l | r |}   \hline
$s$ & $\Phi_\bfmu^J[I_-^s(\bfx)]$  \\ \hline
0 & 1 \\
4  & $3\cdot 2^{-4}$ \\
8 & $29\cdot 2^{-5}$ \\
12 & $69525\cdot 2^{-12}$ \\
16 & $6\,231\,285\cdot 2^{-13}$   \\
\hline
\end{tabular}
\caption{For a smooth four-manifold with $(b_1,b_2^+)=(0,1)$ and odd intersection form, these tables list non-vanishing $\Phi_\bfmu^J[I_-^s(\bfx)]$
  for $0\leq s \leq 17$ and $\bfx=(1,{\bf 0})$. No assumption is made 
  about the value of $b_2$.  The left table is for
  $\mu_1\in \mathbb{Z}$, and the right table is for $\mu_1=\frac{1}{2}+\mathbb{Z}$. For
  $\mu_1\in \mathbb{Z}$, $I_-(\bfx)^s$ is an integral class, while for
  $\mu_1\in \frac{1}{2}+\mathbb{Z}$, $2^s\,I_-(\bfx)^s$ is an integral class.
   The first entry at $s=1$ is fractional, but (we believe) this arises because the 
   moduli space is a stack with nontrivial stabilizer group.}
\label{TableDon3}
\end{table}
\end{center}

\subsubsection*{Even intersection form}
We proceed similarly for the case that the lattice $L$ is even. As in
the discussion of Section \ref{subsec_part}, we choose for the
period point $J=\frac{1}{\sqrt{2}}\,(1,1,{\bf 0})\in L \otimes
\mathbb{R}$. To factor the sum over fluxes $\Psi_\bfmu^J$ in the
presence of the surface observable, we introduce the vector
$C=\frac{1}{\sqrt{2}}\,(1,-1,{\bf 0})\in L \otimes \mathbb{R}$. The
vectors $J$ and $C$ form an orthonormal basis of $\mathbb{I}^{1,1}\otimes \mathbb{R}
\subset L \otimes \mathbb{R}$.
We denote by $\rho_+$ and $\rho_-$ the projections of the elliptic
variable $\bfrho\in L \otimes \mathbb{C}$ to $J$
and $C$,
\be
\rho_+=\sqrt{2}\,B(\bfrho,J),\qquad \rho_-=\sqrt{2}\,B(\bfrho,C).
\ee
With respect to the basis (\ref{lattice_even}), $\bfrho$ reads
\be
\bfrho=(\rho_+,\rho_-,\bfrho_n),
\ee
with $\bfrho_n\in nL_{E_8}\otimes \mathbb{C}$. As in the case of the
partition function (\ref{Psievenfact}), the sum over fluxes
$\Psi_\bfmu^J(\tau,\bar \tau,\bfrho,\bar \bfrho)$
(\ref{FluxesIpm}) factors,
\be
\Psi_\bfmu^J(\tau,\bar \tau,\bfrho, \bar
\bfrho)=\Psi_{\mathbb{I},(\mu_+,\mu_-)}(\tau,\bar \tau,\rho_-,\rho_+,\bar \rho_+)\,\Theta_{nE_8,\bfmu_n}(\tau,\bfrho_n),
\ee
with
\be\label{eq:Psi-mu-nonzero}
\begin{split}
&\Psi_{\mathbb{I},(\mu_+,\mu_-)}(\tau,\bar\tau,\rho_-,\rho_+,\bar \rho_+)=\exp(-\pi y
\,b_+^2)\sum_{\bfk\in \mathbb{I}^{1,1}+(\mu_+,\mu_-)}
\partial_{\bar \tau}(\sqrt{y}(k_1+k_2+b_+)) \\
&\qquad  \times\, (-1)^{k_1 K_2+k_2K_1}\,q^{(k_1-k_2)^2/4}\, \bar
q^{(k_1+k_2)^2/4}\,e^{\pi i \rho _- (k_1-k_2)-\pi i \bar \rho_+ (k_1+k_2) },
\end{split}
\ee
where $b_+=\mathrm{Im}(\rho_+)/y$, and
\be
\Theta_{nE_8,\bfmu_n}(\tau,\bfrho_n)=\sum_{\bfk_n\in nL_{E_8}+\bfmu_n} q^{-\bfk_n^2/2}\,e^{-2\pi i B(\bfrho_n,\bfk_n)}.
\ee
The modular transformations are easily determined if we express
$\Theta_{nE_8,\bfmu_n}(\tau,\bfrho_n)$ in terms of Jacobi theta
series. We define the dual theta series  $\Theta_{D,nE_8,\bfmu_n}$ as
\be
\label{ThetaE8Dual}
\begin{split}
\Theta_{D,nE_8,\bfmu_n}(\tau,\bfrho_{D,n})&=\tau^{-4n}\, e^{-\pi i \rho_{D,n}^2/\tau}\,
\Theta_{nE_8,\bfmu_n}(-1/\tau,\bfrho_{D,n}/\tau)\\
&=\sum_{\bfk_n\in nL_{E_8}} (-1)^{2B(\bfmu_n,\bfk_n)} q^{-\bfk_n^2/2}\,e^{-2\pi i B(\bfrho_{D,n},\bfk_n)}.
\end{split}
\ee

Unlike equation \eqref{eq:Psi-but-zero}, (\ref{eq:Psi-mu-nonzero}) is definitely
nonzero. The series can be decomposed further as
\be
\label{PsibbI}
\begin{split}
&\Psi_{\mathbb{I},(\mu_+,\mu_-)}(\tau,\bar \tau,\rho_-, \rho_+,\bar \rho_+)=\\
& \qquad -i\,(-1)^{\mu_+K_+-\mu_-K_-}\sum_{j=0,1}
\,h_{\mu_{+}+j}(\tau,\bar
\tau,\rho_+,\bar \rho_+)\,t_{\mu_-+j}(\tau,\rho_-),
\end{split}
\ee
with
\be
\begin{split}
h_{\nu}(\tau,\bar \tau,\rho, \bar \rho) & =i\,\exp\!\left(-\pi\, y\,b^2
\right)\sum_{n\in 2\mathbb{Z}+\nu} \partial_{\bar
  \tau}\left(\sqrt{y}(n+b)  \right) \,\bar
q^{n^2/4}\,e^{-\pi i\,\bar \rho\,n},\\
t_{\nu}(\tau,\rho)& =\sum_{n\in
  2\mathbb{Z}+\nu} q^{n^2/4}\,e^{\pi i\,\rho\,n}\\
& = e^{\pi i \rho\,\nu} q^{\nu^2/4}\, \vartheta_{3}(2\tau, \rho+\nu\tau).
\end{split}
\ee
where $b=\mathrm{Im}(\rho)/y$. These functions reduce to those in
(\ref{gtheta}) in the limit $\bfrho \to 0$.

While $\Psi_{\mathbb{I},(\mu_+,\mu_-)}$ is a modular form for
$\Gamma^0(4)$, the functions $h_\nu$ and $t_\nu$ are not. To
continue working with modular forms for $\Gamma^0(4)$,
we rewrite $\Psi_{\mathbb{I},(\mu_+,\mu_-)}$ (\ref{PsibbI}) as
\be
\begin{split}
&\Psi_{\mathbb{I},(\mu_+,\mu_-)}(\tau,\bar \tau, \rho_-,\rho_+,\bar \rho_+)=
-i\,(-1)^{\mu_+K_+-\mu_-K_-} \\
&\qquad  \times \left(g^+_{\mu_+}(\tau,\bar
  \tau,\rho_+,\bar \rho_+)\,\theta^+_{\mu_-}(\tau,\rho_-)+g^-_{\mu_+}(\tau,\bar
  \tau,\rho_+,\bar \rho_+)\,\theta^-_{\mu_-}(\tau,\rho_-)\, \right),
\end{split}
\ee
with
\be
\begin{split}
g^\pm_{\nu}(\tau,\bar \tau,\rho,\bar \rho)&= \tfrac{1}{2}\left( h_\nu(\tau,\bar
\tau,\rho,\bar \rho)\pm
h_{\nu+1}(\tau,\bar \tau,\rho,\bar \rho)\right),\\
\theta^\pm_\nu(\tau,\rho)&= t_\nu(\tau,\rho)\pm t_{\nu+1}(\tau,\rho).
\end{split}
\ee
These functions are modular forms for $\Gamma^0(4)$, which becomes manifest when we express
them in terms of functions we encountered before. We can express
the $g^\pm_\nu$ in terms of $f_\nu$,
\be
\begin{split}
g^+_\nu(\tau,\bar \tau,\rho,\bar \rho)&= \tfrac{1}{2}\,f_\nu(\tau/2,\bar
\tau/2,(\rho+1)/2,(\bar \rho+1)/2),\\
g^-_\nu(\tau,\bar \tau,\rho,\bar \rho)&= \tfrac{1}{2}\,e^{\pi i \nu}\,f_\nu(\tau/2,\bar
\tau/2,\rho/2,\bar \rho/2).
\end{split}
\ee
The $\theta^\pm_\nu$ can be expressed in terms of the Jacobi theta functions $\vartheta_j$ as
\be
\begin{split}
\theta^+_\nu(\tau,\rho)&= \left\{ \begin{array}{ll}
    \vartheta_3(\tau/2,\rho/2), & \qquad \qquad \quad \nu=0\mod
    \mathbb{Z},\\ \vartheta_2(\tau/2,\rho/2),  & \qquad \qquad \quad \nu=\frac{1}{2}\mod
    \mathbb{Z},  \end{array} \right.\\
\theta^-_\nu(\tau,\rho)&=\left\{ \begin{array}{ll} (-1)^\nu\,
    \vartheta_4(\tau/2,\rho/2), & \qquad\,\,  \nu=0\mod
    \mathbb{Z},\\ -\,e^{\pi i \nu}\,\vartheta_1(\tau/2,\rho/2),  &
    \qquad \,\,\nu=\frac{1}{2}\mod
    \mathbb{Z}.  \end{array} \right.
\end{split}
\ee
We define the dual functions as
\be
\label{thetanuD}
\theta^\pm_{D,\nu}(\tau,\rho_D)=(-2i\tau)^{-\frac{1}{2}}\,e^{-\frac{\pi
    i \rho_D^2}{2\tau}}\theta^\pm_{D,\nu}(-1/\tau,\rho_D/\tau).
\ee
These are explicitly given by
\be
\begin{split}
\theta^+_{D,\nu}(\tau,\rho_D)&= \left\{ \begin{array}{ll}
    \vartheta_3(2\tau,\rho_D), & \qquad \qquad \quad \nu=0\mod
    \mathbb{Z},\\ \vartheta_4(2\tau,\rho_D),  & \qquad \qquad \quad \nu=\frac{1}{2}\mod
    \mathbb{Z},  \end{array} \right.\\
\theta^-_{D,\nu}(\tau,\rho_D)&=\left\{ \begin{array}{ll} (-1)^\nu\,
    \vartheta_2(2\tau,\rho_D), & \qquad\,\,  \nu=0\mod
    \mathbb{Z},\\ i\,e^{\pi i \nu}\,\vartheta_1(2\tau,\rho_D),  &
    \qquad \,\,\nu=\frac{1}{2}\mod
    \mathbb{Z}.  \end{array} \right.
\end{split}
\ee

Since the $g^\pm_\nu$ can be expressed in terms of the $f_\nu$, we can determine anti-derivatives $\widehat
G^\pm_\nu$ in terms of $\widehat F_\mu$. Namely,
\be
\begin{split}
\widehat G^+_\nu(\tau,\bar \tau,\rho,\bar \rho)&= \widehat
F_\nu(\tau/2,\bar \tau/2,(\rho+1)/2,(\bar \rho+1)/2), \\
\widehat G^-_\nu(\tau,\bar \tau,\rho,\bar \rho)&= e^{\pi i \nu} \widehat F_\nu(\tau/2,\bar \tau/2,\rho/2,\bar
\rho/2).
\end{split}
\ee
The holomorphic parts of these completed functions are
\be
\begin{split}
G^+_\nu(\tau,\rho)&= F_\nu(\tau/2,(\rho+1)/2),\\
G^-_\nu(\tau,\rho)&= e^{\pi i \nu}\, F_\nu(\tau/2,\rho/2),
\end{split}
\ee
with the $F_\nu$ given by (\ref{F1candidate}) and
(\ref{FmuZ}). We define the dual $\widehat G^\pm_{D,\nu}$ as
\be
\label{hatGD}
\begin{split}
\widehat
G^\pm_{D,\nu}(\tau,\bar \tau,\rho_D,\bar \rho_D)&=-(-2i\tau)^{-\frac{1}{2}}\,e^{\frac{\pi i
    \rho_D^2}{2\tau}}\,\widehat G^\pm_{\nu}(-1/\tau,-1/\bar
\tau,\rho_D/\tau,\bar \rho_D/\bar \tau).
\end{split}
\ee
These evaluate to
\be
\begin{split}
G^+_{D,\nu}(\tau,\rho_D)&=-\frac{1}{2}+\frac{q^{\frac{1}{4}}}{\vartheta_2(2\tau)}\sum_{n\in
\mathbb{Z}} \frac{q^{n(n+1)}}{1-(-1)^{2\nu}w_Dq^{2n+1}},\\
G^-_{D,\nu}(\tau,\rho_D)&=-\frac{e^{\pi i
    \nu}w_D^\frac{1}{2}}{\vartheta_2(2\tau)}\sum_{n\in \mathbb{Z}} \frac{q^{n(n+1)}}{1-(-1)^{2\nu}w_Dq^{2n}}.
\end{split}
\ee

With these expressions, we can present our final expression $\Phi_\bfmu^J[e^{I_-(\bfx)}]$
for four-manifolds with even intersection form,
\be\label{eq:EvenLattSurfaceFinal}
\begin{split}
\Phi_\bfmu^J[e^{I_-(\bfx)}]&=-i\,(-1)^{\mu_+K_+-\mu_-K_-} \sum^3_{s=1} \Phi_{s,\bfmu}^J[e^{I_-(\bfx)}],
\end{split}
\ee
with
\be
\begin{split}
\Phi_{1,\bfmu}^J[e^{I_-(\bfx)}]&=4
\left[ \tilde
  \nu(\tau) \,e^{\bfx^2\,T(u)}\,\Theta_{nE_8,\bfmu_n}(\tau,\bfrho_n)\sum_{\pm} G^\pm_{\mu_+}(\tau,\rho_+)\,\theta^\pm_{\mu_-}(\tau,\rho_-)
\right]_{q^0},\\
\Phi_{2,\bfmu}^J[e^{I_-(\bfx)}] & =2 \left[ \tilde
  \nu_D(\tau)
  \,e^{\bfx^2\,T_D(u_D)}\,\Theta_{D,nE_8,\bfmu_n}(\tau,\bfrho_{D,n})\right.\\
& \left. \qquad \times \sum_{\pm} G^\pm_{D,\mu_+}(\tau,\rho_{D,+})\,\theta^\pm_{D,\mu_-}(\tau,\rho_{D,-})
\right]_{q^0}, \\
\Phi_{3,\bfmu}^J[e^{I_-(\bfx)}] & =2 i\,e^{-2\pi i\bfmu^2}\left[ \tilde
  \nu_D(\tau)
  \,e^{-\bfx^2\,T_D(u_D)}\,\Theta_{D,nE_8,\bfmu_n}(\tau,-i\bfrho_{D,n})
  \right. \\
& \left. \qquad \times \sum_{\pm} G^\pm_{D,\mu_+}(\tau,-i\rho_{D,+})\,\theta^\pm_{D,\mu_-}(\tau,-i\rho_{D,-})
\right]_{q^0}.
\end{split}
\ee
The overall factor 2 for the strong coupling contributions is
due to the factors of $\sqrt{2}$ in (\ref{thetanuD}) and (\ref{hatGD}).

Table \ref{TableDonEven} lists the contribution from the $u$-plane to
Donaldson polynomials for small
instanton number. The expressions confirm that $I_-(\bfx)$ is
an integral class for gauge group SU$(2)$ $\bfmu={\bf 0} \mod \mathbb{Z}$, and
half-integral for $\bfmu \neq {\bf 0} \mod \mathbb{Z}$.

\begin{center}
\begin{table}[]
\centering
\begin{tabular}{| R{1.4cm} | R{13cm}|}   \hline
$s_1+s_2$ &   $P_\bfmu(x_1,x_2)$ for $\bfmu=(0,0)$\\ \hline
1 & \hspace{10.5cm} $-2\,x_1-2\,x_2$\\
5  &  $x_1^5-x_1^4\,x_2+x_1^3\,x_2^2+x_1^2\,x_2^3-x_1\,x_2^4+x_2^5$ \\
9 &  $-40\,x_1^9 +24\, x_1^8\,x_2-12\,x_1^7\,x_2^2+4\,x_1^6\,x_2^3+4\,x_1^3\,x_2^6-12\,x_1^7\,x_2^2+24\,x_1\,x_2^8-40\,x_2^9$ \\
\hline
\end{tabular}
\vspace{.4cm}\\
\begin{tabular}{| R{1.4cm}  | R{13cm} |}   \hline
$s_1+s_2$ & $P_\bfmu(x_1,x_2)$ for $\bfmu=(\frac{1}{2},0)$  \\ \hline
1 & $-x_1+2\,x_2$ \\
5  &  $\frac{31}{16}\,x_1^5-\frac{7}{4}\,x_1^4\,x_2+x_1^3\,x_2^2$ \\
9 &\tiny{
$-\frac{757}{256}\,x_1^9-\frac{465}{128}\,x_1^8\,x_2+\frac{699}{64}\,x_1^7\,x_2^2-\frac{305}{32}\,x_1^6\,x_2^3+\frac{243}{16}\,x_1^5\,x_2^4-\frac{81}{8}\,x_1^4\,x_2^5+\frac{27}{4}x_1^3\,x_2^6-\frac{9}{2}x_1^2\,x_2^7+3\, x_1\,x_2^8-2\,x_2^9$}
\\
\hline
\end{tabular}
\vspace{.4cm}\\
\begin{tabular}{| R{1.4cm}  | R{13cm} |}   \hline
$s_1+s_2$ & $P_\bfmu(x_1,x_2)$ for $\bfmu=(\frac{1}{2},\frac{1}{2})$  \\ \hline
3 & $\frac{13}{4}\,x_1^3-\frac{3}{4}\,x_1^2\,x_2-\frac{3}{4}\,x_1\,x_2^2+\frac{13}{4}\,x_2^3$ \\
7  & \small{ $-\frac{143}{32}\,x_1^7-\frac{275}{32}\,x_1^6\,x_2+\frac{229}{32}\,x_1^5\,x_2^2-\frac{71}{32}\,x_1^4\,x_2^3-\frac{71}{32}\,x_1^3\,x_2^4+\frac{229}{32}\,x_1^2\,x_2^5-\frac{275}{32}\,x_1\,x_2^6 -\frac{143}{32}\,x_2^7$} \\
\hline
\end{tabular}
\caption{ Let $M$ be a smooth four-manifold
  with $(b_1,b_2^+,b_2^-)=(0,1,1)$, even intersection form, $K_{1,2}
  =2\mod 4$ and period point $J$ (\ref{Jeven}). Examples of such
  manifolds are $S^2\times S^2$, and the Hirzebruch surfaces
  $\mathbb{F}_n$ with $n$ even. Let $\bfx_1=(x_1,0)$
  and $\bfx_2=(0,x_2)\in L\otimes \mathbb{R}$ in the
  basis (\ref{lattice_even}).
The tables list the non-vanishing polynomials $P_\bfmu(x_1,x_2)=\sum_{s_1,s_2}\Phi_\bfmu^J[ I_-(\bfx_1)^{s_1}
  I_-(\bfx_2)^{s_2}  ]$ with $s_1+s_2\leq 9$, and 't Hooft flux
  $\bfmu=(\mu_1,\mu_2)=(0,0)$, $(\frac{1}{2},0)$ and
  $(\frac{1}{2},\frac{1}{2})\mod \mathbb{Z}^2$ . The polynomials for $(\mu_1,\mu_2)=(0,\frac{1}{2})$
  follow from those for $(\mu_1,\mu_2)=(\frac{1}{2},0)$ by the exchange $x_1\leftrightarrow x_2$. }
\label{TableDonEven}
\end{table}
\end{center}

\section{Asymptology of the $u$-plane integral}
\label{asympbehavior}

Up to this point we have treated the $u$-plane integral
$\Phi_\bfmu^J[e^{2pu +I_-(\bfx)}]$ as a formal generating series in
the homology elements $p$ and $\bfx$. However, one might ask if the integral
actually expresses a well-defined function on the homology of the four-manifold $M$.
In other words, one might ask if the formal series is in fact convergent. The contribution
of the Seiberg-Witten invariants is a finite sum and hence in fact defines an entire
function on $H_*(M,\IC)$. Therefore the Donaldson-Witten partition function is a well-defined
function on the homology if and only if the $u$-plane is a well-defined function.
If that were the case then one could explore interesting questions such as the analytic
structure of the resulting partition function, which, in turn, might signal interesting
physical effects. In this Section we will explore that question,
starting with the point observable in Subsection \ref{pointobs}.
We will find strong evidence that in fact the $u$-plane integral is indeed
an entire function of $p$.

The situation for $\bfx$ is less clear, since the numerical results
are more limited. We discuss in Subsection \ref{pointobs} that the
results do suggest that $\Phi_\bfmu^J[e^{2pu
  +I_-(\bfx)}]$ is also an entire function $\bfx$. As a
step towards understanding the analytic structure in $\bfx$, we will consider in Subsection
 \ref{asymp_behavior} the magnitude of the integrand in the
weak-coupling limit. Although the integral is independent of $\alpha$ 
we will see that the integrand behaves best when $\alpha=1$.

\subsection{Asymptotic growth of point observables}
\label{pointobs}
We will analyze the dependence of the contribution from the $u$-plane
$\Phi^J_\bfmu[e^{2p\,u}]$ to the correlation function
$\left<e^{2p\,u}\right>^J_\bfmu$.  Due to the
exponential divergence of $u$ (\ref{utau}) for $\tau\to i\infty$, the divergence of
$e^{2p\,u}$ is doubly exponential. The $u$-plane integral thus
formally diverges. The discussion of Section \ref{regularization} does not provide an 
immediate definition of such divergent expressions. On the other hand, the
vev of the exponentiated point observable $e^{2p\,u}$ should be
understood as a generating function of correlation functions, and we
can define $\Phi_\bfmu^J[e^{2p\,u}]$ as
\be
\Phi_\bfmu^J[e^{2p\,u}]=\sum_{\ell=0}^\infty \frac{(2p)^\ell}{\ell!} \,\Phi_\bfmu^J[u^\ell].
\ee
As discussed in Section \ref{uplanemock}, there is no problem evaluating
$\Phi_\bfmu^J[u^\ell]$ using the definition of Section \ref{regularization}.

We consider the case
of odd lattices, and the period point $J$ (\ref{Jodd}). Modifying (\ref{Phiepu}), we express
$\Phi_\bfmu^J[u^\ell]$ as
\be
\begin{split}
\Phi_\bfmu^J[u^\ell]=\frac{1}{192}\int_0^4
d\tau\,\frac{H^{(2)}(\tau)}{\,\eta(\tau)^{9}}\,\vartheta_4(\tau)^{12}\,u(\tau)^\ell-\frac{1}{96}\int_0^1
d\tau\,\frac{H^{(2)}(\tau)}{\,\eta(\tau)^{9}}\,\vartheta_2(\tau)^{12}\,u_D(\tau)^\ell,\\
\end{split}
\ee
We have replaced here $F$ by $\frac{1}{24}H^{(2)}$, since its
completion is an equally good choice of anti-derivative.  It is
straightforward to determine $\Phi_\bfmu^J[u^\ell]$ using this expression.
We list in Table \ref{TableDonGrowth} values of $\Phi_\bfmu^J[u^\ell]$ for
various large values of $\ell$.
\begin{center}
\begin{table}[]
\centering
\begin{tabular}{| r | r |r|r|}   \hline
$\ell$ & $\Phi_\bfmu^J[u^\ell]$ & $\ell\,\Phi_\bfmu^J[u^\ell]$ &
$\log(\ell)\,\ell\,\Phi_\bfmu^J[u^\ell]$ \\ \hline
0 & 1 & 0 & - \\
100  & $5.02\times 10^{-3}$ & 0.502 & 2.3131  \\
200 & $2.25\times 10^{-3}$ & 0.450 & 2.3834   \\
300 & $1.40\times 10^{-3}$ & 0.421 & 2.4032    \\
400 & $1.01\times 10^{-3}$ & 0.402 & 2.4095   \\
500 & $7.76\times 10^{-4}$ &  0.388 &2.4105   \\
600 & $6.28\times 10^{-4}$ & 0.377 & 2.4090   \\
700 & $5.25\times 10^{-4}$ & 0.367 & 2.4062    \\
800 & $4.49\times 10^{-4}$ & 0.359 & 2.4028   \\
900 & $3.92\times 10^{-4}$ & 0.353 & 2.3990   \\
1000 & $3.47\times 10^{-4}$ & 0.347 & 2.3951 \\
2000 & $1.55\times 10^{-4}$ & 0.310 & 2.3574 \\
3000 & $9.69\times 10^{-5}$ & 0.291 & 2.3273   \\
4000 & $6.94\times 10^{-5}$ & 0.278 & 2.3030   \\
\hline
\end{tabular}
\caption{Table with various data on the asymptotics of
  $\Phi_\bfmu^J[u^\ell]$ for large $\ell$. }
\label{TableDonGrowth}
\end{table}
\end{center}
Before giving evidence that the $\Phi_\bfmu^J[e^{2p\,u}]$ is an entire
function of $p$, let us discuss the integrand in more detail. We first write $\Phi_\bfmu^J[u^\ell]$ as an integral from 0 to 1:
\be
\begin{split}
\Phi_\bfmu^J[u^\ell]&=\frac{1}{192}\int_0^1
d\tau\,\frac{H^{(2)}(\tau)}{\,\eta(\tau)^{9}}
\left(\,(1+(-1)^\ell)\,u(\tau)^\ell\,\vartheta_4(\tau)^{12}\right.\\
&\qquad \left. - (1+(-1)^\ell)\,u(\tau-1)^\ell\, \vartheta_3(\tau)^{12}
+2\,u_D(\tau)^\ell\,\vartheta_2(\tau)^{12}\right),\\
\end{split}
\ee
We can express the integrand in a ${\rm SL}(2,\mathbb{Z})$ invariant form. To this end, note
\be
\begin{split}
u(\tau)&=-u(\tau-2)=\frac{\vartheta_4^2(\vartheta_2^4+\vartheta_3^4)}{8\,\eta^{6}}\\
u(\tau-1)&=i\frac{\vartheta_3^2(-\vartheta_2^4+\vartheta_4^4)}{8\,\eta^{6}}\\
u_D(\tau)&=\frac{\vartheta_2^2(\vartheta_4^4+\vartheta_3^4)}{8\,\eta^{6}}\\
\end{split}
\ee
For $\ell$ even, we find thus that  $\Phi_\bfmu^J$ can be expressed as
\be
\Phi_\bfmu^J[u^\ell]=\frac{1}{96}\left[
  \frac{H^{(2)}(\tau)}{8^\ell\,\eta(\tau)^{9+6\ell}}\,Q_\ell(\tau)\right]_{q^0},
 \ee
where $Q_\ell$ is the weight $6+3\ell$ modular form defined by
\be
Q_\ell(\tau)=
    \vartheta_4^{12+2\ell}(\vartheta_2^4+\vartheta_3^4)^\ell-(-1)^{\ell/2}\vartheta_3^{12+2\ell}(\vartheta_4^4-\vartheta_2^4)^\ell+\vartheta_2^{12+2\ell}(\vartheta_4^4+\vartheta_3^4)^\ell.
\ee
The first few terms are
\be
\label{Qellexp}
Q_\ell(\tau)=\left\{
\begin{array}{ll}
8(5\ell-6)\,q^{\frac{1}{2}}+\dots, & \quad \ell=0\mod 4,\\
2+(528-1496\ell+400\,\ell^2)\,q+\dots, &\quad  \ell=2\mod 4.
\end{array}\right.
\ee

The most straightforward way of trying to establish the large $\ell$
asymptotics is by a saddle point analysis. Expressing $u^\ell$ as
$u_D(-1/\tau)^\ell\sim e^{32\,\ell\,e^{-\frac{2\pi i}{\tau}}}$, we
find that, to first approximation, the saddle point is at $\tau_*=\frac{2\pi
  i}{\log(-32\,\ell)}$ for large $\ell$. We find that the contribution of this saddle
point to $\left|\Phi^J_\bfmu[u^\ell] \right|$ behaves as $C/\ell$ for some constant
$C$. We leave an investigation into the difference between the
saddle point contribution and Table \ref{TableDonGrowth} for another
occasion.

Let us explore the consequences of the large $\ell$ asymptotics for $\Phi_\bfmu^J[e^{2pu}]$. We deduce
from the Table \ref{TableDonGrowth} that $\Phi_\bfmu^J[u^\ell]$ also
decreases faster than $C\,(\ell+1)^{-1}$.  This
estimate strongly suggests that radius of convergence for
$\sum_{\ell\geq 0}p^\ell\,\Phi_\bfmu^J[u^\ell]/\ell!$ is infinite
and thus that $\Phi_\bfmu^J[e^{2pu}]$ is an entire
function of $p$. Moreover, we
can easily bound $|\Phi_\bfmu^J[e^{2pu}]|$ for real $p$ by
\be
\left|\,\Phi_\bfmu^J[e^{2pu}]\,\right|< C\,\frac{\sinh(2p)}{2p}.
\ee
The exponentials in $\sinh(2p)$ resemble the
SW contribution at $u=\pm \Lambda^2$. Comparing with the SW-simple
type expression (\ref{SWsimpletype}), we see that the SW contribution
to $\left< u^\ell \right>$ is $O(1)$, while the $u$-plane contribution
is subleading.

While we have focused in this Subsection on the point observables, the
behavior of $\Phi_\bfmu^J[I^s_-(\bfx)]$ for large $s$ is equally if
not more interesting. We list in Table \ref{TableSurfaceAsymp} some numerical data for
 $\log(\Phi[I_-(\bfx)^s])/s$. These data, while admittedly limited, do give the impression that the
 asymptotic growth of $\Phi^J_\bfmu[I_-(\bfx)^s]$ is bounded by $e^{\alpha s\,\log(s)}$ for some
positive constant $\alpha$, and that $\alpha<1$. Assuming that this is
the correct behavior for large $s$, the radius of convergence for $x$
of $\Phi^J_\bfx[e^{I_-(\bfx)}]=\sum_{s\geq 0}
\Phi^J_\bfmu[I_-(\bfx)^s]/s!$ is infinite, implying that
$\Phi^J_\bfx[e^{I_-(\bfx)}]$ is entire in $\bfx$. We hope to get back
to the asymptotics of these correlators in future work.

\begin{center}
\begin{table}[]
\centering
\begin{tabular}{| R{1cm} | R{3.2cm} | R{5cm}|}   \hline
$s$ & $\log(\Phi^J_\bfmu[I_-(\bfx)^s])/s$ & $\log(\Phi^J_\bfmu[I_-(\bfx)^s])/(s\log(s))$  \\ \hline
17 & 0.4612 & 0.1628 \\
37 & 0.9396 & 0.2602 \\
57 &   1.2079  & 0.2987\\
77 &   1.3925 & 0.3206\\
97 &   1.5326 & 0.3350\\
\hline
\end{tabular}
\vspace{.4cm}\\
\begin{tabular}{| R{1cm} | R{3.2cm} | R{5cm}|}   \hline
$s$ & $\log(\Phi^J_\bfmu[I_-(\bfx)^s])/s$ & $\log(\Phi^J_\bfmu[I_-(\bfx)^s])/(s\log(s))$  \\ \hline
20 & 0.5541 & 0.1849 \\
40 & 0.9880 & 0.2678 \\
60 &   1.2395  & 0.3027\\
80 &   1.4157 & 0.3231\\
100 &   1.5509 & 0.3368\\
\hline
\end{tabular}
\caption{For a four-manifold with odd intersection form, these tables list data for non-vanishing $\Phi_\bfmu^J[I_-^s(\bfx)]$
  for $ s \leq 100$ in steps of 20, and with $\bfx=(1,{\bf 0})$. The top table is for
$\mu_1\in \mathbb{Z}$ and the bottom table for $\mu_1\in \frac{1}{2}+\mathbb{Z}$.}
\label{TableSurfaceAsymp}
\end{table}
\end{center}

\subsection{Weak coupling limit of the integrand}
\label{asymp_behavior}

As a first step towards understanding the asymptotic behavior of the series in $\bfx$  we investigate here the
growth of the integrand of the $u$-plane integral in the weak coupling region.  In order to do this it is
useful to recall that one can add to the surface observable
 the operator $I_+(\bfx)$ discussed in
section \ref{subsec_correlation}  with an arbitrary
coefficient $\alpha$. Since  $I_+(\bfx)$ is $\cal{Q}$ exact  such an addition
does not modify the resulting integral. (Because the integral is subtle and formally
divergent this statement requires careful justification, but it turns out to be correct \cite{Korpas:2019ava}.)
 In this way, we can interpolate between
$\alpha=0$ \cite{Moore:1997pc} and $\alpha=1$ \cite{Korpas:2017qdo}.
While the result is independent of $\alpha$, the dependence of the integrand is worth
exploring in more detail, in particular the behavior in the weak
coupling limit. In this limit,
$\frac{du}{da}$ diverges as $q^{-\frac{1}{8}}=e^{-i\frac{\theta}{8}+\frac{\pi y}{4}}$. As a
result, the elliptic variable $\bfrho=\frac{\bfx}{2\pi} \frac{du}{da}$
diverges. This subsection studies this divergence as function of $\alpha$.

The $u$-plane integral with observable $e^{I_-+\alpha\,I_+}$ results
in a modified sum over fluxes $\Psi^J_{\bfmu,\alpha}$. This sum is defined as in
(\ref{FluxesIpm}), but with $\bar \bfrho$ replaced by $\alpha \bar
\bfrho$,  and reads
\be
\label{tildePsiJ}
\begin{split}
\Psi^J_{\bfmu,\alpha}(\tau,\bar \tau, \bfrho,\bar \bfrho)=&e^{\frac{\pi (\bfrho_+-\alpha\bar \bfrho_+)^2}{2y}} \sum_{\bfk\in
  L + \bfmu} \partial_{\bar \tau}
(\sqrt{2y}B(\bfk+\frac{\bfrho-\alpha \bar \bfrho}{2iy}, J)\,(-1)^{B(\bfk, K)} \\
&\times q^{-\bfk_-^2/2} \bar q^{\bfk_+^2/2}\, e^{-2\pi i B(\bfrho, \bfk_-)-2\pi i \alpha B(\bar \bfrho,\bfk_+)}.
\end{split}
\ee
By completing the squares in the exponent, we can write this as
\be
\label{tildePsiJ2}
\begin{split}
\Psi^J_{\bfmu,\alpha}(\tau,\bar \tau, \bfrho,\bar \bfrho)&=e^{\frac{\pi (\bfrho_+-\alpha \bar \bfrho_+)^2}{2y}+\pi i \tau
   \bfb_-^2+\pi i \bar \tau \alpha^2 \bfb_+^2} \sum_{\bfk\in
  L + \bfmu} \partial_{\bar \tau} (\sqrt{2y}B(\bfk+\frac{\bfrho-\alpha \bar \bfrho}{2iy}, J)\,(-1)^{B(\bfk, K)} \\
&\times q^{-(\bfk+ \bfb)_-^2/2} \bar q^{(\bfk+\alpha \bfb)_+^2/2} e^{-2\pi i B(a, \bfk_-)- 2\pi i B(\alpha a, \bfk_+)},
\end{split}
\ee
where $\bfb={\rm Im}(\bfrho)/y$ as before. The sum over $\bfk\in L+\bfmu$ is dominated by the $\bfk$ for which
$-(\bfk+\bfb)_-^2+(\bfk+\bfb)_+^2$ is minimized. For a generic choice
of period point $J$, there is only one $\bfk\in L+\bfmu$ which
minimizes this quantity. The leading asymptotic behavior is
given by the exponential multiplying the sum. This evaluates to
\be
\label{Psi_dexp}
\left|\Psi^J_{\bfmu,\alpha} \right|\sim e^{-\pi y \bfb^2+\frac{\pi(1-\alpha)^2}{2y} |\bfrho_+^2|}.
\ee
Thus we see that $\alpha= 1$ is special, since for this choice the exponent
is negative definite for $\bfx^2>0$. Moreover, for large $y$,
$\Psi^J_{\bfmu,\alpha}$ will only remain finite in the domain
$\varphi=\mathrm{Re}(\tau)\in [0,4]$ for this choice of
parameters. The double exponential divergence of the exponentiated
surface observable is therefore mitigated at $\alpha=1$.

Let us make a rough estimate for the magnitude of the $u$-plane integral using (\ref{Psi_dexp}),
\be
\begin{split}
\left|\Phi_\bfmu^J[e^{I_-(\bfx)+I_+(\bfx)}]\,\right|&\sim\int dy\wedge d\varphi\,e^{\frac{3}{4}\pi y-\frac{\pi \bfx^2}{y} e^{\frac{1}{2}\pi y}
  \sin(\frac{\pi}{4}\varphi)^2}\\
&=\int dy\, e^{\frac{3}{4}\pi y -\frac{\pi \bfx^2}{2y} e^{\frac{1}{2}\pi y}} \int_{-\frac{1}{2}}^{\frac{7}{2}} d\varphi\,e^{\frac{\pi \bfx^2}{2y} e^{\frac{1}{2}\pi y} \cos(\frac{\pi}{2}\varphi)},
\end{split}
\ee
where we only consider terms which are non-vanishing in the limit for $y\to \infty$. The integral over $\varphi$ is a Bessel function $I_0(z)$
with $z=\frac{\pi \bfx^2}{2y} e^{\frac{1}{2}\pi
  y}$, which behaves for large $z$ as $e^z/\sqrt{2\pi z}$. This leads
to a single exponential divergence, $\int dy\, e^{\frac{1}{4}\pi y}$,
which can be treated as discussed before. We leave a more detailed analysis including the dependence on $\bfmu$ and
possible cancellations in the integral along the interval $\varphi\in [0,4]$ for
future work.

\subsection*{Acknowledgments}
We thank Johannes Aspman, Aliakbar Daemi, Elias Furrer, Lothar
G\"ottsche, Jeff Harvey, John Morgan, Tom Mrowka and Hiraku Nakajima for discussions.
GK would like to thank the Institute Of Pure and Applied Mathematics of
UCLA and the Physics Department of the National and Kapodistrian
University of Athens for hospitality. JM would like to thank the
NHETC, Rutgers University for hospitality. JM is supported by the
Laureate Award 15175 “Modularity in Quantum Field Theory and Gravity” of the Irish Research Council.
GM and IN are supported by the US Department of Energy under grant DE-SC0010008.


\appendix


\section{Modular forms and theta functions}
\label{app_mod_forms}
In this appendix we collect a few essential aspects of the theory of
modular forms, Siegel-Narain theta functions and indefinite theta functions. For more comprehensive treatments we refer the reader
to the available literature. See for example \cite{Serre,Zagier92,Bruinier08}.

\subsection*{Modular groups}
The modular group $\operatorname{SL}(2,\mathbb{Z})$ is the group of integer matrices
with unit determinant
\be
\operatorname{SL}(2,\mathbb{Z})=\left\{ \left. \left( \begin{array}{ccc}
a & b   \\
c & d  \end{array} \right) \right| a,b,c,d\in \BZ; \, ad-bc=1\right\}.
\ee
We introduce moreover the congruence subgroup $\Gamma^0(n)$
\begin{equation}
\label{Gamma04}
\Gamma^0(n) = \left\{  \left.\left( \begin{array}{ccc}
a & b   \\
c & d  \end{array} \right) \in \text{SL}(2,\BZ) \right| b = 0 \text{ mod } n \right\}.
\end{equation}

\subsection*{Eisenstein series}
We let $\tau\in \mathbb{H}$ and define $q=e^{2\pi i \tau}$. Then the Eisenstein series $E_k:\mathbb{H}\to \mathbb{C}$ for even $k\geq 2$ are defined as the $q$-series
\be
\label{Ek}
E_{k}(\tau)=1-\frac{2k}{B_k}\sum_{n=1}^\infty \sigma_{k-1}(n)\,q^n,
\ee
with $\sigma_k(n)=\sum_{d|n} d^k$ the divisor sum. For $k\geq 4$, $E_{k}$ is a modular form of
$\operatorname{SL}(2,\mathbb{Z})$ of weight $k$. In other words, it transforms under $\operatorname{SL}(2,\mathbb{Z})$ as
\be
E_k\!\left( \frac{a\tau+b}{c\tau+d}\right)=(c\tau+d)^kE_k(\tau).
\ee
Note that the space of modular forms of ${\rm SL}(2,\BZ)$ forms a ring that is generated precisely by $E_4(\TT)$ and $E_6(\TT)$. On the other hand $E_2$ is a quasi-modular form, which means that the $\operatorname{SL}(2,\mathbb{Z})$ transformation of $E_2$ includes a shift in addition to the weight,
\be
\label{E2trafo}
E_2\!\left(\frac{a\tau+b}{c\tau+d}\right) =(c\tau+d)^2E_2(\tau)-\frac{6i}{\pi}c(c\tau+d).
\ee

\subsection*{Dedekind eta function}
The Dedekind eta function $\eta:\mathbb{H}\to\mathbb{C}$ is defined as
\be
\eta(\tau)=q^{\frac{1}{24}}\prod_{n=1}^\infty (1-q^n).
\ee
It is a modular form of weight $\frac{1}{2}$ under SL$(2,\BZ)$ with a
non-trivial multiplier system. It transforms under the generators of
SL$(2,\BZ)$ as\footnote{For an unambiguous value of the square root, we define the phase of $z\in \mathbb{C}^*$ by $\log z:=\log|z|+i\,\mathrm{arg}(z)$ with $-\pi<\arg(z)\leq \pi$.}
\be
\begin{split}
&\eta(-1/\tau)=\sqrt{-i\tau}\,\eta(\tau),\\
&\eta(\tau+1)=e^{\frac{\pi i}{12}}\, \eta(\tau).
\end{split}
\ee

\subsection*{Jacobi theta functions}
The four Jacobi theta functions $\vartheta_j:\mathbb{H}\times
\mathbb{C}\to \mathbb{C}$, $j=1,\dots,4$, are defined as
\be
\label{Jacobitheta}
\begin{split}
&\vartheta_1(\tau,v)=i \sum_{r\in
  \mathbb{Z}+\frac12}(-1)^{r-\frac12}q^{r^2/2}e^{2\pi i
  rv}, \\
&\vartheta_2(\tau,v)= \sum_{r\in
  \mathbb{Z}+\frac12}q^{r^2/2}e^{2\pi i
  rv},\\
&\vartheta_3(\tau,v)= \sum_{n\in
  \mathbb{Z}}q^{n^2/2}e^{2\pi i
  n v},\\
&\vartheta_4(\tau,v)= \sum_{n\in
  \mathbb{Z}} (-1)^nq^{n^2/2}e^{2\pi i
  n v}.
\end{split}
\ee

We let $\vartheta_j(\tau,0)=\vartheta_j(\tau)$ for $j=2,3,4$. These have the following transformations for modular inversion
\be
\begin{split}
&\vartheta_2(-1/\tau)=\sqrt{-i\tau}\,\vartheta_4(\tau),\\
&\vartheta_3(-1/\tau)=\sqrt{-i\tau}\,\vartheta_3(\tau),\\
&\vartheta_4(-1/\tau)=\sqrt{-i\tau}\,\vartheta_2(\tau),
\end{split}
\ee
and for the shift
\be
\begin{split}
&\vartheta_2(\tau+1)=e^{2\pi i /8}\,\vartheta_2(\tau),\\
&\vartheta_3(\tau+1)=\vartheta_4(\tau),\\
&\vartheta_4(\tau+1)=\vartheta_3(\tau).
\end{split}
\ee

Their transformations under the generators of $\Gamma^0(4)$ are
\be
\label{Jacobitheta_trafos}
\begin{split}
&\vartheta_2(\tau+4)=-\vartheta_2(\tau),\qquad
\vartheta_2\!\left(\frac{\tau}{\tau+1}\right)=\sqrt{\tau+1}\,\vartheta_3(\tau),  \\
&\vartheta_3(\tau+4)=\vartheta_3(\tau),\qquad
\vartheta_3\!\left(\frac{\tau}{\tau+1}\right)=\sqrt{\tau+1}\,\vartheta_2(\tau),  \\
&\vartheta_4(\tau+4)=\vartheta_4(\tau),\qquad
\vartheta_4\!\left(\frac{\tau}{\tau+1}\right)=e^{-\frac{\pi
    i}{4}}\sqrt{\tau+1}\,\vartheta_4(\tau).  \\
\end{split}
\ee
Two useful identities are
\be
\label{thetaeta}
\begin{split}
&\vartheta_2\,\vartheta_3\,\vartheta_4=2\,\eta^3,\\
&\vartheta_2^4 + \vartheta_4^4=\vartheta_3^4.\\
\end{split}
\ee

\section{Siegel-Narain theta function}
\label{SNtheta}
Siegel-Narain theta functions form a large class of theta functions of
which the Jacobi theta functions are a special case. For our
applications in the main text, it is sufficient to consider
Siegel-Narain theta functions for which the associated lattice
$L$ is a uni-modular lattice with signature
$(1,n-1)$ (or a Lorentzian lattice). We denote the bilinear form by
$B(\bfx,\bfy)$ and the quadratic form by $B(\bfx,\bfx)\equiv Q(\bfx)\equiv \bfx^2 $.
Let $K$ be a characteristic vector of $L$, such that
$ Q(\bfk) + B(\bfk, K) \in 2\mathbb{Z}$ for each $\bfk\in L$.

Given an element $J\in L\otimes \mathbb{R}$ with $Q(J)=1$, we
may decompose the space $L\otimes \mathbb{R}$ in a positive
definite subspace $L_+$ spanned by $J$, and a negative definite
subspace $L_-$, orthogonal to $L_+$. The projections of a
vector $\bfk\in L$ to $L_+$ and $L_-$ are then given by
\be
\label{k+k-}
\bfk_+=B(\bfk, J)\,  J, \qquad \qquad \bfk_{-} = \bfk-\bfk_+.
\ee

Given this notation, we can introduce the Siegel-Narain theta function of our
interest $\Psi^J_\bfmu[\CK]:\mathbb{H}\to \IC$, as
\be
\label{PsiJ}
\begin{split}
\Psi^J_\bfmu[\CK](\tau,\bar \tau)=&\sum_{\bfk\in
  L + \bfmu} \CK(\bfk)\,(-1)^{B(\bfk, K)} q^{-\bfk_-^2/2} \bar q^{\bfk_+^2/2} \\
\end{split}
\ee
where $\bfmu\in L/2$ and $\CK: L\otimes \mathbb{R} \to \mathbb{C}$ is a summation
kernel. Let us be slightly more generic and include  the  
elliptic variables which are relevant for the Donaldson
observables. We define
\be
\label{PsiJell}
\begin{split}
\Psi^J_\bfmu[\CK](\tau,\bar \tau,\bfz, \bar \bfz)&=e^{-2\pi y\, \bfb_+^2}\sum_{\bfk\in
  L + \bfmu} \CK(\bfk)\,(-1)^{B(\bfk, K)} q^{-\bfk_-^2/2} \bar q^{\bfk_+^2/2} \\
  & \quad \times \exp\left( -2\pi i B(\bfz, \bfk_-) - 2\pi i B(\bar \bfz, \bfk_+) \right),
\end{split}
\ee
with $\bfb=\mathrm{Im}(\bfz)/y$.

The modular properties of $\Psi^J_\bfmu[\CK]$ depend on the kernel $\CK$. The modular transformations under the ${\rm SL}(2,\BZ)$ generators for $\Psi^J_\bfmu[1]$ are
\be
\begin{split} \label{SL2Z}
\Psi^J_{\bfmu+K/2}[1](\tau+1,\bar{\tau}+1, \bfz, \bar{\bfz}) &= e^{\pi i (\bfmu^2-K^2/4)} \Psi_{\bfmu+K/2}[1](\tau,\bar \tau, \bfz + \bfmu, \bar{\bfz} + \bfmu), \\
\Psi^J_{\bfmu + K/2}[1]\left(-1/\tau, -1/\bar{\tau},
 \bfz/\tau, \bar{\bfz}/\bar \tau\right) &=
(-i\tau)^{\frac{n-1}{2}}(i\bar{\tau})^\frac{1}{2} \exp(-\pi i
\bfz^2/\tau+\pi i K^2/2)  \\
 & \quad \times  (-1)^{B(\bfmu, K)}\,\Psi^J_{K/2}[1](\tau, \bar{\tau}, \bfz-\bfmu,\bar \bfz -\bfmu)
\end{split}
\ee
For the case of the partition function, we set the elliptic variables
$\bfz, \bar \bfz$ to zero. Using the above ${\rm SL}(2,\BZ)$
transformations and Poisson resummation one may verify that
$\Psi^J_\bfmu[1]$
is a modular form for the congruence subgroup $\Gamma^0(4)$. The
transformations under the generators of this group read
\be
\label{Psitrafos}
\begin{split}
&\Psi^J_\bfmu[1]\!\left( \frac{\tau}{\tau+1},
  \frac{\bar \tau}{\bar \tau+1} \right)=(\tau+1)^{\frac{n-1}{2}}(\bar
\tau+1)^\frac{1}{2}\exp\!\left(\tfrac{\pi i}{4}K^2\right)
\Psi^J_{\bfmu}[1](\tau,\bar \tau),\\
&\Psi^J_\bfmu[1](\tau+4,\bar \tau +4)=e^{2\pi i
  B(\bfmu,K)}\,\Psi_\bfmu[1](\tau,\bar \tau),
\end{split}
\ee
where we have set $\bfz= \bar \bfz = 0$.
Transformations for other kernels appearing in the main text are
easily determined from these expressions.

\section{Indefinite theta functions for uni-modular lattices of
  signature $(1,n-1)$}
\label{Zwegers_theta}

In this appendix we discuss various aspects of the theory of indefinite theta
functions and their modular completion. We assume that the corresponding
lattice $L$ is unimodular and of signature
$(1,n-1)$, and use the notation introduced in Appendix \ref{SNtheta}.
To define the indefinite theta function $\Theta_\bfmu^{JJ'}$, we let
$\bfmu\in L/2$ and choose a vector $J\in L\otimes
\mathbb{R}$ and a vector $J'\in L$, such that
\begin{enumerate}[(i)]
\item $J$ is positive definite, $Q(J)=1$,
\item $J'$ is a null-vector, $Q(J')=0$,
\item $B(J,J')>0$,
\item  $B(\bfk,J')\neq 0$ for all $\bfk\in L +\bfmu$.
\end{enumerate}
The indefinite theta function $\Theta_\bfmu^{JJ'}(\tau,\bfz)$ is then
defined as
\be
\label{indeftheta}
\begin{split}
\Theta^{JJ'}_{\bfmu}\!(\tau,\bfz)=&\sum_{\bfk\in L+\bfmu}
\tfrac{1}{2}\Big[
\sgn(B(\bfk,J))-\sgn(B(\bfk,J'))\Big]\,(-1)^{B(\bfk,K)}
q^{-\bfk^2/2}e^{-2\pi i B(\bfk,\bfz)}.
\end{split}
\ee
The kernel within the straight brackets ensures that the sum over the
indefinite lattice is convergent, since it vanishes on positive
definite vectors \cite{ZwegersThesis}. This is also the case if both
$J$ and $J'$ are positive definite, without the need to impose condition (iv). One may
start from this situation and obtain the conditions above by taking the limit that $J'$ approaches a null
vector. The indefinite theta series $\Theta^{JJ'}_{\bfmu}$ can also be
defined, for $\bfmu$ which do not satisfy requirement (iv) above, but this requires more care.

We can express $\Theta^{JJ'}_{\bfmu}$ also as
$\Psi_\bfmu^J[\CK]$ (\ref{PsiJ}) with the kernel
\be
\CK(\bfk)= \tfrac{1}{2}\Big[
\sgn(B(\bfk,J))-\sgn(B(\bfk,J'))\Big] \,e^{2\pi y \bfk_+^2+4\pi y B(\bfk,\bfb)},
\ee
where $\bfb=\mathrm{Im}(\bfz)/y$.

While the sum over $L$ is convergent, $\Theta^{JJ'}_{\bfmu}$   only
transforms as a modular form after  addition of certain non-holomorphic
terms. References \cite{ZwegersThesis, MR2605321} explain that the
modular completion $\widehat \Theta^{JJ'}_\bfmu$ of
$\Theta^{JJ'}_\bfmu$ is obtained by substituting (rescaled) error
function for the sgn-function in (\ref{indeftheta}). We let
$E(u):\mathbb{R}\to (-1,1)$ be defined as
\begin{equation}
\label{Eerrorfunction}
E(u) = 2\int_0^u e^{-\pi t^2}dt = \text{Erf}(\sqrt{\pi}u).
\end{equation}
The completion
$\widehat \Theta^{JJ'}_\bfmu$ then transforms as a modular form of
weight $n/2$, and is explicitly given by
\be
\label{hatTheta}
\begin{split}
\widehat \Theta^{JJ'}_{\bfmu}\!(\tau,\bar \tau,\bfz,\bar \bfz)&=\sum_{\bfk\in L+\mu}
\tfrac{1}{2}\left( E(\sqrt{2y}\,B(\bfk+\bfb, \underline
  J))-\sgn(B(\bfk, J'))\right)\\
&\quad \times (-1)^{B(\bfk,K)} q^{-\bfk^2/2}e^{-2\pi
  i B(\bfk,\bfz)}.
\end{split}
\ee
Note that in the limit $y\to \infty$, $E(\sqrt{2y}\,u)$
approaches the $\sgn$-function of (\ref{indeftheta}),
$$\lim_{y\to \infty} E(\sqrt{2y}\,u)=\sgn(u).$$

The transformation properties under SL$(2,\BZ)$ follow from Chapter 2 of Zwegers'
thesis \cite{ZwegersThesis} or Vign\'eras \cite{Vigneras:1977}. One
finds
\be
\label{indef_theta_mod}
\begin{split}
&\widehat \Theta^{JJ'}_{\bfmu+K/2}(\tau+1,\bar \tau+1,\bfz,\bar
\bfz)=e^{\pi i (\bfmu^2-K^2/4)}\,
\widehat\Theta^{JJ'}_{\bfmu+K/2}(\tau,\bar \tau,\bfz+\bfmu,\bar \bfz+\bfmu),\\
&\widehat\Theta^{JJ'}_{\bfmu+K/2}(-1/\tau,-1/\bar \tau,\bfz/\tau,\bar
\bfz/\bar \tau)=i(-i\tau)^{n/2} \exp\!\left(-\pi
  i \bfz^2/\tau+\pi i K^2/2\right)\\
&\qquad \qquad\qquad \qquad \qquad \times (-1)^{B(\bfmu,K)}\,
\widehat\Theta^{JJ'}_{K/2}(\tau,\bar \tau,\bfz-\bfmu,\bar \bfz-\bfmu).
\end{split}
\ee
The argument of $E$ in (\ref{hatTheta}) depends only on the imaginary part of
   $\bfz$ in and is
   valued in $\mathbb{R}$. Reference \cite{Moore:2017cmm} demonstrates that
   if one formally sets $\bar \bfz=0$ such that the argument of $E$ is
   complex-valued, the modular
   properties of $\widehat \Theta^{JJ'}_{\bfmu}$ remain unchanged.

When $\bfz=0$, we set $\widehat \Theta^{JJ'}_{\bfmu}(\tau,\bar
\tau,0,0)=\widehat \Theta^{JJ'}_{\bfmu+K/2}(\tau,\bar \tau)$.
 One finds for the action of the generators on $\widehat
\Theta^{JJ'}_{\bfmu}(\tau,\bar \tau)$
\be
\label{theta_comp_mod}
\begin{split}
&\widehat\Theta^{JJ'}_{\bfmu}\!\left(\frac{\tau}{\tau+1},\frac{\bar \tau}{\bar
    \tau+1}\right)=(\tau+1)^{n/2} \exp\!\left(\tfrac{\pi
    i}{4} K^2\right) \widehat\Theta^{JJ'}_{\bfmu}(\tau,\bar \tau).\\
&\widehat \Theta^{JJ'}_{\bfmu}(\tau+4,\bar \tau+4)=e^{2\pi i
  B(\bfmu,K)}\, \widehat\Theta^{JJ'}_{\bfmu}(\tau,\bar \tau).
\end{split}
\ee
For our application, the  $\bar \tau$-derivative of $\widehat \Theta^{JJ'}_\bfmu$ is of particular
interest. This gives the ``shadow''\footnote{Indefinite theta
  functions can often be expressed as product of a mock modular form
  and modular form, in other words they are a mixed mock modular
  form. We therefore use the notion of ``shadow'' slightly differently from its
  definition for mock modular forms \cite{MR2605321}.} of
$\Theta^{JJ'}_\bfmu$, whose modular properties are easier to
determine than those of $\Theta^{JJ'}_\bfmu$. We obtain here
\be
\label{shadow}
\begin{split}
\partial_{\bar \tau} \widehat
\Theta^{JJ'}_\bfmu(\tau,\bar \tau)=& \Psi^J_\bfmu[\CK_0](\tau,\bar \tau),
\end{split}
\ee
with $\Psi^J_\bfmu$ (\ref{PsiJ}) the same function discussed in
Appendix \ref{SNtheta} and $\CK_0$,
\be
\CK_0=\tfrac{i}{2\sqrt{2y}}\,B(\bfk, {J}).
\ee

\subsection*{An example}
\begin{figure}[h]
\captionsetup{singlelinecheck=off}
\includegraphics[scale=2.5]{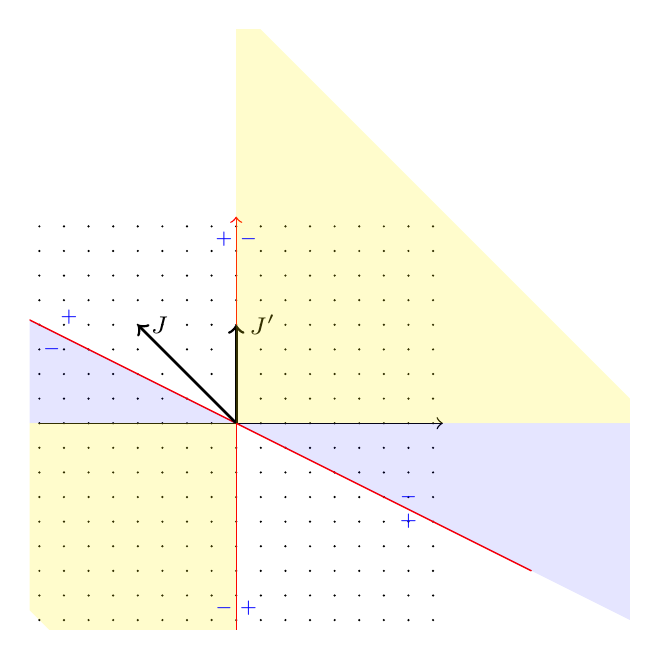}
\centering
\caption{The lattice $L \cong \BZ^{1,1}$ with quadratic form  $-\left(\protect\begin{smallmatrix}
  1 & 1\\
  1 & 0
\protect\end{smallmatrix}\right)$. The negative definite
  set of this lattice is the union of the yellow and purple
  regions. For the given choices of $J$ and $J'$, only the lattice vectors
  in the yellow area contribute to the sum in the indefinite theta function.}
\label{CONES}
\end{figure}

Let us now specialize to an example which is useful in Section
\ref{uplanemock}. We consider a two-dimensional lattice $L \cong
\BZ^{1,1}$ with quadratic form $-\bigl(   \begin{smallmatrix}
  1 & 1\\
  1 & 0
\end{smallmatrix} \bigr)$. The positive and negative definite cones of
this lattices are illustrated in Figure
\ref{CONES}. The upper-right component of the negative cone for this lattice is
generated by the vectors $(0,1)$ and $(2,-1)$. Any linear combination of these
vectors with positive definite coefficients will be negative
definite. We choose the vectors $J$ and $J'$ as follows:  $J=(-1,1)$ and
$J'=(0,1)$. For $\bfk = (n, \ell)$, the kernel in (\ref{indeftheta})
becomes $({\rm sgn}(\ell) + {\rm sgn}(n))$. Then the only elements of
$L$ which contribute to the theta series are those in the two
yellow areas in Figure \ref{CONES}.

For the characteristic vector we choose $K=(0,-1)$, while we choose
for $\bfmu=\frac{1}{2}(1,1)$. With these choices, $\Theta^{JJ'}_\bfmu$
  becomes the following $q$-series,
\be
\label{indefthetaeta}
\begin{split}
\Theta^{JJ'}_\bfmu(\tau)=&\sum_{n,\ell\in \mathbb{Z}+\frac{1}{2}} \tfrac{1}{2}\left(
  \sgn(\ell)+\sgn(n)\right)\,(-1)^{n}\,q^{\frac{1}{2}n^2+n\ell} \\
=&-i\sum_{n\in \mathbb{Z}} \frac{(-1)^nq^{\frac{1}{2}n^2-\frac{1}{8}}}{1-q^{n-\frac{1}{2}}},
\end{split}
\ee
where we performed the geometric sum over $\ell$ on the second
line. The first part of this Appendix discussed that
$\Theta^{JJ'}_\bfmu$ can be completed by replacing $\sgn(\ell)$ in (\ref{indeftheta}) by $E(\sqrt{2y}\,\ell)$, where $E$ is the rescaled error function defined in
(\ref{Eerrorfunction}). We can write $E(\sqrt{2y}\,\ell)$ as
$\sgn(\ell)$ plus a non-holomorphic period integral,
\be
E(\sqrt{2y}\,\ell)=\sgn(\ell)+i\ell\,q^{\frac{1}{2}\ell^2} \int_{-\bar \tau}^{i\infty}
\frac{e^{\pi i \ell^2 w}}{\sqrt{-i(w+\tau)}}dw.
\ee
The completion can then be written as
\be
\widehat \Theta^{JJ'}_\bfmu(\tau)=\Theta^{JJ'}_\bfmu(\tau)+\tfrac{1}{2}\vartheta_4(\tau)\,\int_{-\bar
\tau}^{i\infty} \frac{\eta(w)^3}{\sqrt{-i(w+\tau)}}dw,
\ee
and transforms as a non-holomorphic modular form for $\Gamma^0(4)$ as discussed in
Appendix \ref{Zwegers_theta}. We thus find that
$F=-i\,\Theta^{JJ'}_\bfmu/\vartheta_4$ is the
holomorphic part of $\widehat F$ in Section \ref{uplanemock}.

We  conclude this appendix by deriving $F_D$, which is the holomorphic part of $\widehat F_{D}(\tau,\bar \tau)=-(-i\tau)^{-\frac{1}{2}}\widehat F(-1/\tau,-1/\bar \tau)$. We are instructed by (\ref{indef_theta_mod}) to determine \linebreak[4] $\widehat \Theta^{JJ'}_{K/2}(\tau,-\bfmu+K/2)$ with $\bfmu=\frac{1}{2}(1,1)$. Its holomorphic part reads
\be 
\begin{split}
\Theta^{JJ'}_{K/2}(\tau,-\bfmu+K/2)&=-i\sum_{n\in \mathbb{Z}\atop \ell
  \in \mathbb{Z}+\frac{1}{2}}
(\sgn(\ell)+\sgn(n))\,(-1)^{\ell-\frac{1}{2}}\,q^{\frac{1}{2}n^2+n\ell}\\
&=-i\sum_{n\in \mathbb{Z}}\frac{q^{\frac{1}{2}(n+\frac{1}{2})^2-\frac{1}{8}}}{1+q^n}.
\end{split}
\ee
This gives for $F_D$
\be
F_D(\tau)=\frac{-1}{\vartheta_2(\tau)}\,\sum_{n\in \mathbb{Z}} \frac{q^{\frac{1}{2}(n+\frac{1}{2})^2-\frac{1}{8}}}{1+q^n}.
\ee

\section{The Appell-Lerch sum}
\label{subsecZwmu}
We recall the definition and properties of the Appell-Lerch sum.
We will denote this function by $M(\tau,u,v)$ rather then the more
common $\mu(\tau,u,v)$ to avoid a class of notation with the 't Hooft fluxes. We will mostly follow the exposition of
Zwegers \cite{ZwegersThesis}.

For fixed $\tau$, the Appell-Lerch function is a function of two
complex variables $M:(\mathbb{C}\backslash\{\mathbb{Z}\tau+\mathbb{Z}\})^2 \to
\mathbb{C}$, defined as
\be
\label{muv}
M(\tau,u,v):=M(u,v)=\frac{e^{\pi i u}}{\vartheta_1(\tau,v)}\sum_{n\in
  \mathbb{Z}} \frac{(-1)^n q^{n(n+1)/2}e^{2\pi i n v}}{1-e^{2\pi i u}q^n}.
\ee
It has single order poles at $\mathbb{Z}\tau+\mathbb{Z}$ for both $u$
and $v$.

We list a number of useful properties, whose proofs can be found in \cite{ZwegersThesis}:
\begin{enumerate}
\item Periodicity of $u$ and $v$:
\be
M(u+1,v)=M(u,v+1)=-M(u,v).
\ee
\item Quasi-periodicity of $M$ under simultaneous translations of $u$ and $v$. For $u,v,u+z,v+z\neq
\mathbb{Z}\tau+\mathbb{Z}$, $M$ satisfies
\be
\label{mushifts}
M(u+z,v+z)-M(u,v)=\frac{i\,\eta^3\,\vartheta_1(u+v+z)\,\vartheta_1(z)}{\vartheta_1(u)\,\vartheta_1(v)\,\vartheta_1(u+z)\,\vartheta_1(v+z)}.
\ee
This relation can be demonstrated by showing that the periodicity,
zeroes and poles of the variable $z$ are identical on the left and right hand side.
\item Inversion of the elliptic arguments leaves $\mu$ invariant:
\be
\label{Inversion}
M(-u,-v)=M(u,v)
\ee
\item $M$ is symmetric under exchange of $u$ and $v$:
\be
M(v,u)=M(u,v).
\ee
Note that this relation follows from (\ref{mushifts}) and (\ref{Inversion})
using $z=-u-v$.
\end{enumerate}

A further property of $M$ is that $M$   transforms as a Jacobi form
after the addition of a suitable non-holomorphic function $R$, which
is analytic in its arguments. It is defined explicitly as
\be
\label{muR}
R(\tau, \bar \tau, u,\bar u):=R(u)=\sum_{n\in \mathbb{Z}+\frac{1}{2}} \left( \sgn(n) -
  \mathrm{Erf}\!\left( (n+a)\sqrt{2\pi y}
     \right) \right) (-1)^{n-\frac{1}{2}}\,e^{-2\pi i un}q^{-n^2/2},
\ee
where $a=\mathrm{Im}(u)/y$, and $\mathrm{Erf}(t)$ is the error function
\be
\label{Eerror}
\mathrm{Erf}(t)=\frac{2}{\sqrt{\pi}} \int_0^t e^{- u^2}du.
\ee
The anti-holomorphic derivative of $R(\tau,\bar \tau, u,\bar u)$ is
\be
\label{dbarR}
\partial_{\bar \tau} R(\tau,\bar \tau,u,\bar u)=-2e^{-2\pi\,y\,a^2}\sum_{n\in \mathbb{Z}+\frac{1}{2}} \partial_{\bar \tau}\!\left(
  \sqrt{2y}\,(n+a)\right)\,(-1)^{n-\frac{1}{2}}\,e^{-2\pi i \bar u n}\,\bar q^{n^2/2}.
\ee

Addition of this function to $M$ provides a function $\widehat M$,
which transforms as a weight $\frac{1}{2}$ Jacobi form. This
non-holomorphic completion $\widehat M$ of $M$ is explicitly given
by
\be
\label{hatmuv}
\begin{split}
\widehat M(\tau,\bar \tau,u,\bar u,v,\bar v)= M(\tau,u,v)+\frac{i}{2}
R(\tau,\bar \tau, u-v,\bar u - \bar v).
\end{split}
\ee
This function transforms under ${\rm SL}(2,\mathbb{Z})$ as
\be
\label{mucomplete}
\begin{split}
&\widehat
M\!\left(\frac{a\tau+b}{c\tau+d}, \frac{a\bar \tau+b}{c\bar
    \tau+d},\frac{u}{c\tau+d},\frac{\bar u}{c\bar
    \tau+d},\frac{v}{c\tau+d}, \frac{\bar v}{c\bar \tau+d}\right)=\\
&\qquad \qquad \qquad \qquad \qquad \qquad \varepsilon(\gamma)^{-3} (c\tau+d)^{\frac{1}{2}} e^{-\pi i c(u-v)^2/(c\tau+d)}\,\widehat
M(\tau,\bar \tau,u,\bar u,v,\bar v),
\end{split}
\ee
where $\varepsilon(\gamma)$ is the multiplier system of the Dedekind
$\eta$ function. The anti-holomorphic derivative of $\widehat M$ is given by
\be
\begin{split}
&\partial_{\bar \tau} \widehat M(\tau,\bar \tau,u,\bar u,v,\bar v)= \\
&\qquad -i \left(\partial_{\bar \tau}
  \sqrt{2y}\right) e^{-2\pi (a-b)^2} \sum_{n\in \mathbb{Z}+\frac{1}{2}} (n+a-b)
(-1)^{n-\frac{1}{2}} \bar q^{n^2/2}e^{-2\pi i (\bar u-\bar v) n},
\end{split}
\ee
where $a=\mathrm{Im}(u)/y$ and $b=\mathrm{Im}(v)/y$, and we hope there
is no confusion with the $a,b,c,d$ used in Equation (\ref{mucomplete}).

\providecommand{\href}[2]{#2}\begingroup\raggedright\endgroup


\begin{thebibliography}{10}

\bibitem{Witten:1994ev}
E.~Witten, \emph{{Supersymmetric Yang-Mills theory on a four manifold}},
  \href{http://dx.doi.org/10.1063/1.530745}{\emph{J. Math. Phys.} {\bf 35}
  (1994) 5101--5135}, [\href{https://arxiv.org/abs/hep-th/9403195}{{\tt
  hep-th/9403195}}].

\bibitem{Witten:1994cg}
E.~Witten, \emph{{Monopoles and four manifolds}},
  \href{http://dx.doi.org/10.4310/MRL.1994.v1.n6.a13}{\emph{Math. Res. Lett.}
  {\bf 1} (1994) 769--796}, [\href{https://arxiv.org/abs/hep-th/9411102}{{\tt
  hep-th/9411102}}].

\bibitem{Moore:1997pc}
G.~W. Moore and E.~Witten, \emph{{Integration over the u plane in Donaldson
  theory}}, {\emph{Adv. Theor. Math. Phys.} {\bf 1} (1997) 298--387},
  [\href{https://arxiv.org/abs/hep-th/9709193}{{\tt hep-th/9709193}}].

\bibitem{LoNeSha}
A.~Losev, N.~Nekrasov and S.~L. Shatashvili, \emph{{Issues in topological gauge
  theory}}, \href{http://dx.doi.org/10.1016/S0550-3213(98)00628-2}{\emph{Nucl.
  Phys.} {\bf B534} (1998) 549--611},
  [\href{https://arxiv.org/abs/hep-th/9711108}{{\tt hep-th/9711108}}].

\bibitem{Marino:1998bm}
M.~Marino and G.~W. Moore, \emph{{The Donaldson-Witten function for gauge
  groups of rank larger than one}},
  \href{http://dx.doi.org/10.1007/s002200050494}{\emph{Commun. Math. Phys.}
  {\bf 199} (1998) 25--69}, [\href{https://arxiv.org/abs/hep-th/9802185}{{\tt
  hep-th/9802185}}].

\bibitem{Pestun:2007rz}
V.~Pestun, \emph{{Localization of gauge theory on a four-sphere and
  supersymmetric Wilson loops}},
  \href{http://dx.doi.org/10.1007/s00220-012-1485-0}{\emph{Commun. Math. Phys.}
  {\bf 313} (2012) 71--129}, [\href{https://arxiv.org/abs/0712.2824}{{\tt
  0712.2824}}].

\bibitem{Shapere:2008zf}
A.~D. Shapere and Y.~Tachikawa, \emph{{Central charges of N=2 superconformal
  field theories in four dimensions}},
  \href{http://dx.doi.org/10.1088/1126-6708/2008/09/109}{\emph{JHEP} {\bf 09}
  (2008) 109}, [\href{https://arxiv.org/abs/0804.1957}{{\tt 0804.1957}}].

\bibitem{Witten:1988ze}
E.~Witten, \emph{{Topological Quantum Field Theory}},
  \href{http://dx.doi.org/10.1007/BF01223371}{\emph{Commun. Math. Phys.} {\bf
  117} (1988) 353}.

\bibitem{Seiberg:1994rs}
N.~Seiberg and E.~Witten, \emph{{Electric - magnetic duality, monopole
  condensation, and confinement in N=2 supersymmetric Yang-Mills theory}},
  \href{http://dx.doi.org/10.1016/0550-3213(94)90124-4,
  10.1016/0550-3213(94)00449-8}{\emph{Nucl. Phys.} {\bf B426} (1994) 19--52},
  [\href{https://arxiv.org/abs/hep-th/9407087}{{\tt hep-th/9407087}}].

\bibitem{Gottsche:1996aoa}
L.~Gottsche and D.~Zagier, \emph{{Jacobi forms and the structure of Donaldson
  invariants for 4-manifolds with $b_2^+=1$}},
  \href{https://arxiv.org/abs/alg-geom/9612020}{{\tt alg-geom/9612020}}.

\bibitem{Marino:1998rg}
M.~Marino and G.~W. Moore, \emph{{Donaldson invariants for nonsimply connected
  manifolds}}, \href{http://dx.doi.org/10.1007/s002200050611}{\emph{Commun.
  Math. Phys.} {\bf 203} (1999) 249},
  [\href{https://arxiv.org/abs/hep-th/9804104}{{\tt hep-th/9804104}}].

\bibitem{Malmendier:2008db}
A.~Malmendier and K.~Ono, \emph{{SO(3)-Donaldson invariants of $\mathbb{P}^2$
  and Mock Theta Functions}},
  \href{http://dx.doi.org/10.2140/gt.2012.16.1767}{\emph{Geom. Topol.} {\bf 16}
  (2012) 1767--1833}, [\href{https://arxiv.org/abs/0808.1442}{{\tt
  0808.1442}}].

\bibitem{Malmendier:2010ss}
A.~Malmendier, \emph{{Donaldson invariants of $\mathbb{P}^1 \times
  \mathbb{P}^1$ and Mock Theta Functions}},
  \href{http://dx.doi.org/10.4310/CNTP.2011.v5.n1.a5}{\emph{Commun. Num. Theor.
  Phys.} {\bf 5} (2011) 203--229}, [\href{https://arxiv.org/abs/1008.0175}{{\tt
  1008.0175}}].

\bibitem{Griffin:2012kw}
M.~Griffin, A.~Malmendier and K.~Ono, \emph{{SU(2)-Donaldson invariants of the
  complex projective plane}},
  \href{http://dx.doi.org/10.1515/forum-2013-6013}{\emph{Forum Math.} {\bf 27}
  (2015) 2003--2023}, [\href{https://arxiv.org/abs/1209.2743}{{\tt
  1209.2743}}].

\bibitem{ZwegersThesis}
S.~P. Zwegers, \emph{Mock Theta Functions}.
\newblock PhD thesis, 2008.

\bibitem{MR2605321}
D.~Zagier, \emph{Ramanujan's mock theta functions and their applications (after
  {Z}wegers and {O}no-{B}ringmann)}, {\emph{Ast\'erisque} (2009) Exp. No. 986,
  vii--viii, 143--164 (2010)}.

\bibitem{Korpas:2017qdo}
G.~Korpas and J.~Manschot, \emph{{Donaldson-Witten theory and indefinite theta
  functions}}, \href{http://dx.doi.org/10.1007/JHEP11(2017)083}{\emph{JHEP}
  {\bf 11} (2017) 083}, [\href{https://arxiv.org/abs/1707.06235}{{\tt
  1707.06235}}].

\bibitem{Moore:2017cmm}
G.~W. Moore and I.~Nidaiev, \emph{{The Partition Function Of Argyres-Douglas
  Theory On A Four-Manifold}},  \href{https://arxiv.org/abs/1711.09257}{{\tt
  1711.09257}}.

\bibitem{Korpas:2019ava}
G.~Korpas, J.~Manschot, G.~W. Moore and I.~Nidaiev, \emph{{Renormalization and
  BRST symmetry in Donaldson-Witten theory}},
  \href{http://dx.doi.org/10.1007/s00023-019-00835-x}{\emph{Annales Henri
  Poincar{\'e}} (2019) }, [\href{https://arxiv.org/abs/1901.03540}{{\tt
  1901.03540}}].

\bibitem{1603.03056}
K.~Bringmann, N.~Diamantis and S.~Ehlen, \emph{Regularized inner products and
  errors of modularity},
  \href{http://dx.doi.org/10.1093/imrn/rnw225}{\emph{International Mathematics
  Research Notices} {\bf 2017} (2017) 7420--7458}.

\bibitem{kronheimer1995}
P.~B. Kronheimer and T.~S. Mrowka, \emph{Embedded surfaces and the structure of
  donaldson's polynomial invariants},
  \href{http://dx.doi.org/10.4310/jdg/1214456482}{\emph{J. Differential Geom.}
  {\bf 41} (1995) 573--734}.

\bibitem{Grassi:2019txd}
A.~Grassi, Z.~Komargodski and L.~Tizzano, \emph{{Extremal Correlators and
  Random Matrix Theory}},  \href{https://arxiv.org/abs/1908.10306}{{\tt
  1908.10306}}.

\bibitem{Nekrasov:2003vi}
N.~A. Nekrasov, \emph{{Localizing gauge theories}},  in \emph{{Mathematical
  physics. Proceedings, 14th International Congress, ICMP 2003, Lisbon,
  Portugal, July 28-August 2, 2003}}, pp.~645--654, 2003.

\bibitem{Bershtein:2015xfa}
M.~Bershtein, G.~Bonelli, M.~Ronzani and A.~Tanzini, \emph{{Exact results for $
  \mathcal{N} $ = 2 supersymmetric gauge theories on compact toric manifolds
  and equivariant Donaldson invariants}},
  \href{http://dx.doi.org/10.1007/JHEP07(2016)023}{\emph{JHEP} {\bf 07} (2016)
  023}, [\href{https://arxiv.org/abs/1509.00267}{{\tt 1509.00267}}].

\bibitem{Seiberg:1994aj}
N.~Seiberg and E.~Witten, \emph{{Monopoles, duality and chiral symmetry
  breaking in N=2 supersymmetric QCD}},
  \href{http://dx.doi.org/10.1016/0550-3213(94)90214-3}{\emph{Nucl. Phys.} {\bf
  B431} (1994) 484--550}, [\href{https://arxiv.org/abs/hep-th/9408099}{{\tt
  hep-th/9408099}}].

\bibitem{Laba05}
J.~Labastida and M.~Marino, \emph{{Topological quantum field theory and four
  manifolds}}, vol.~25.
\newblock Springer, Dordrecht, 2005,
  \href{http://dx.doi.org/10.1007/1-4020-3177-7}{10.1007/1-4020-3177-7}.

\bibitem{MooreNotes2017}
G.~W. Moore, \emph{{Lectures On The Physical Approach To Donaldson And
  Seiberg-Witten Invariants}}, {\emph{Item 78 at
  http://www.physics.rutgers.edu/~gmoore/} (2017) }.

\bibitem{DONALDSON1990257}
S.~Donaldson, \emph{Polynomial invariants for smooth four-manifolds},
  \href{http://dx.doi.org/http://dx.doi.org/10.1016/0040-9383(90)90001-Z}{\emph{Topology}
  {\bf 29} (1990) 257 -- 315}.

\bibitem{Donaldson90}
S.~K. Donaldson and P.~B. Kronheimer, \emph{The geometry of four-manifolds /
  S.K. Donaldson and P.B. Kronheimer}.
\newblock Clarendon Press ; Oxford University Press Oxford : New York, 1990.

\bibitem{Labastida:1991qq}
J.~M.~F. Labastida and P.~M. Llatas, \emph{{Topological matter in
  two-dimensions}},
  \href{http://dx.doi.org/10.1016/0550-3213(92)90596-4}{\emph{Nucl. Phys.} {\bf
  B379} (1992) 220--258}, [\href{https://arxiv.org/abs/hep-th/9112051}{{\tt
  hep-th/9112051}}].

\bibitem{Wu:1952}
W.-T. Wu, \emph{{Sur les classes caract\'eristiques des structures fibr\'ees
  sph\'eriques}}, {\emph{Actualit\'es Schi. Ind.} {\bf 1183} (1952) }.

\bibitem{Scorpan}
A.~Scorpan, \emph{The Wild World of 4-Manifolds}.
\newblock Wiley Classics Library. American Mathematical Society, 2005.

\bibitem{Yoshioka1994}
K.~Yoshioka, \emph{The betti numbers of the moduli space of stable sheaves of
  rank2 on p2.}, {\emph{Journal fur die reine und angewandte Mathematik} {\bf
  453} (1994) 193--220}.

\bibitem{Manschot:2014cca}
J.~Manschot, \emph{{Sheaves on $\mathbb{P}^2$ and generalized Appell
  functions}}, \href{http://dx.doi.org/10.4310/ATMP.2017.v21.n3.a3}{\emph{Adv.
  Theor. Math. Phys.} {\bf 21} (2017) 655--681},
  [\href{https://arxiv.org/abs/1407.7785}{{\tt 1407.7785}}].

\bibitem{Barth}
W.~Barth, K.~Hulek, C.~Peters and A.~van~de Ven, \emph{Compact Complex
  Surfaces}.
\newblock Springer-Verlag Berlin Heidelberg, 2~ed., 2004.

\bibitem{mcduff:1996}
D.~McDuff and D.~Salamon, \emph{A survey of symplectic manifolds with $b^+=1$},
  {\emph{Tr. J. of Mathematics} {\bf 20} (1996) 47--60}.

\bibitem{park_2004}
J.~Park, \emph{Non-complex symplectic 4-manifolds with
  $\lowercase{b}_{2}^+=1$},
  \href{http://dx.doi.org/10.1112/S0024609303002893}{\emph{Bulletin of the
  London Mathematical Society} {\bf 36} (2004) 231--240}.

\bibitem{Park2005}
J.~Park, \emph{Simply connected symplectic 4-manifolds with b2+=1 and c12=2},
  \href{http://dx.doi.org/10.1007/s00222-004-0404-1}{\emph{Inventiones
  mathematicae} {\bf 159} (Mar, 2005) 657--667}.

\bibitem{Baldridge:2005}
S.~Baldridge, \emph{New symplectic 4-manifolds with $b_+=1$}, {\emph{Math.
  Annalen} {\bf 333} (2005) 633--643}.

\bibitem{baulieu1989}
L.~Baulieu and I.~M. Singer, \emph{The topological sigma model}, {\emph{Comm.
  Math. Phys.} {\bf 125} (1989) 227--237}.

\bibitem{Witten:1995gf}
E.~Witten, \emph{{On S duality in Abelian gauge theory}},
  \href{http://dx.doi.org/10.1007/BF01671570}{\emph{Selecta Math.} {\bf 1}
  (1995) 383}, [\href{https://arxiv.org/abs/hep-th/9505186}{{\tt
  hep-th/9505186}}].

\bibitem{Matone:1995rx}
M.~Matone, \emph{{Instantons and recursion relations in N=2 SUSY gauge
  theory}}, \href{http://dx.doi.org/10.1016/0370-2693(95)00920-G}{\emph{Phys.
  Lett.} {\bf B357} (1995) 342--348},
  [\href{https://arxiv.org/abs/hep-th/9506102}{{\tt hep-th/9506102}}].

\bibitem{Dixon:1990pc}
L.~J. Dixon, V.~Kaplunovsky and J.~Louis, \emph{{Moduli dependence of string
  loop corrections to gauge coupling constants}},
  \href{http://dx.doi.org/10.1016/0550-3213(91)90490-O}{\emph{Nucl. Phys.} {\bf
  B355} (1991) 649--688}.

\bibitem{Harvey:1995fq}
J.~A. Harvey and G.~W. Moore, \emph{{Algebras, BPS states, and strings}},
  \href{http://dx.doi.org/10.1016/0550-3213(95)00605-2}{\emph{Nucl. Phys.} {\bf
  B463} (1996) 315--368}, [\href{https://arxiv.org/abs/hep-th/9510182}{{\tt
  hep-th/9510182}}].

\bibitem{Malmendier:2012zz}
A.~Malmendier and K.~Ono, \emph{{Moonshine and Donaldson invariants of
  $\mathbb{P}^2$}},
  \href{http://dx.doi.org/10.4310/CNTP.2012.v6.n4.a1}{\emph{Commun. Number
  Theory Phys.} {\bf 6} (2012) 759--770},
  [\href{https://arxiv.org/abs/1207.5139}{{\tt 1207.5139}}].

\bibitem{HarveyINI2012}
J.~A. Harvey, ``{Permutation groups, Mock modular forms, K3 surfaces and
  Moonshine, Lectures at Isaac Newton Institute}.''
  \url{https://www.newton.ac.uk/seminar/20120126140015001}, 2012.

\bibitem{HarveyGGI2015}
J.~A. Harvey, ``{(Mock) Modular Forms and Applications to String Theory},
  {L}ectures at {LACES, GGI}.''
  \url{https://www.youtube.com/watch?v=CSmKdMc3a64}, 2015.

\bibitem{Kachru:2016nty}
S.~Kachru, \emph{{Elementary introduction to Moonshine}},  2016.
\newblock \href{https://arxiv.org/abs/1605.00697}{{\tt 1605.00697}}.

\bibitem{HarveyICTP2016}
J.~A. Harvey, ``Moonshine and {S}tring {T}heory, {L}ectures at {ICTP} {S}pring
  {S}chool on {S}uperstring {T}heory and {R}elated {T}opics.''
  \url{https://www.youtube.com/watch?v=NPI6bAI_Rdc}, 2016.

\bibitem{Anagiannis:2018jqf}
V.~Anagiannis and M.~C.~N. Cheng, \emph{{TASI Lectures on Moonshine}},
  \href{http://dx.doi.org/10.22323/1.305.0010}{\emph{PoS} {\bf TASI2017} (2018)
  010}, [\href{https://arxiv.org/abs/1807.00723}{{\tt 1807.00723}}].

\bibitem{Cheng:2013wca}
M.~C.~N. Cheng, J.~F.~R. Duncan and J.~A. Harvey, \emph{{Umbral Moonshine and
  the Niemeier Lattices}},  \href{https://arxiv.org/abs/1307.5793}{{\tt
  1307.5793}}.

\bibitem{Eguchi:2010ej}
T.~Eguchi, H.~Ooguri and Y.~Tachikawa, \emph{{Notes on the K3 Surface and the
  Mathieu group $M_{24}$}},
  \href{http://dx.doi.org/10.1080/10586458.2011.544585}{\emph{Exper. Math.}
  {\bf 20} (2011) 91--96}, [\href{https://arxiv.org/abs/1004.0956}{{\tt
  1004.0956}}].

\bibitem{Dabholkar:2012nd}
A.~Dabholkar, S.~Murthy and D.~Zagier, \emph{{Quantum Black Holes, Wall
  Crossing, and Mock Modular Forms}},
  \href{https://arxiv.org/abs/1208.4074}{{\tt 1208.4074}}.

\bibitem{ellingsrud1995wall}
G.~Ellingsrud and L.~G{\"o}ttsche, \emph{Wall-crossing formulas, bott residue
  formula and the donaldson invariants of rational surfaces}, {\emph{Quart. J.
  Math. Oxford Ser.} {\bf 49} (1998) 307--329},
  [\href{https://arxiv.org/abs/alg-geom/9506019}{{\tt alg-geom/9506019}}].

\bibitem{Semikhatov:2003uc}
A.~M. Semikhatov, A.~Taormina and I.~{\relax Yu}. Tipunin, \emph{{Higher level
  Appell functions, modular transformations, and characters}},
  \href{http://dx.doi.org/10.1007/s00220-004-1280-7}{\emph{Commun. Math. Phys.}
  {\bf 255} (2005) 469}, [\href{https://arxiv.org/abs/math/0311314}{{\tt
  math/0311314}}].

\bibitem{Manschot:2011dj}
J.~Manschot, \emph{{BPS invariants of $\CN=4$ gauge theory on Hirzebruch
  surfaces}}, \href{http://dx.doi.org/10.4310/CNTP.2012.v6.n2.a4}{\emph{Commun.
  Num. Theor. Phys.} {\bf 6} (2012) 497--516},
  [\href{https://arxiv.org/abs/1103.0012}{{\tt 1103.0012}}].

\bibitem{Vafa:1994tf}
C.~Vafa and E.~Witten, \emph{{A Strong coupling test of S duality}},
  \href{http://dx.doi.org/10.1016/0550-3213(94)90097-3}{\emph{Nucl. Phys.} {\bf
  B431} (1994) 3--77}, [\href{https://arxiv.org/abs/hep-th/9408074}{{\tt
  hep-th/9408074}}].

\bibitem{Yoshioka:1994}
K.~Yoshioka, \emph{{The Betti numbers of the moduli space of stable sheaves of
  rank 2 on $\mathbb{P}^2$}}, {\emph{J. Reine Angew. Math} {\bf 453} (1994)
  193--220}.

\bibitem{Bringmann:2010sd}
K.~Bringmann and J.~Manschot, \emph{{From sheaves on $\mathbb{P}^2$ to a
  generalization of the Rademacher expansion}}, {\emph{Am. J. of Math.} {\bf
  135} (2013) 1039--1065}, [\href{https://arxiv.org/abs/1006.0915}{{\tt
  1006.0915}}].

\bibitem{Gottsche:1996}
L.~G\"ottsche, \emph{Modular forms and {D}onaldson invariants for 4-manifolds
  with $b_2^+= 1$}, {\emph{Journal of the American Mathematical Society} {\bf
  9} (1996) 827--843}, [\href{https://arxiv.org/abs/alg-geom/9506018}{{\tt
  alg-geom/9506018}}].

\bibitem{Serre}
J.~P. Serre, \emph{A course in arithmetic}.
\newblock Graduate Texts in Mathematics, no. 7, Springer, New York, 1973.

\bibitem{Zagier92}
D.~Zagier, \emph{Introduction to modular forms; From Number Theory to Physics}.
\newblock Springer, Berlin (1992), pp. 238-291, 1992.

\bibitem{Bruinier08}
G.~H. J.H.~Bruinier, G. van der~Geer and D.~Zagier, \emph{The 1-2-3 of Modular
  Forms}.
\newblock Springer-Verlag Berlin Heidelberg, 2008,
  \href{http://dx.doi.org/10.1007/978-3-540-74119-0}{10.1007/978-3-540-74119-0}.

\bibitem{Vigneras:1977}
M.-F. Vign\'eras, \emph{{S\'eries th\^eta des formes quadratiques
  ind\'efinies}}, {\emph{Springer Lecture Notes} {\bf 627} (1977) 227 -- 239}.

\end{thebibliography}
\end{document}